\documentclass[10pt,twocolumn,showpacs,preprintnumbers,amsmath,amssymb,aps,prb,longbibliography,superscriptaddress]{revtex4-1}
\usepackage{mathrsfs}
\usepackage{graphicx}
\usepackage{dcolumn}
\usepackage{bm}
\usepackage{amsmath}
\usepackage{amsfonts}
\usepackage{color}
\usepackage{hyperref}

\begin{document}

\title{Conserved Quantities from Entanglement Hamiltonian}

\author{Biao Lian}
\affiliation{Department of Physics, Princeton University, Princeton, New Jersey 08544, USA}

\begin{abstract}
We show that the subregion entanglement Hamiltonians of excited eigenstates of a quantum many-body system are approximately linear combinations of subregionally (quasi)local approximate conserved quantities, with relative commutation errors $\mathcal{O}\left(\frac{\text{subregion boundary area}}{\text{subregion volume}}\right)$. By diagonalizing an entanglement Hamiltonian superdensity matrix (EHSM) for an ensemble of eigenstates, we can obtain these conserved quantities as the EHSM eigen-operators with nonzero eigenvalues. For free fermions, we find the number of nonzero EHSM eigenvalues is cut off around the order of subregion volume, and some of their EHSM eigen-operators can be rather nonlocal, although subregionally quasilocal. In the interacting XYZ model, we numerically find the nonzero EHSM eigenvalues decay roughly in power law if the system is integrable, with the exponent $s\approx 1$ ($s\approx 1.5\sim 2$) if the eigenstates are extended (many-body localized). For fully chaotic systems, only two EHSM eigenvalues are significantly nonzero, the eigen-operators of which correspond to the identity and the subregion Hamiltonian.
\end{abstract}

\date{\today}

\maketitle

\section{Introduction}

Conserved quantities significantly affect the integrability and non-equilibrium dynamics of a quantum many-body system, for instance, they may lead the system to equilibrate into a non-thermal state described by a generalized Gibbs ensemble \cite{rigol2007,rigol2008,cassidy2011,caux2012,caux2013,vidmar2016,dymarsky2019}. 
In quantum systems, the generic belief is that only local and quasilocal conserved quantities contribute to the quantum integrability. However, unlike classical systems, it is not clear how many (quasi)local conserved quantities are needed for a quantum system to be integrable. Moreover, it is difficult to identify all the local and quasilocal conserved quantities even for exactly solvable models \cite{tetelman1981,grabowski1995,ilievski2015,nozawa2020}, and it is unclear which of them contribute to the quantum integrability.

Two frequently employed indicators distinguishing between chaotic and integrable systems are the level spacing statistics (LSS) \cite{berry1977,bohigas1984} and the eigenstate thermalization hypothesis (ETH) \cite{jensen1985,deutsch1991,srednicki1994,srednicki1999,dalessio2016}. However, neither LSS nor ETH can give much information (e.g., conserved quantities) about a quantum system which is not fully chaotic. Another feature of quantum chaos is the Lyapunov exponent in the out-of-time-ordered correlation \cite{maldacena2016,murthy2019b}. This, however, usually requires certain large flavor limit, for instance in the Sachdev-Ye-Kitaev type models \cite{sachdev1992fk,polchinski2016xgd,maldacena2016hyu,kitaev2017awl,lian2019syk}.

Here we ask, given a set of many-body eigenstates of a quantum system, can one obtain the (quasi)local conserved quantities and tell the integrability of the system? Previous studies show that if a Hamiltonian is strictly local, it can be recovered (up to local conserved quantities) if an exact single eigenstate is known \cite{qi2019}. Besides, ETH suggests that the subregion entanglement Hamiltonian of fully chaotic systems resembles the physical subregion Hamiltonian \cite{garrison2018,murthy2019,lu2019}. For generic systems, one expects other conserved quantities may also contribute to the entanglement Hamiltonian \cite{murthy2019,lu2019}, but which conserved quantities contribute has not been carefully studied. In this letter, we show that the subregion entanglement Hamiltonians of excited eigenstates of a quantum system are the linear combinations of \emph{subregionally (quasi)local} approximate conserved quantities, with relative mutual commutation errors $\mathcal{O}\left(\frac{\text{subregion boundary area}}{\text{subregion volume}}\right)$. We define an entanglement Hamiltonian superdensity matrix (EHSM) for a given ensemble of eigenstates, and show that the eigen-operators of EHSM with nonzero eigenvalues resemble the subregionally (quasi)local conserved quantities. For free fermions \cite{anderson1958}, we find the number of nonzero EHSM eigenvalues is proportional to the subregion volume, with the coefficient depending on whether the free fermion eigenstates are extended or localized. In particular, for extended free fermions, we reveal that the entanglement Hamiltonians contain a set of rather nonlocal conserved quantities, although they still satisfies the definition of subregional quasilocality. 
We further study the interacting 1D XYZ model with or without disorders (within sizes calculable), for which we find the $n$-th largest EHSM eigenvalue decays as $n^{-s}$ if the system is integrable. The exponent $s\approx 1$ if the many-body eigenstates are delocalized, and $s\approx 1.5\sim 2$ if the system shows many-body localization (which is arguably integrable) \cite{basko2006,gornyi2005,oganesyan2007,marko2008,pal2010,serbyn2013,huse2014,chandran2015,ros2015}. If the system is fully chaotic, only two EHSM eigenvalues are significantly nonzero, corresponding to the only two (quasi)local subregion conserved quantities: the identity and the physical Hamiltonian as suggested by the ETH. We conjecture that the conserved quantities in EHSM are those governing the quantum integrability behaviors of a system.

The rest of the paper is organized as follows. In Sec. \ref{sec:II}, we give the arguments and criteria for subregionally quasilocal conserved quantities in the eigenstate entanglement Hamiltonians. In Sec. \ref{sec:III}, we define the EHSM for calculating the conserved quantities. In Sec. \ref{sec:IV}, we investigate the conserved quantities in the eigenstate entanglement Hamiltonians of free fermion models in different spatial dimensions, and verify the validity of our generic criterion of subregional quasilocality. Sec. \ref{sec:V} is devoted to an exact diagonalization study of the EHSM and conserved quantities of the interacting XYZ model, which has both quantum integrable and chaotic phases. Lastly, we summarize and discuss the possible generalization to time-evolution problems in Sec. \ref{sec:VI}.


\section{Approximate Conserved Quantities in the Entanglement Hamiltonian}\label{sec:II}

\begin{figure}[tbp]
\begin{center}
\includegraphics[width=3.4in]{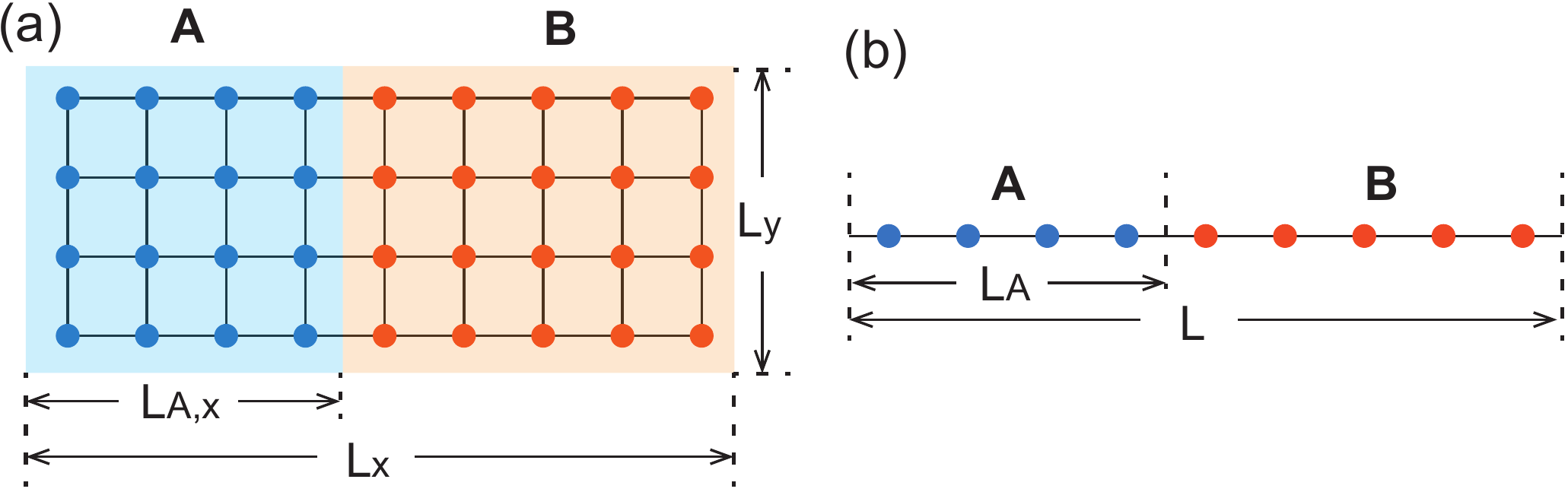}
\end{center}
\caption{Illustration of a system in (a) a 2D square lattice and (b) a 1D lattice divided into two subregions $A$ and $B$. 
}
\label{fig1}
\end{figure}

We shall consider quantum systems in lattices, and assume each lattice site has a finite Hilbert space dimension $d$. Consider a system in a finite real space region with $L$ sites, which has a Hilbert space dimension $N=d^L$. Assume the system has a local $N\times N$ many-body Hamiltonian $H$ in this region, and has eigenstates $|\alpha\rangle$:
\begin{equation}
H|\alpha\rangle=E_\alpha|\alpha\rangle\ ,
\end{equation}
where $E_\alpha$ is the energy of eigenstate $|\alpha\rangle$ ($1\le\alpha\le N$). We divide this region into two subregions $A$ and $B$ with number of sites $L_A$ and $L_B=L-L_A$ (Fig. \ref{fig1}), which have Hilbert space dimensions $N_A=d^{L_A}$ and $N_B=d^{L_B}$, respectively. We denote the boundary number of sites between $A$ and $B$ as $l_{AB}$. In $D$ spatial dimensions, if the linear size of the system is of order $l_x$, one generically has $L_A,L_B\propto l_x^D$, and $l_{AB}\propto l_x^{D-1}$. The Hamiltonian $H$ can then be divided into 
\begin{equation}\label{eq-H-HAB}
H=H_A\otimes I_B+I_A\otimes H_B+H_{AB}, 
\end{equation}
where $I_A$ and $I_B$ are the identity matrix in $A$ and $B$ subregions, $H_A\otimes I_B$ ($I_A\otimes H_B$) contains all the product terms with supports within subregion $A$ ($B$) (including the identity term $I_A\otimes I_B$), while $H_{AB}$ denotes all the product terms with supports across subregions $A$ and $B$. Here a product term is defined as the product $\prod_{j\in \mathcal{S}} O_{j}$ of \emph{traceless} on-site operators $O_{j}$ (hence the identity operator is not included) of a set of sites $j\in\mathcal{S}$, and the set of sites $\mathcal{S}$ is called the \emph{support}. Note that $\text{tr}(H_{AB})=0$. Therefore, $H_A$ and $H_B$ can be understood as the bulk Hamiltonian of subregions $A$ and $B$, while $H_{AB}$ is the boundary coupling between subregions $A$ and $B$. 


For a given eigenstate $|\alpha\rangle$ of the entire system, the reduced density matrix in subregion $A$ is
\begin{equation}\label{eq-rhoA}
\rho_{A}(\alpha)=\text{tr}_B |\alpha\rangle\langle \alpha| =e^{-H_{E}^A(\alpha)}\ .
\end{equation}
$H_E^A(\alpha)$ is the \emph{entanglement Hamiltonian} \cite{lihui2008} of eigenstate $|\alpha\rangle$. 

Assume the eigenbasis of the two subregions are defined by $H_A|\alpha_A,A\rangle=E_{\alpha_A}^A|\alpha_A,A\rangle$ and $H_B|\alpha_B,B\rangle=E_{\alpha_B}^B|\alpha_B,B\rangle$, where $E_{\alpha_A}^A$ and $E_{\alpha_B}^B$ are the eigenenergies, and $1\le \alpha_A\le N_A, 1\le \alpha_B\le N_B$. If the eigenstate $|\alpha\rangle$ of the entire system under the subregion eigenbasis has wavefunction
\begin{equation}\label{eq-alpha-u}
|\alpha\rangle=\sum_{\alpha_A,\alpha_B} u_{\alpha,\alpha_A,\alpha_B} |\alpha_A,A\rangle\otimes|\alpha_B,B\rangle\ ,
\end{equation}
the elements of $\rho_A(\alpha)$ will be 
\begin{equation}\label{eq:rhoalpha}
\langle \alpha_A,A|\rho_A(\alpha)|\alpha_A',A\rangle=\sum_{\alpha_B=1}^{N_B} u_{\alpha,\alpha_A,\alpha_B}u_{\alpha,\alpha_A',\alpha_B}^*\ .
\end{equation}
In the below, we examine the relation between $H_E^A(\alpha)$ and conserved quantities.


\subsection{Fully chaotic systems}

We start by considering fully many-body chaotic systems with a local Hamiltonian, for which the ETH holds. Approximately, the coupling $H_{AB}$ in the subregion eigenbasis $|\alpha_A,A\rangle\otimes|\alpha_B,B\rangle$ will have matrix elements $(H_{AB})_{\alpha_A'\alpha_B';\alpha_A\alpha_B}=\delta_{\alpha_A,\alpha_A'}\delta_{\alpha_B,\alpha_B'}E^{(\text{d})}(E_{\alpha_A}^A, E_{\alpha_B}^B)+h^{\text{(off)}}_{\alpha_A'\alpha_B';\alpha_A\alpha_B}$, where $E^{(\text{d})}$ is the diagonal term, while $h^{\text{(off)}}_{\alpha_A'\alpha_B';\alpha_A\alpha_B}$ is a random off-diagonal matrix decaying exponentially in $|E_{\alpha_A}^A-E_{\alpha_A'}^A|$ and $|E_{\alpha_B}^B-E_{\alpha_B'}^B|$ \cite{dalessio2016}. When the boundary size $l_{AB}\ll L_A,L_B$, one can show that the wavefunction of an excited state $|\alpha\rangle$ approximately satisfies (App. \ref{app:chaotic})
\begin{equation}
u_{\alpha,\alpha_A,\alpha_B}u^*_{\alpha,\alpha_A',\alpha_B'} \propto  \delta_{\alpha_A,\alpha_A'}\delta_{\alpha_B,\alpha_B'} \delta(E_\alpha-E_{\alpha_A}^A-E_{\alpha_B}^B-E^{(\text{d})}) 
\end{equation}
under the random average of $h^{\text{(off)}}$, which determines the reduced density matrix $\rho_A(\alpha)$ (by Eq. (\ref{eq:rhoalpha})). The width of the delta function $\propto l_{AB}$, while generically $E_{\alpha_A}^A\propto L_A$, $E_{\alpha_B}^B\propto L_B$, and $E^{(\text{d})}\propto l_{AB}$. 


When the boundary size $l_{AB}\ll L_A\ll L_B$, as studied in literature \cite{deutsch1991,srednicki1994,srednicki1999,garrison2018,murthy2019,lu2019} and re-derived in App. \ref{app:chaotic}, the entanglement Hamiltonian of subregion $A$ of excited states $|\alpha\rangle$ reads approximately (up to boundary terms)
\begin{equation}\label{eq-KA-chaotic}
H_E^A(\alpha)\approx \beta_A^{(0)}(\alpha)I_A +\beta_A^{(1)}(\alpha)(H_A-E_{av}^AI_A)\ ,
\end{equation}
where $E_{av}^A=\text{tr}(H_A)/N_A$ is the average energy of the subregion Hamiltonian $H_A$. In other words, $H_E^A(\alpha)$ resembles the physical Hamiltonian $H_A$, which is the only local subregion conserved quantity for the fully chaotic system. The coefficients to the zeroth order of $\frac{l_{AB}}{L_A}$ are given by $\beta^{(0)}_A(\alpha)=\log \left[\frac{N_A\Omega(E_\alpha)}{\Omega_B(E_\alpha-E^A_{av})}\right]$, and $\beta_A^{(1)}(\alpha)\approx\frac{\text{d}\log \Omega_B(E)}{\text{d}E}|_{E=E_\alpha-E^A_{av}}$, where $\Omega(E)$ and $\Omega_B(E)$ are the normalized densities of states of Hamiltonians $H_A\otimes I_B+I_A\otimes H_B$ and $H_B$, respectively. 


\subsection{Generic systems}

For a generic quantum system which is not fully chaotic, we assume there are linearly independent Hermitian conserved quantities $Q^{(n)}$ ($n\ge0$) satisfying 
\begin{equation}
[Q^{(m)},Q^{(n)}]=[H,Q^{(n)}]=0\ . 
\end{equation}
The Hamiltonian $H$ is the linear combination of some $Q^{(n)}$, and the energy eigenstates $|\alpha\rangle$ can be simultaneously eigenstates of $Q^{(n)}$. Without loss of generality, we define $Q^{(0)}=I$ as the identity matrix, and assume $(Q^{(m)}, Q^{(n)})=0$ if $m\neq n$ (which indicates $\text{tr}(Q^{(n)})=0$ for $n\ge1$). 

For later convenience, we define 
\begin{equation}
(M,M')=\text{tr}(M^\dag M')\ ,\qquad ||M||=\sqrt{(M,M)}\ ,
\end{equation}
as the \emph{Frobenius} (\emph{Hilbert-Schmidt}) \emph{inner product} of operators (matrices) $M$ and $M'$, and the \emph{Frobenius} \emph{norm} of operator $M$, respectively.

Generically, there are always $N=d^L$ conserved quantities given by the linear combinations of eigenstate projection operators $|\alpha\rangle\langle\alpha|$, most of which are nonlocal. To characterize their locality, similar to Eq. (\ref{eq-H-HAB}), we decompose each $Q^{(n)}$ ($n\ge1$) as
\begin{equation}
Q^{(n)}=Q^{(n)}_A\otimes I_B+I_A\otimes Q^{(n)}_B+Q^{(n)}_{AB}\ ,
\end{equation}
where $Q^{(n)}_A\otimes I_B$ ($I_A\otimes Q^{(n)}_B$) consists of product terms with supports in subregion $A$ ($B$), while $Q^{(n)}_{AB}$ contains product terms with supports across subregions $A$ and $B$.
We then define a conserved quantity $Q^{(n)}$ as \emph{subregionally quasilocal} in subregion $A$ if and only if it satisfies 
\begin{equation}\label{eq:sublocal-cond}
\frac{||Q^{(n)}_{AB}||}{||Q^{(n)}_{A}\otimes I_B||}=\mathcal{O} \left(\sqrt{\frac{l_{AB}}{L_A}}\right) \ ,
\end{equation}
when $l_{AB}\ll L_A,L_B$ ($\mathcal{O}(x)$ denotes up to order $x$), and similarly for subregion $B$. For an extended local conserved quantity $Q^{(n)}$, Eq. (\ref{eq:sublocal-cond}) can be seen by noting that $Q^{(n)}_{AB}$ consists of order $l_{AB}$ local terms, while $Q^{(n)}_{A}\otimes I_B$ contains order $L_A$ local terms. Instead, if $Q^{(n)}$ is a localized conserved quantity in $A$, the error $\frac{||Q^{(n)}_{AB}||}{||Q^{(n)}_{A}\otimes I_B||}$ will be exponentially small ($\sim e^{-cL_A/l_{AB}}$), and Eq. (\ref{eq:sublocal-cond}) will be an overestimate.

If Eq. (\ref{eq:sublocal-cond}) (or similar condition for subregion $B$) is satisfied, we can treat $Q_A^{(n)}$ ($Q_B^{(n)}$) as approximate conserved quantities in subregions $A$ ($B$) when $l_{AB}\ll L_A,L_B$, respectively. Then, similar to the argument of Eq. (\ref{eq-KA-chaotic}) for fully chaotic systems, we can argue that (App. \ref{app:multipleQ}) the subregion entanglement Hamiltonian $H_E^A(\alpha)$ is approximately given by (up to boundary terms)

\begin{equation}\label{eq-KA-conserv}
H_E^A(\alpha)\approx \beta^{(0)}_A(\alpha)I_A+\sum_{n\in\text{Loc}}\beta_A^{(n)}(\alpha)Q_A^{(n)} \ ,
\end{equation}
where $n\in\text{Loc}$ runs over all subregionally quasilocal conserved quantities in $A$ ($n>0$), and the coefficients $\beta^{(n)}_A(\alpha)$ are estimated in App. \ref{app:multipleQ} Eq. (\ref{seq:betacoefficients}).

The subregional quasilocality of Eq. (\ref{eq:sublocal-cond}) is equivalent to the following relative commutation error requirement (App. \ref{sec:non-comm}): $\forall \ n,m\in\text{Loc}$ contributing to Eq. (\ref{eq-KA-conserv}),
\begin{equation}\label{eq-QA-criterion}
\frac{|| [H_A,Q^{(n)}_{A}] ||}{||H_A Q^{(n)}_{A}||}\sim \frac{|| [Q^{(n)}_{A},Q^{(m)}_{A}] ||}{||Q^{(n)}_{A} Q^{(m)}_{A}||}=\mathcal{O}\left( \frac{l_{AB}}{L_A}\right)\ .
\end{equation}
We conjecture Eq. (\ref{eq-QA-criterion}) is the generic criterion for $Q^{(n)}_{A}$ to contribute to Eq. (\ref{eq-KA-conserv}). If $Q^{(n)}_{A}$ is a localized conserved quantity in subregion $A$, Eq. (\ref{eq-QA-criterion}) is an overestimation, and the error will be exponentially small ($\sim e^{-cL_A/l_{AB}}$). Compared to Eq. (\ref{eq:sublocal-cond}), the criterion of Eq. (\ref{eq-QA-criterion}) is sometimes more convenient, since it only involves operators within subregion $A$.

While $H_E^A(\alpha)$ as summation of local conserved quantities has been proposed in literature \cite{deutsch1991,srednicki1994,srednicki1999,murthy2019,lu2019}, here we emphasize on two key observations which are not discussed before: (i) The contributing conserved quantities $Q^{(n)}_{A}$ in Eq. (\ref{eq-KA-conserv}) only approximately mutually commute up to Eq. (\ref{eq-QA-criterion}); (ii) they only need be subregionally quasilocal, which could be rather nonlocal in subregion $A$. This can be explicitly seen in the free fermion example discussed in Sec. \ref{sec:free-fermion} below.



\section{Entanglement Hamiltonian Superdensity Matrix}\label{sec:III}

Eq. (\ref{eq-KA-conserv}) allows us to numerically recover the subregionally (quasi)local conserved quantities $Q_A^{(n)}$ from a set of entanglement Hamiltonians of full system eigenstates. Note that an entanglement Hamiltonian $H_E^A(\alpha)$ can be regarded as a vector $|H_E^A(\alpha))$ in the linear space of $N_A\times N_A$ matrices. Given the entanglement Hamiltonians of eigenstates $|\alpha\rangle$ in an ensemble $\Xi$, we can define an \emph{entanglement Hamiltonian superdensity matrix} (EHSM) of size $N_A^2\times N_A^2$:
\begin{equation}\label{eq-EHSM1}
R_A=\sum_{\alpha\in\Xi}\frac{w_\alpha}{N_A}|H_E^A(\alpha)) ( H_E^A(\alpha)|,
\end{equation}
where $w_\alpha > 0$ is the weight of state $|\alpha\rangle$ ($\sum_{\alpha\in\Xi}w_\alpha=1$). We can then diagonalize the EHSM $R_A$ into
\begin{equation}\label{eq-EHSM2}
R_A=\sum_{n\ge0} p_{A,n}|\overline{Q}_A^{(n)})( \overline{Q}_A^{(n)}|\ ,
\end{equation}
where $p_{A,n}\ge0$ is the $n$-th eigenvalue ($n\ge0$) of $R_A$ (in descending order), and $\overline{Q}_A^{(n)}$ is the normalized eigen-operator satisfying $( \overline{Q}_A^{(m)},\overline{Q}_A^{(n)})=\text{tr}(\overline{Q}_A^{(m)\dag}\overline{Q}_A^{(n)})=\delta_{mn}$. We expect $\overline{Q}_A^{(n)}$ with $p_{A,n}>0$ to resemble the normalized subregionally (quasi)local conserved quantities in subregion $A$. An extensive conserved quantity $Q_A^{(n)}$ in physical units will scale as $Q_A^{(n)}\sim \sqrt{N_AL_A}\  \overline{Q}_A^{(n)}$.

The EHSM $R_A$ in Eq. (\ref{eq-EHSM1}) is a huge matrix to diagonalize. However, if the number of known eigenstates (the size of ensemble $\Xi$) is much smaller than the size of matrix $R_A$, namely, $N_\Xi\ll N_A^2$, the matrix $R_A$ will only have a rank up to $N_\Xi$. Accordingly, $R_A$ can be easily diagonalized by diagnalizing a much smaller $N_\Xi\times N_\Xi$ correlation matrix 
\begin{equation}
K_{A,\alpha\beta}=\frac{\sqrt{w_\alpha w_\beta}}{N_A}( H_E^A(\alpha),H_E^A(\beta))\ . 
\end{equation}
It can be proved (App. \ref{app:EHSMdiag}) that $R_A$ and $K_A$ have exactly the same nonzero eigenvalues $p_{A,n}$, and each eigenvector $v_{n}^A$ of $K_A$ (satisfying $K_Av_{n}^A=p_{A,n}v_{n}^A$) corresponds to a normalized eigen-operator of $R_A$ of the same eigenvalue: 
\begin{equation}
\overline{Q}_A^{(n)}=\frac{1}{\sqrt{N_A p_{A,n}}}\sum_\alpha \sqrt{w_\alpha}v_{n,\alpha}^A H_E^A(\alpha) \ .
\end{equation}
In the below, we study the EHSM eigenvalues and eigenoperators of several models.


\section{Conserved quantities of free fermions}\label{sec:free-fermion}\label{sec:IV}

In this section, we investigate the conserved quantities in the entanglement Hamiltonians of free fermion eigenstates. Free fermion models are many-body integrable by solving their single-particle spectra. We consider the Anderson model \cite{anderson1958} in both 2D square lattice and 1D lattice as shown in Fig. \ref{fig1}, with the Hamiltonian
\begin{equation}\label{eq-freefermion-H}
H=-t\sum_{\langle ij\rangle}(c_{\mathbf{r}_i}^\dag c_{\mathbf{r}_j}+h.c.)+\sum_{j} \mu_j c_{\mathbf{r}_j}^\dag c_{\mathbf{r}_j}\ ,
\end{equation}
where $t$ (real) is the nearest neighbor hopping, $\mu_j$ is an on-site random potential within an interval $[-W,W]$, and periodic boundary condition is imposed. Generically, the entanglement Hamiltonian of free fermion eigenstates $|\alpha\rangle$ in suregion $A$ takes the fermion bilinear form\cite{peschel2003}
\begin{equation}
H_E^A(\alpha)=\gamma_{A}(\alpha)I_A+\sum_{ij\in A}\kappa_{A,ij}(\alpha)c^\dag_{\mathbf{r}_i}c_{\mathbf{r}_j}\ ,
\end{equation}
where $\gamma_{A}(\alpha)$ and matrix $\kappa_{A,ij}(\alpha)$ can be calculated from the correlation matrix $\mathcal{C}_{A,ij}(\alpha)=\langle \alpha|c^\dag_{\mathbf{r}_i}c_{\mathbf{r}_j}|\alpha\rangle$ (see App. \ref{sec:EHSM-free}). We first diagonalize the single-particle Hamiltonian in Eq. (\ref{eq-freefermion-H}), then randomly choose an ensemble $\Xi$ of $N_\Xi=1000$ many-body Fock eigenstates $|\alpha\rangle$ of the entire system, with weight $w_\alpha=\frac{1}{N_\Xi}$, and diagonalize the EHSM of their entanglement Hamiltonians to extract the subregionally quasilocal conserved quantities. Generically, we find the EHSM eigenvalues $p_{A,n}$ drops to zero at $n=zL_A$ for some coefficient $z$, as we will show below.

\subsection{1D free femions}

In 1D, we take the total system size of the free fermion model as $L=500$. For different $A$ subregion volumes $L_A=10,30,50$, we diagonalize the EHSM of an ensemble $\Xi$ of $N_\Xi=1000$ randomly chosen Fock eigenstates. Generically, we find the leading eigenvalue $p_{A,0}$ corresponds to the identity operator $\overline{Q}_A^{(0)}\approx I_A/\sqrt{N_A}$. The other EHSM eigen-operators depend on whether the free fermions are extended or localized.

\begin{figure}[tbp]
\begin{center}
\includegraphics[width=3.4in]{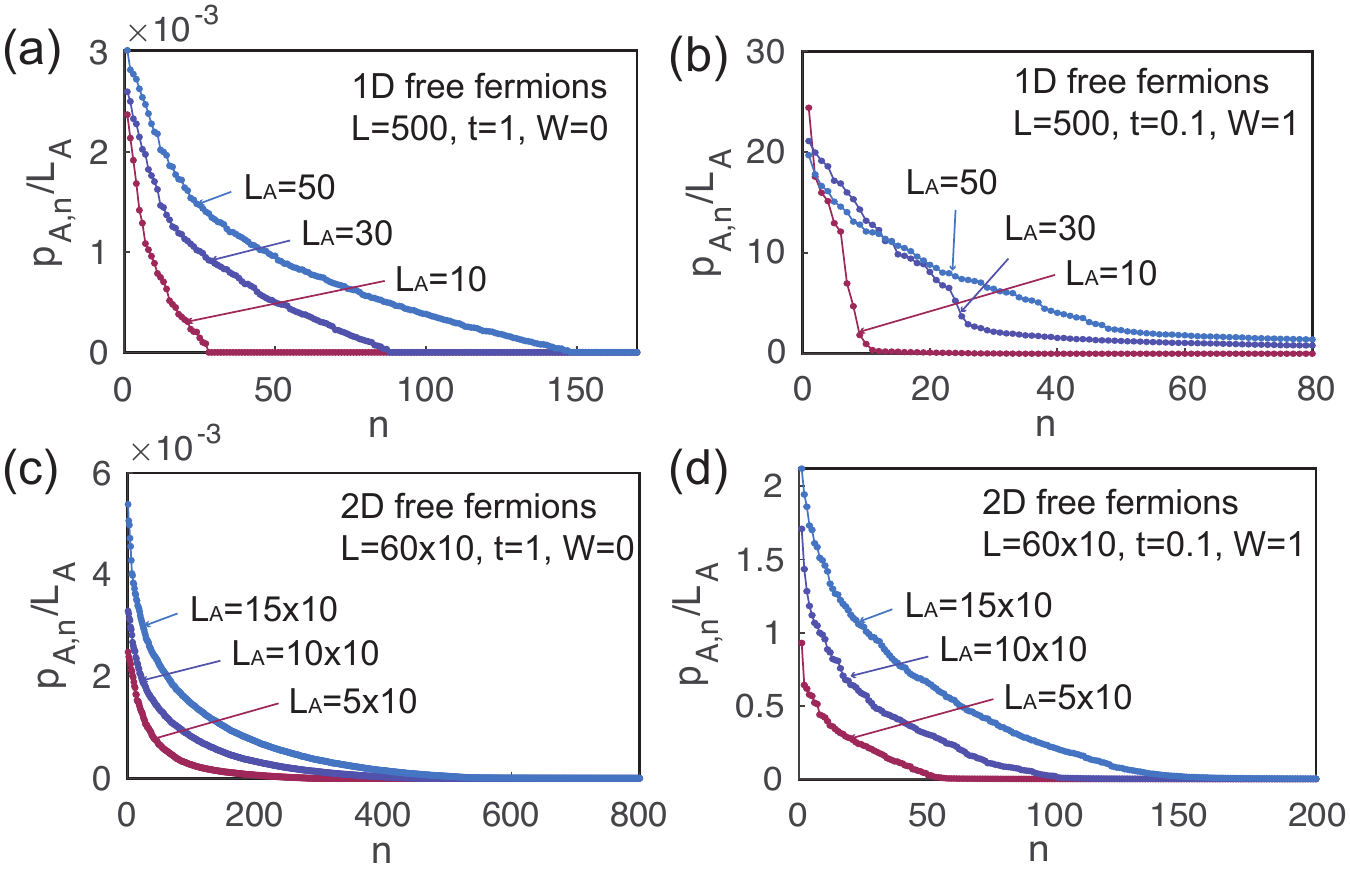}
\end{center}
\caption{The EHSM eigenvalues $p_{A,n}$ for free fermions (parameters given in panels), calculated for $N_\Xi=1000$ randomly chosen eigenstates. The model is on (see Fig. \ref{fig1}) (a)-(b) a 1D lattice with total size $L=500$ and $A$ subregion size $L_A=10,30,50$; (c)-(d) a 2D square lattice with total size $L_x=60$, $L_y=10$, $L=L_xL_y=600$ and $A$ subregion volume $L_A=L_{A,x}L_y=50,100,150$.
}
\label{fig2}
\end{figure}

\begin{figure}[tbp]
\begin{center}
\includegraphics[width=3.4in]{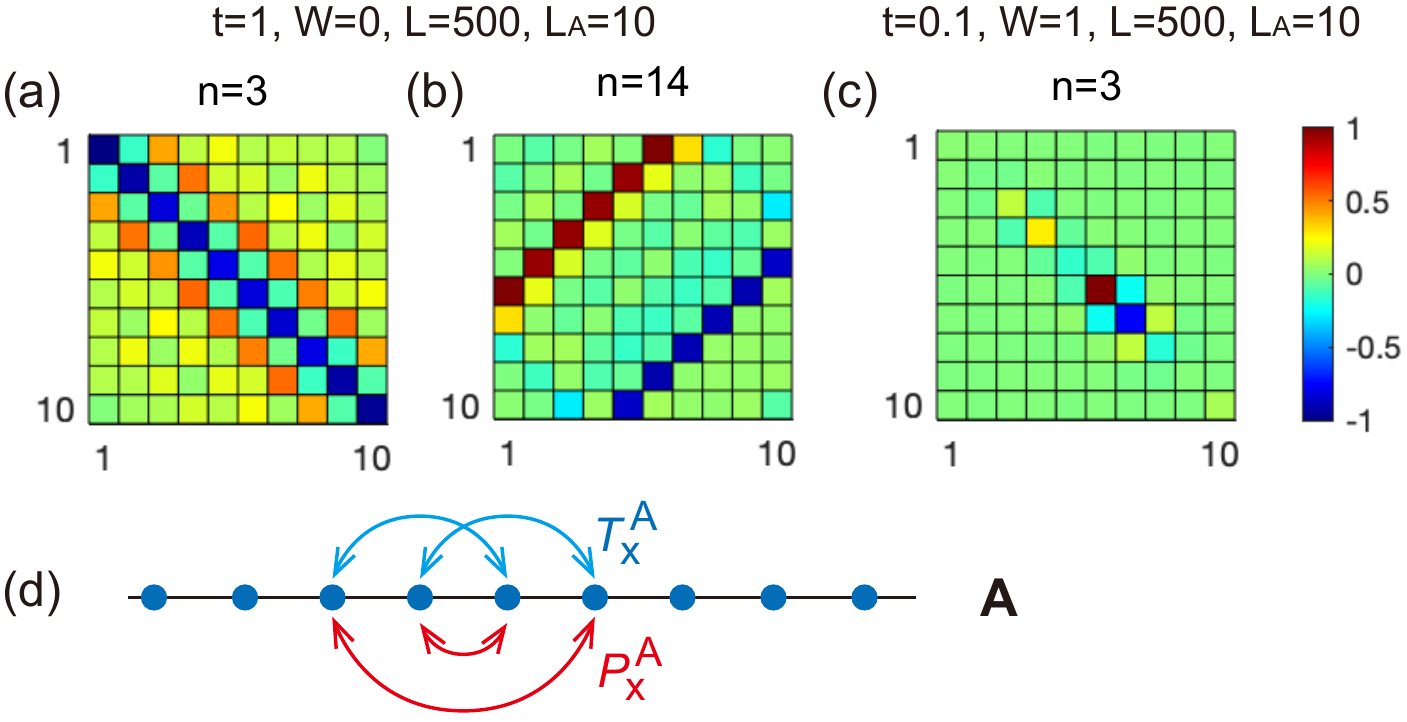}
\end{center}
\caption{The EHSM eigen-operators of 1D free fermions ($L=500, L_A=10$) take the fermion bilinear form of $\overline{Q}_A^{(n)}=\gamma_A^{(n)}I_A+ \sum_{i,j\in A}\kappa_{A,ij}^{(n)}c^\dag_{x_i} c_{x_j}$, and the values of matrices $\kappa_{A,ij}^{(n)}$ (normalized by its largest element) for given model parameters are plotted in (a)-(c), where the axis are sites $x_i$ and $x_j$. The parameters for (a)-(b) are $t=1,W=0$ (extended fermions), and for (c) are $t=0.1,W=1$ (localized fermions). (d) illustrates the hoppings in the conserved quantities $T_x^A$ in Eq. (\ref{eq:TA}) and $P_x^A$ in Eq. (\ref{eq:PA}), respectively.
}
\label{fig1DQn}
\end{figure}

When the 1D single-fermion wavefunctions are extended, the EHSM eigenvalues of subregion $A$ are as shown in Fig. \ref{fig2}(a), where we have set $t=1$ and $W=0$. We find $p_{A,n}$ for 1D extended fermions drops to zero around $n=3L_A$, indicating the presence of $3L_A$ approximately conserved quantities. This cutoff remains robust for week disorders $W$, provided the Anderson localization length is larger than the subregion $A$ size $L_A$. Numerically, as shown in Fig. \ref{fig1DQn}(a)-(b), the $3L_A$ eigen-operators $\overline{Q}_A^{(n)}$ ($n>0$) with nonzero EHSM eigenvalues $p_{A,n}$ are approximately linear combinations of the following $3L_A$ operators:
\begin{equation}\label{eq:TA}
T^A_{x}=\sum_{x_i,x_i+x\in A} (c^\dag_{x_i+x}c_{x_i}+c^\dag_{x_i}c_{x_i+x} )\ ,
\end{equation}
with $0\le x< L_{A}$, and 
\begin{equation}\label{eq:PA}
P^A_{x}=\sum_{x_i,x-x_i\in A} c^\dag_{x-x_i}c_{x_i} \ ,
\end{equation}
with $2\le x\le 2L_{A}$. As shown in App. \ref{sec:EHSM-free-1D}, they are indeed approximate conserved quantities in subregion $A$ (which has open boundaries) satisfying the criterion of Eq. (\ref{eq-QA-criterion}). $T^A_x$ are the Fourier transforms of the single-particle momenta. $P^A_x$ come from the Fourier transform of the hoppings between momentum $k$ and $-k$ fermion states, which are conserved since the fermion energy does not change under $k\rightarrow -k$. Remarkably, unlike the naive expectation that only local conserved quantities contribute to the entanglement Hamiltonian, the $2L_A$ operators $P^A_{x}$ are fairly nonlocal within subregion $A$, with non-decaying hoppings between two sites with a fixed center (Fig. \ref{fig1DQn}(d)). However, $P^A_{x}$ are still subregionally quasilocal, satisfying the criterion of Eq. (\ref{eq-QA-criterion})). 


Where all the 1D single-particle eigenstates are strongly localized, the EHSM eigenvalues $p_{A,n}$ are as shown in Fig. \ref{fig2}(b) ($t=0.1$, $W=1$), which drops significantly towards zero around a cutoff $n=L_A$. Accordingly, except for $\overline{Q}_A^{(0)}\approx I_A/\sqrt{N_A}$, the eigen-operators $\overline{Q}_A^{(n)}$ ($0< n< L_A$) give approximately the occupation number operators (approximately $c_{x_i}^\dag c_{x_i}$) of the $L_A$ localized single-particle eigenstates (Fig. \ref{fig1DQn}(c), see also App. \ref{sec:EHSM-free-1D}).

\subsection{2D free femions}

We further examine a 2D system with a fixed total system size $L=L_xL_y$, with $L_x=60$ and $L_y=10$. The subregions are defined as shown in Fig. \ref{fig1}(a), with different $A$ subregion volumes $L_A=L_{A,x}L_y=10L_{A,x}$, and $L_{A,x}=5,10,15$. Again, generically, we find $\overline{Q}_A^{(0)}\approx I_A/\sqrt{N_A}$, and the rest EHSM eigen-operators are different for extended/localized fermions.

When the 2D single-fermion wavefunctions are extended (or when the localization length is larger than the linear size of subregion $A$), the EHSM eigenvalues $p_{A,n}$ are as shown in Fig. \ref{fig2}(c) (where we set $t=1$ and $W=0$). We find the eigenvalues $p_{A,n}$ also drops to zero at some cutoff $n=zL_A$. However, different from the 1D case where $z$ is fixed at $3$, in the 2D case here we find $z$ depends on the aspect ratio of subregion $A$: 
\begin{equation}
z\rightarrow
\begin{cases}
& 7\ ,\qquad (\frac{L_{A,x}}{L_y}\ll 1) \\
& 3\ ,\qquad (\frac{L_{A,x}}{L_y}\gg 1)
\end{cases}
\end{equation}
Generically, $3<z<7$. By both numerical and analytical investigations (see App. \ref{sec:EHSM-free-2D}), we arrive at the following explanation for this aspect ratio dependent behavior: First, similar to the 1D case, there always exists $3L_A$ approximately mutually commuting conserved quantities (i.e., satisfying Eq. (\ref{eq-QA-criterion})) given by:
\begin{equation}\label{eq-TP-free-2D}
\begin{split}
&T_{x,y}^A\approx \sum_{x_i,x+x_i\in A} (c^\dag_{x_i+x,y_i+y}c_{x_i,y_i}+c^\dag_{x_i,y_i+y}c_{x_i+x,y_i} )\ , \\
&P_{x,y}^{A,1}\approx \sum_{x_i,x-x_i\in A} c^\dag_{x-x_i,y_i+y}c_{x_i,y_i}\ ,
\end{split}
\end{equation}
where $0\le x\le L_{A,x}, 0\le y <L_y$ for $T_{x,y}^A$, and $2\le x\le 2L_{A,x}, 0\le y <L_y$ for $P_{x,y}^{A,1}$. Note that $P_{x,y}^{A,1}$ is nonlocal in the $x$ direction. In addition, one can show that there are other $4L_A$ operators which approximately commute with $H_A$, given by
\begin{equation}\label{eq-TP-free-2D-2}
\begin{split}
&P_{x,y}^{A,2}\approx \sum_{x_i,x-x_i\in A} c^\dag_{x_i+x,y-y_i}c_{x_i,y_i}\ , \\
&P_{x,y}^{A,3}\approx \sum_{x_i,x-x_i\in A} c^\dag_{x-x_i,y-y_i}c_{x_i,y_i}\ ,
\end{split}
\end{equation}
where $-L_{A,x}< x< L_{A,x},\ 0\le y<L_y$ for $P_{x,y}^{A,2}$, and $2\le x\le 2L_{A,x},\ 0\le y<L_y$ for $P_{x,y}^{A,3}$. Note that $P_{x,y}^{A,2}$ is nonlocal in the $y$ direction, while $P_{x,y}^{A,3}$ is nonlocal in both $x$ and $y$ directions. However, the $4L_A$ operators in Eq. (\ref{eq-TP-free-2D-2}) do not always approximately commute with the $3L_A$ operators in Eq. (\ref{eq-TP-free-2D}), and one can show their relative commutation errors are around $\mathcal{O}(\sqrt{\frac{L_{A,x}}{L_y}}\frac{l_{AB}}{L_A})$. Therefore, when $\frac{L_{A,x}}{L_y}\ll 1$, all the $7L_A$ operators in Eqs. (\ref{eq-TP-free-2D}) and (\ref{eq-TP-free-2D-2}) satisfy the criterion of Eq. (\ref{eq-QA-criterion}). Conversely, when $\frac{L_{A,x}}{L_y}\gg 1$, only the $3L_A$ conserved quantities in Eq. (\ref{eq-TP-free-2D}) satisfy the criterion of Eq. (\ref{eq-QA-criterion}). This example shows remarkably the validity of the criterion Eq. (\ref{eq-QA-criterion}).

Where the 2D single-particle wavefunctions are strongly localized, the story is similar to the 1D case. As shown in Fig. \ref{fig2}(d) ($t=0.1$, $W=1$),  the EHSM eigenvalues $p_{A,n}$ drops to zero at $n=L_A$. The corresponding eigen-operators $\overline{Q}_A^{(n)}$ ($0< n< L_A$) are again approximately the number operators (approaching $c_{\mathbf{r}_i}^\dag c_{\mathbf{r}_i}$) of the $L_A$ localized single-particle eigenstates.

\subsection{Generic observation}

Generically, if we denote the free fermion entanglement Hamiltonians as $H_{E}^A(\alpha)=\sum_{n\le z L_A} \overline{\beta}^{(n)}_A(\alpha) \overline{Q}_A^{(n)}$ in terms of normalized eigen-operators $\overline{Q}_A^{(n)}$, the free fermion EHSM spectra $p_{A,n}$ cutoff behaviors can be roughly fitted by assuming a standard deviation 
\begin{equation}
\sigma^{(n)}\propto (1-\frac{n}{zL_A})^{r}
\end{equation}
for $\overline{\beta}^{(n)}_A(\alpha)$ among all the eigenstates $|\alpha\rangle$, where $r\ge0$ (see App. \ref{sec:fitting-r-free}). From Fig. \ref{fig2}, we find $r\approx 0.5D$ for $D$-dimensional extended free fermions, and $r\approx 0.5(D-1)$ for $D$-dimensional localized free fermions.

Besides, we conjecture the subregionally quasilocal conserved quantities $\overline{Q}_A^{(n)}$ found here may play a role in the \emph{eigenstate typicality} of free fermions \cite{lai2015,tian2021breakdown}.


\section{Conserved quantities of Interacting models: the XYZ model}\label{sec:V}

As an example of EHSM of interacting many-body systems, we study the traceless 1D spin $1/2$ XYZ model Hamiltonian in a magnetic field (with periodic boundary condition):
\begin{equation}\label{eq-XYZ}
H=\sum_{j=1}^L\Big[\sum_{\nu=x,y,z}J_\nu\sigma_{j,\nu}\sigma_{j+1,\nu} +(\mathbf{B}+\delta\mathbf{B}_{j})\cdot\bm{\sigma}_j\Big]\ ,
\end{equation}
where $\sigma_{j,\nu}$ are the spin Pauli matrices on site $j$, $J_\nu$ are the neighboring spin interactions, $\mathbf{B}$ is a uniform magnetic field, and $\delta\mathbf{B}_j$ is a random magnetic field with components $\delta B_{j,\nu}\in[-B_{R,\nu},B_{R,\nu}]$ ($\nu=x,y,z$). We perform exact diagonalization (ED) of the model for $L=14$ sites, and study the EHSMs of subregion $A$ with $L_A\le 7$ (Fig. \ref{fig1}(b), see App. \ref{app:XYZ} for details). 

\begin{figure}[tbp]
\begin{center}
\includegraphics[width=3.4in]{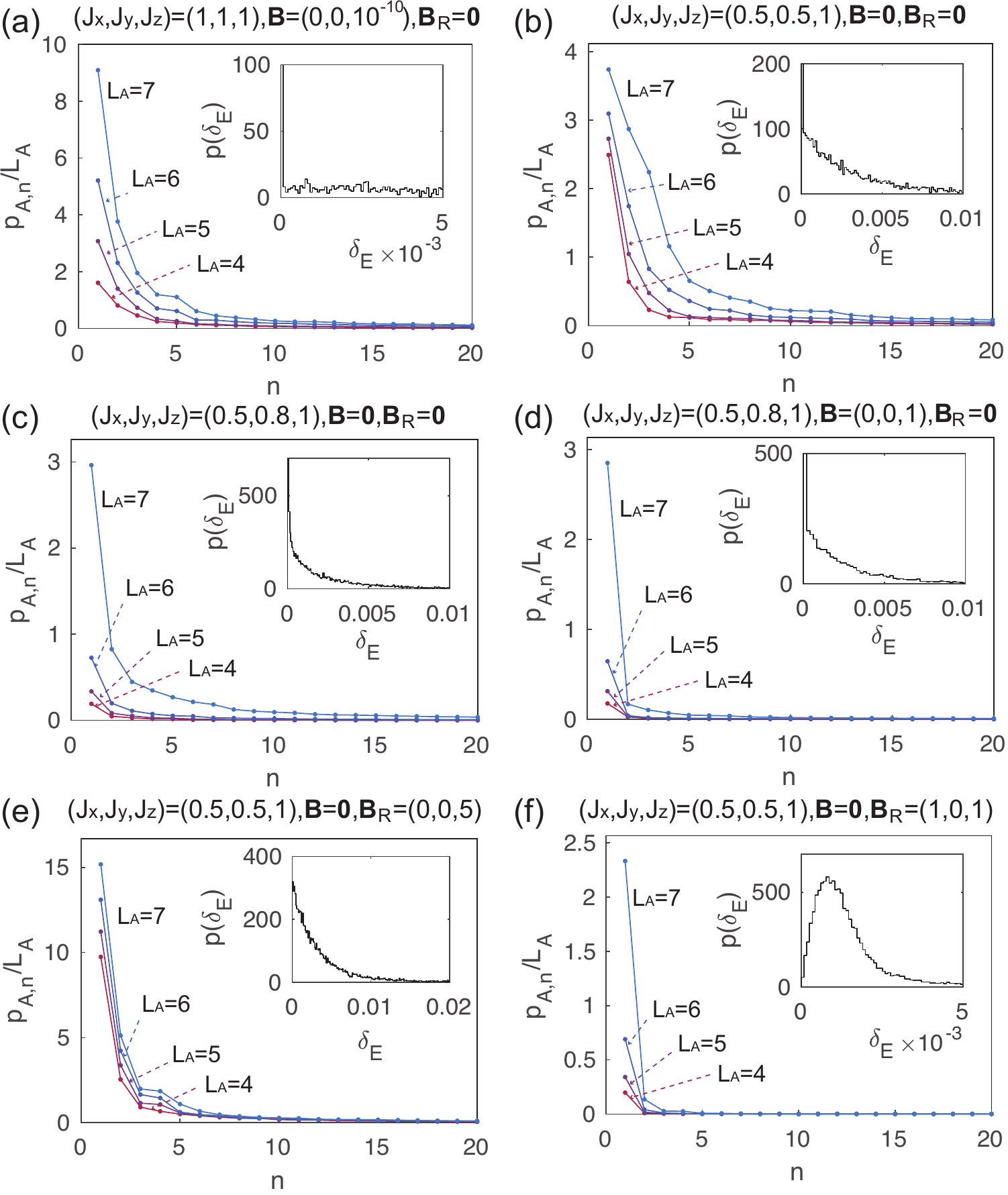}
\end{center}
\caption{ The EHSM eigenvalues $p_{A,n}$ of the 1D XYZ model in a magnetic field, which are calculated for the ensemble $\Xi$ of all the eigenstates $|\alpha\rangle$ with equal weights $w_\alpha$. The full system size is $L=14$, and the subregion size $L_A$ and model parameters are given in each panel. The insets show the LSS of ensemble $\Xi$ (Poisson in (a)-(e), and Wigner-Dyson in (f)).
}
\label{fig3}
\end{figure}

\begin{figure}[tbp]
\begin{center}
\includegraphics[width=3.4in]{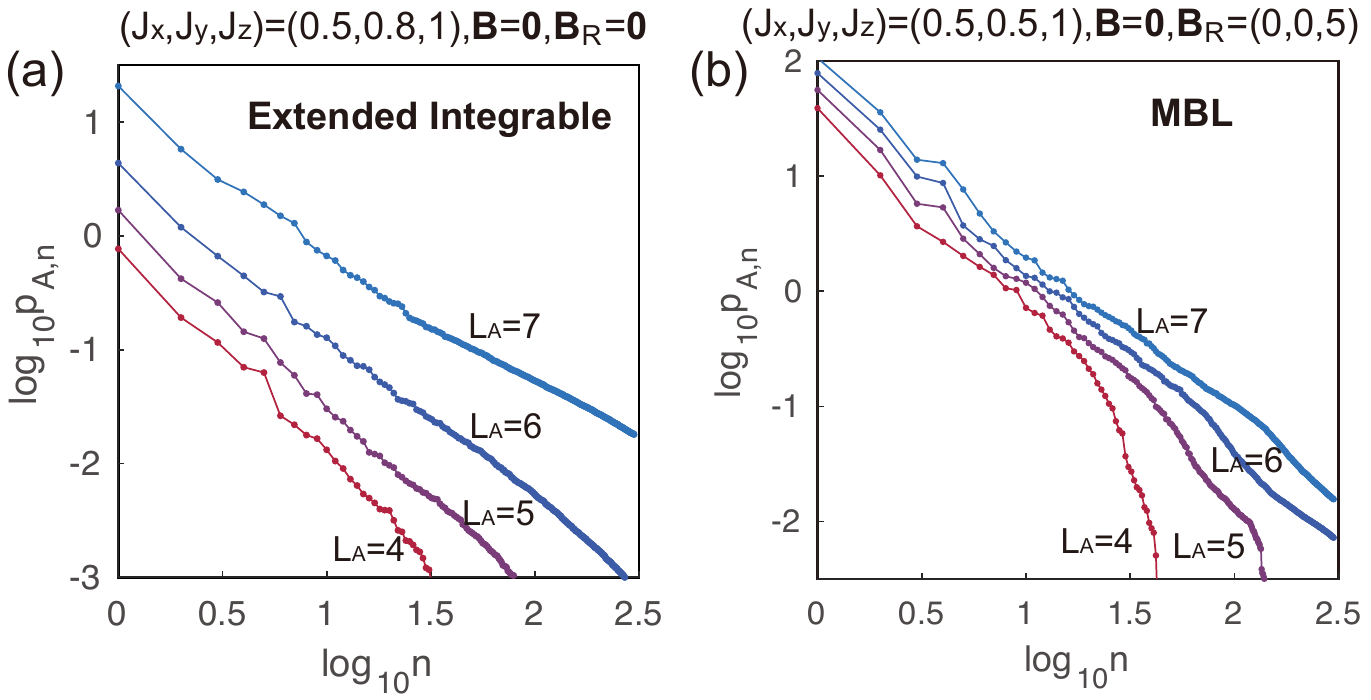}
\end{center}
\caption{ The log-log plot ($\log_{10} p_{A,n}$ vs. $\log_{10}n$) of Fig. \ref{fig3}(b) (panel (a) here) and (e) (panel (b) here), respectively. The results show that $p_{A,n}\propto n^{-s}$, where the exponent $s\approx 1$ for extended integrable phases (panel (a)), and $s\approx 1.5\sim 2$ for MBL phases (panel (b)).
}
\label{fig-mlogpXYZ}
\end{figure}

In Fig. \ref{fig3}, we calculate the EHSM eigenvalues $p_{A,n}$ for the ensemble $\Xi$ of all the $N$ eigenstates $|\alpha\rangle$ with equal weights $w_\alpha=\frac{1}{N}$, with parameters labeled in each panel. The insets show the LSS of the ensemble $\Xi$. The largest eigenvalue $p_{A,0}$ is not shown, which always dominantly gives $\overline{Q}_A^{(0)}\approx I_A/\sqrt{N_A}$. 

In Fig. \ref{fig3}(a)-(c) where $\mathbf{B}=\mathbf{B}_R=\mathbf{0}$, the XYZ model is exactly solvable \cite{baxter1973a,baxter1973b} (thus integrable), and we find the EHSM eigenvalues approximately decaying as power law (see Fig. \ref{fig-mlogpXYZ}(a)):
\begin{equation}\label{eq:pAdecay}
p_{A,n}\propto n^{-s}\ ,\qquad s\approx 1\ .
\end{equation}
For the XYZ model (three $J_\nu$ unequal) in a uniform magnetic field $\mathbf{B}$ shown in Fig. \ref{fig3}(d), $p_{A,n}$ also decays approximately as $n^{-s}$ (for $n>2$) with $s\approx 1$, indicating the existence of subregionally quasilocal conserved quantities, although exactly local conserved quantities are proved non-existing \cite{shiraishi2019}. 

Fig. \ref{fig3}(e) shows the EHSM of XXZ model ($J_x=J_y\neq J_z$) with a $\hat{\mathbf{z}}$ direction random field $\mathbf{B}_R$, which is in the many-body localization (MBL) phase \cite{basko2006,gornyi2005,oganesyan2007,marko2008,pal2010}. In this case, we also find  $p_{A,n}$ approximately decays in power law, but with a larger exponent (Fig. \ref{fig-mlogpXYZ}(b)):
\begin{equation}\label{eq:pAdecay2}
p_{A,n}\propto n^{-s}\ ,\qquad s\approx 1.5\sim2\ .
\end{equation}
We expect the eigen-operators to give the MBL localized conserved quantities \cite{serbyn2013,huse2014,chandran2015,ros2015}, which makes the system (approximately) integrable. 

Note that the power-law decaying $p_{A,n}\propto n^{-s}$ of the integrable XYZ models (which may be subject to finite size effects) is different from the cutoff behavior of nonzero $p_{A,n}$ of free fermion models in Fig. \ref{fig2}. Here it indicates a standard deviation $\sigma^{(n)}\propto n^{-s/2}$ for the coefficient $\overline{\beta}^{(n)}_A(\alpha)$ in $H_{E}^A(\alpha)=\sum_{n} \overline{\beta}^{(n)}_A(\alpha) \overline{Q}_A^{(n)}$ among all eigenstates $|\alpha\rangle$ (App. \ref{sec:fitting-s-XYZ}).

Lastly, for the XXZ model with a random direction field $\mathbf{B}_R$ shown in Fig. \ref{fig3}(f), the LLS shows the Wigner-Dyson distribution, 
indicating full quantum chaos. In this case, only the leading two eigenvalues $p_{A,0}$ and $p_{A,1}$ are significantly nonzero, which correspond to linear combinations of the only two local conserved quantities $I_A$ and $H_A$, in agreement with ETH (Eq. (\ref{eq-KA-chaotic})).

\begin{figure}[tbp]
\begin{center}
\includegraphics[width=3.4in]{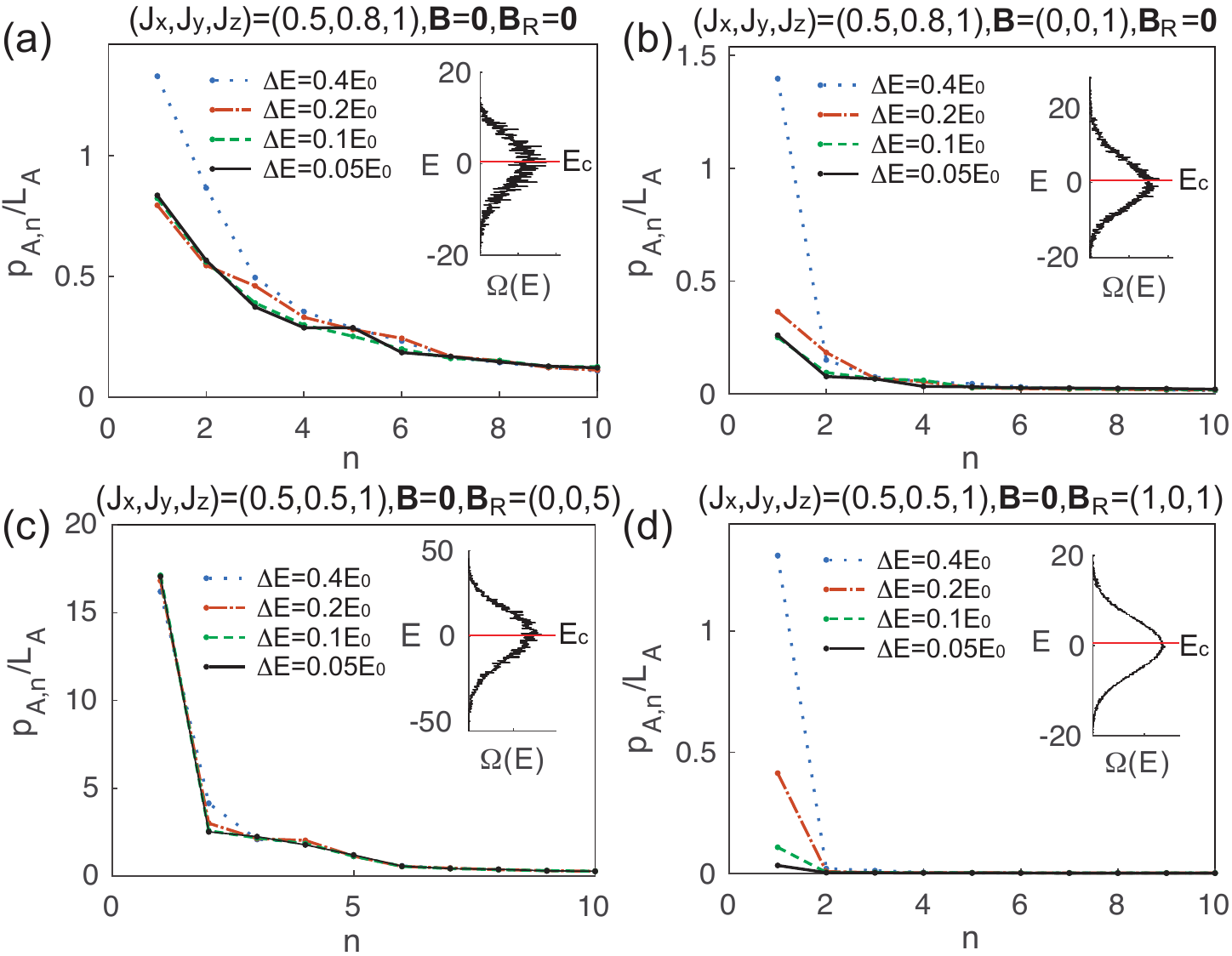}
\end{center}
\caption{The EHSM eigenvalues $p_{A,n}$ ($n\ge1$) of the 1D XYZ model for an ensemble $\Xi$ of all eigenstates in the energy interval $[E_c-\frac{\Delta E}{2},E_c+\frac{\Delta E}{2}]$, where $\Delta E$ is varied. The full system (subregion) size is $L=14$ ($L_A=7$). $E_0$ is the range of the energy spectrum of the entire system. The insets show the density of states $\Omega(E)$ of the system and $E_c$.
}
\label{fig4}
\end{figure}

We now take a closer look at the eigen-operators $\overline{Q}_A^{(n)}$, which roughly commute with subregion Hamiltonian $H_A$ by Eq. (\ref{eq-QA-criterion}) (App. \ref{app:XYZ-commu}). Generically, the eigen-operator of the largest eigenvalue $p_{A,0}$ is quite accurately $\overline{Q}_A^{(0)}\propto I_A$. In Fig. \ref{fig3}(a),(b),(e) which possess a $\hat{\mathbf{z}}$ direction spin rotational symmetry, $\overline{Q}_A^{(1)}$ and $\overline{Q}_A^{(2)}$ are approximately linear combinations of $H_A$ and $\sum_{j}\sigma_{j,z}$ (App. \ref{app:XYZ} Tab. \ref{Tab-XXZ}). 
For Fig. \ref{fig3}(c),(d),(f), we find dominantly $\overline{Q}_A^{(1)}\propto H_A$. The higher eigen-operators $\overline{Q}_A^{(n)}$ are generically less local (App. \ref{app:XYZ-Qn}), or even fairly nonlocal within subregion $A$ although still subregionally quasilocal, similar to $P^A_{x,y}$ for the extended free fermions. In the zero field XXZ model (Fig. \ref{fig3}(a),(b)), we find 
\begin{equation}
\overline{Q}_A^{(3)}\approx \sum_{j,\ell,\nu}\zeta_\nu(l)\sigma_{\nu,j}\sigma_{\nu,j+l}+\zeta' \sum_j\sigma_{z,j}, 
\end{equation}
with the functions $\zeta_{\nu}(l)$ decaying with $l$, and $\zeta'$ is some constant. $\overline{Q}_A^{(4)}$ has a large overlap with the known support-4 local conserved quantity $P_4$ (definition in Eq. (\ref{seq:P4})) of XXZ model \cite{tetelman1981,grabowski1995} (App. \ref{app:XYZ-Qn}). In contrast, all $\overline{Q}_A^{(n)}$ has zero overlap with the known support-3 local conserved quantity $P_3$ (definition in Eq. (\ref{seq:P3})) \cite{tetelman1981,grabowski1995}. 

We can also calculate the EHSM for a microcanonical ensemble $\Xi$ consisting of all the eigenstates $|\alpha\rangle$ with energies $E_{\alpha}\in [E_c-\frac{\Delta E}{2},E_c+\frac{\Delta E}{2}]$ with equal weights $w_\alpha$, for some center energy $E_c$, to characterize the integrability of the system near energy $E_c$. Fig. \ref{fig4} (a)-(c) show the cases with subregionally quasilocal conserved quantities other than $I_A$ and $H_A$, where the nonzero EHSM eigenvalues $p_{A,n}$ ($n\ge1$) asymptotically approach nonzero constants as $\Delta E\rightarrow 0$. In sharp contrast, in the fully chaotic case where the ETH holds, we find $p_{A,n}\rightarrow 0$ for all $n\ge1$ when $\Delta E\rightarrow 0$. This is because all the entanglement Hamiltonians are given by Eq. (\ref{eq-KA-chaotic}) with $E_\alpha=E_c$ and thus equal, leading to only one nonzero EHSM eigenvalue $p_{A,0}$ and the corresponding eigen-operator 
\begin{equation}
\overline{Q}_A^{(0)}\propto \beta^{(0)}_AI_A+\beta^{(1)}_AH_A\ .
\end{equation}

\section{Discussion}\label{sec:VI}

Sometimes only the time $\tau$ evolution $|\psi(\tau)\rangle$ of a non-eigenstate $|\psi(0)\rangle$ within time $T$ is known. In this case, one can define approximate ``eigenstates" 
\begin{equation}
|\widetilde{\alpha}\rangle_T=\frac{1}{\mathcal{N}_T(\alpha)}\int_0^{T}d\tau e^{i\widetilde{E}_{\alpha}\tau}|\psi(\tau)\rangle 
\end{equation}
for a set of random energies $\{\widetilde{E}_\alpha\}$, where $\mathcal{N}_T(\alpha)$ is the normalization factor. One can then diagonalize the EHSM in subregion $A$ of these states. With a time $T$ power-law in system size, we find such a calculation still yields a similar EHSM spectrum as Fig. \ref{fig3}, and the second eigen-operator $Q_A^{(1)}$ reproduces the subregion Hamiltonian $H_A$ well (App. \ref{sec:Tstate}). However, to accurately retrieve conserved quantities other than the Hamiltonian $H_A$, this method may require an exponentially long time $T\sim \mathcal{O}(d^L)$.

We have seen that if an ensemble of $N_\Xi$ excited eigenstates are known, subregionally (quasi)local conserved quantities including the Hamiltonian can be extracted as eigen-operators of their subregion EHSM with eigenvalues $p_{A,n}>0$. For free fermions, the nonzero $p_{A,n}$ has a cutoff proportional to the subregion volume. For the interacting XYZ models, $p_{A,n}$ decays as $n^{-s}$ if integrable, while only $p_{A,0}$ and $p_{A,1}$ are significantly nonzero if fully chaotic. One future question is to understand the power-law EHSM spectrum in interacting integrable models, and which conserved quantities contribute. This might be studied more analytically from the Bethe Ansatz \cite{bethe1931} eigenstates of 1D solvable models, which allow much larger system sizes than ED. Another future question is whether terms in the EHSM eigen-operators not commuting with $H_A$ (Eq. (\ref{eq-QA-criterion})) are located near the subregion boundary. Moreover, how the EHSM eigen-operators affect the nonequilibrium evolution in a subregion is to be understood.

\begin{acknowledgments}
\emph{Acknowledgments}. The author is grateful to conversations with Yichen Hu, Abhinav Prem, and especially the insightful discussion with David Huse. The author is also thankful to the comments from the referees which helps improve this paper. The author acknowledges support from the Alfred P. Sloan Foundation.

\end{acknowledgments}

\begin{widetext}

\appendix


\section{Entanglement Hamiltonian of a fully chaotic system}\label{app:chaotic}

In this section, we study the properties of eigenstate wavefunctions of the Hamiltonian in main text Eq. (2) when the system is fully quantum chaotic. The $N\times N$ Hamiltonian ($N=d^L$ is the Hilbert space dimension of the full system) in the main text Eq. (2) is of the form:
\begin{equation}\label{seq-H-HAB}
H=H_0+H_{AB}\ ,\qquad H_0=H_A\otimes I_B+I_A\otimes H_B\ .
\end{equation}
Here the subregions $A$ and $B$ have sizes (number of sites) $L_A$ and $L_B$, respectively, and the total number of sites is $L=L_A+L_B$. Besides, we denote the number of sites on the boundary between subregions $A$ and $B$ as $l_{AB}$. 

For a local Hamiltonian and $l_{AB}\ll L_A,L_B$ (i.e., large system sizes), $H_0$ is dominant, and $H_{AB}$ can be treated as a perturbation. 
We adopt the subregion energy eigenstate direct product basis $|\alpha_A,A;\alpha_B,B\rangle=|\alpha_A,A\rangle\otimes|\alpha_B,B\rangle$, where $H_A|\alpha_A,A\rangle=E_{\alpha_A}^A|\alpha_A,A\rangle$ and $H_B|\alpha_B,B\rangle=E_{\alpha_B}^B|\alpha_B,B\rangle$ ($1\le\alpha_A\le N_A$, $1\le\alpha_B\le N_B$, with $N_A=d^{L_A}$, $N_B=d^{L_B}$ being the Hilbert space dimensions of subregions $A$ and $B$, $N_AN_B=N$). In this basis, both $H_A$ and $H_B$ are diagonal, and thus $H_0$ is diagonal, with eigenvalues $E_{\alpha_A}^A+E_{\alpha_B}^B$. 

The eigenstates $|\alpha\rangle$ ($1\le \alpha\le N$) of the entire Hamiltonian $H$ under the subregion energy eigenbasis then have wavefunctions of the form
\begin{equation}\label{seq-alpha-u}
|\alpha\rangle=\sum_{\alpha_A,\alpha_B} u_{\alpha,\alpha_A,\alpha_B} |\alpha_A,A\rangle\otimes|\alpha_B,B\rangle\ .
\end{equation}
The elements of the reduced density matrix $\rho_A(\alpha)$ can thus be expressed as 
\begin{equation}
\langle \alpha_A,A|\rho_A(\alpha)|\alpha_A',A\rangle=\sum_{\alpha_B} u_{\alpha,\alpha_A,\alpha_B}u_{\alpha,\alpha_A',\alpha_B}^*\ .
\end{equation}

\subsection{Estimation of the boundary term $H_{AB}$ from ETH}


If the system is fully quantum chaotic, one expects ETH to hold. Since the Hamiltonian is local, we expect 
\begin{equation}
H_{AB}=\sum_m O_m^AO_m^B 
\end{equation}
is the sum over a set of local terms $O_m^AO_m^B$, where $O_m^A$ is supported in subregion $A$ and $O_m^B$ is supported in subregion $B$. The number of $m$ indices is proportional to the boundary size $l_{AB}$. According to ETH \cite{jensen1985,deutsch1991,srednicki1994,srednicki1999,dalessio2016}, their matrix elements can be estimated as
\begin{equation}
\begin{split}
&\langle \alpha_A',A|O_m^A|\alpha_A,A\rangle =O^A_m(\overline{E}_A)\delta_{\alpha_A,\alpha_A'}+ e^{-S_A(\overline{E}_A)/2}f^A_m(\overline{E}_A,\omega_A)r^{A,m}_{\alpha_A,\alpha_A'}\ ,\\
& \langle \alpha_B',B|O_m^B|\alpha_B,B\rangle =O^B_m(\overline{E}_B)\delta_{\alpha_B,\alpha_B'}+ e^{-S_B(\overline{E}_B)/2}f^B_m(\overline{E}_B,\omega_B)r^{B,m}_{\alpha_B,\alpha_B'}\ ,
\end{split}
\end{equation}
where $\overline{E}_A=(E_{\alpha_A}^A+E_{\alpha_A'}^A)/2$ and $\overline{E}_B=(E_{\alpha_B}^B+E_{\alpha_B'}^B)/2$ are the average energies, $\omega_A=E_{\alpha_A}^A-E_{\alpha_A'}^A$ and $\omega_B=E_{\alpha_B}^B-E_{\alpha_B'}^B$ are energy differences, while $S_A(\overline{E}_A)$ and $S_B(\overline{E}_B)$ are the entropies in each subregion at the average energies. $r^{A,m}_{\alpha_A,\alpha_A'}$ and $r^{B,m}_{\alpha_B,\alpha_B'}$ are random matrices with a root mean square for each element being $1$. We note that for chaotic systems satisfying the ETH, $e^{-S_A(\overline{E}_A)/2}\propto 1/\sqrt{N_A}$, and $e^{-S_B(\overline{E}_B)/2}\propto 1/\sqrt{N_B}$. The function $f_m^{A,B}(E,\omega)$ decay exponentially as $e^{-|\omega|/\omega_0}$ at large $\omega$ (comparable to $\omega_0$), and is smooth at small $\omega$ (the values scale as $\sqrt{L}$), where $\omega_0$ is independent of system size (i.e., of order $1$ in the expansion with respect to system size $L_{A,B}$) \cite{dalessio2016}. Therefore, we find $H_{AB}$ (contributed by order $l_{AB}$ number of local terms $O_m^AO_m^B$) under the basis $|\alpha_A,A\rangle\otimes|\alpha_B,B\rangle$ consists of a diagonal part and an off-diagonal part:
\begin{equation}\label{seq-HAB-ETH}
\begin{split}
&H_{AB}= H_{AB}^{(\text{d})}+h^{(\text{off})}\ ,\\
&(H_{AB}^{\text{d}})_{\alpha_A'\alpha_B';\alpha_A\alpha_B}=\delta_{\alpha_A,\alpha_A'}\delta_{\alpha_B,\alpha_B'}E^{(\text{d})}(E_{\alpha_A}^A, E_{\alpha_B}^B)\ ,\\
&h^{(\text{off})}_{\alpha_A'\alpha_B';\alpha_A\alpha_B}= \delta_{\alpha_A,\alpha_A'} r^{B}_{\alpha_B,\alpha_B'} \frac{\lambda_B(\overline{E}_B,\omega_B)}{\sqrt{N_B}}+ \delta_{\alpha_B,\alpha_B'} r^{A}_{\alpha_A,\alpha_A'} \frac{\lambda_A(\overline{E}_A,\omega_A)}{\sqrt{N_A}}   + \frac{\lambda_{AB}(\overline{E}_A,\omega_A, \overline{E}_B,\omega_B)}{\sqrt{N}} r^{AB}_{\alpha_A\alpha_B,\alpha_A'\alpha_B'}\ ,
\end{split}
\end{equation}
where all the $r$ matrices are random matrices with the root mean square of each element being $1$, while the functions $\lambda_A,\lambda_B,\lambda_{AB}$ are proportional to $\sqrt{L_A l_{AB}},\sqrt{L_B l_{AB}},\sqrt{L_AL_B l_{AB}}$ and decay as $e^{-|\omega_A|/\omega_{A,0}}$, $e^{-|\omega_B|/\omega_{B,0}}$ and $e^{-|\omega_A|/\omega_{A,0}-|\omega_B|/\omega_{B,0}}$, with $\omega_{A,0}$ and $\omega_{B,0}$ independent of system sizes $L_A$ and $L_B$. To the lowest order, we can approximately assume the diagonal part takes the form of
\begin{equation}\label{seq-Ep-approx}
E^{(\text{d})}(E_{\alpha_A}^A, E_{\alpha_B}^B)\approx \frac{l_{AB}}{\epsilon_0^{AB}}(\frac{E_{\alpha_A}^A}{L_A}-\epsilon_0^A)(\frac{E_{\alpha_B}^B}{L_B}-\epsilon_0^B)\ ,
\end{equation}
where $l_{AB}$ is the boundary area (number of sites on the boundary), while $\epsilon_0^{AB}$, $\epsilon_0^{A}$ and $\epsilon_0^{B}$ are of order $1$ energies and are asymptotically independent of system sizes $L_A$ and $L_B$. Since we have defined that each product term in the boundary term $H_{AB}$ is traceless, we have 
\begin{equation}\label{seq-epAB}
\epsilon_0^A=\frac{1 }{N_A L_A} \sum_{\alpha_A=1}^{N_A} E_{\alpha_A}^A=\frac{E^A_{av}}{L_A}\ ,\qquad  \epsilon_0^B=\frac{1 }{N_B L_B} \sum_{\alpha_B=1}^{N_B} E_{\alpha_B}^B=\frac{E^B_{av}}{L_B}\ ,
\end{equation}
where $E^A_{av}$ and $E^B_{av}$ are the mean values of $E_{\alpha_A}^A$ and $E_{\alpha_B}^B$, respectively. 
Note that for systems with delocalized eigenstates (e.g., the fully chaotic systems considered here), the energy range of $E_{\alpha_A}^A$ ($E_{\alpha_B}^B$) generically scale linearly with $L_A$ ($L_B$), so $\epsilon_0^{A}$ and $\epsilon_0^{B}$ are of order $1$. 

This yields a correlation for the matrix elements of the off-diagonal Hermitian part $h^{\text{off}}$ (averaged over the random $r$ matrices in Eq. (\ref{seq-HAB-ETH})):
\begin{equation}\label{seq-HAB-off}
\langle h^{\text{off}}_{\alpha_A'\alpha_B';\alpha_A\alpha_B}\rangle=0\ ,\qquad \langle h^{\text{off}}_{\alpha_A'\alpha_B';\alpha_A\alpha_B}h^{\text{off}}_{\alpha_A\alpha_B;\alpha_A'\alpha_B'}\rangle=\frac{\lambda_{AB}^2}{N}+ \delta_{\alpha_A,\alpha_A'} \frac{\lambda_B^2}{N_B} + \delta_{\alpha_B,\alpha_B'} \frac{\lambda_A^2}{N_A}\ ,
\end{equation}
where we have omitted the variables of the functions $\lambda_{AB}$, $\lambda_A$ and $\lambda_B$ for simplicity. Note that here the bra and ket stands for the average over the random $r$ matrices in Eq. (\ref{seq-HAB-ETH}).

\subsection{Derivation of the entanglement Hamiltonian for eigenstates}

To find the properties of the eigenstate wavefunctions for determining the entanglement Hamiltonian, we treat the off-diagonal part $h^{\text{off}}$ of $H_{AB}$ as fluctuating quantum fields obeying Eq. (\ref{seq-HAB-off}), and define the statistically averaged Green's function:
\begin{equation}
G_{\alpha_A\alpha_B;\alpha_A'\alpha_B'}(\omega)=\langle\Big( \langle\alpha_A,A;\alpha_B,B| \frac{1}{\omega-H}|\alpha_A',A;\alpha_B',B\rangle\Big)\rangle=\delta_{\alpha_A\alpha_A'}\delta_{\alpha_B\alpha_B'}G_{\alpha_A\alpha_B}(\omega)\ ,
\end{equation}
where the outer bra and ket stand for the statistical average over all possible random $h^{\text{off}}$ matrices satisfying Eq. (\ref{seq-HAB-off}). The fact that $G_{\alpha_A\alpha_B;\alpha_A'\alpha_B'}(\omega)\propto \delta_{\alpha_A\alpha_A'}\delta_{\alpha_B\alpha_B'}$ can be seen by noting that $G_{\alpha_A\alpha_B;\alpha_A'\alpha_B'}(\omega)$ should be invariant under flipping of any basis $|\alpha_A,A\rangle\rightarrow -|\alpha_A,A\rangle$ or $|\alpha_B,B\rangle\rightarrow -|\alpha_B,B\rangle$, given that the matrix elements of $h^{\text{off}}$ are random with zero mean. In the large $N_A,N_B$ limit, by treating $h^{\text{off}}$ as a matrix quantum field, one can show that the Green's function $G_{\alpha_A\alpha_B}(\omega)$ satisfy the Schwinger-Dyson (SD) equation:
\begin{equation}\label{seq-SD1}
G_{0,\alpha_A\alpha_B}(\omega)^{-1}=G_{\alpha_A\alpha_B}(\omega)^{-1}+\Sigma_{\alpha_A\alpha_B}(\omega)\ ,
\end{equation}
where the unperturbed Green's function $G_{0,\alpha_A\alpha_B}(\omega)$ and the self energy $\Sigma_{\alpha_A\alpha_B}(\omega)$ are given by
\begin{equation}\label{seq-SD2}
G_{0,\alpha_A\alpha_B}(\omega)=\frac{1}{\omega-E_{\alpha_A}^A-E_{\alpha_B}^B-E^{(\text{d})}(E_{\alpha_A}^A, E_{\alpha_B}^B)},\quad  \Sigma_{\alpha_A\alpha_B}(\omega)=\sum_{\alpha_A'\alpha_B'} \langle h^{\text{off}}_{\alpha_A'\alpha_B';\alpha_A\alpha_B}h^{\text{off}}_{\alpha_A\alpha_B;\alpha_A'\alpha_B'}\rangle G_{\alpha_A'\alpha_B'}(\omega).
\end{equation}
Eqs. (\ref{seq-SD1}) and (\ref{seq-SD2}) then gives a self-consistent equation
\begin{equation}
\Sigma_{\alpha_A\alpha_B}(\omega)=\frac{1}{N}\sum_{\alpha_A'\alpha_B'} \frac{\lambda_{AB}^2 +N_A\lambda_B^2\delta_{\alpha_A,\alpha_A'} + N_B\lambda_A^2\delta_{\alpha_B,\alpha_B'}}{\omega-E_{\alpha_A'}^A-E_{\alpha_B'}^B-E^{(\text{d})}(E_{\alpha_A'}^A, E_{\alpha_B'}^B)-\Sigma_{\alpha_A'\alpha_B'}(\omega)}\ .
\end{equation}
In the large $L_A$ and $L_B$ limit, $h^{\text{off}} $ is much smaller than $H_A$ and $H_B$, thus Eq. (\ref{seq-SD2}) implies that the self energy $\Sigma_{\alpha_A\alpha_B}(\omega)$ is much smaller than $E_{\alpha_A'}^A$ and $E_{\alpha_B'}^B$. To the leading order of $\frac{l_{AB}}{L_A}$ and $\frac{l_{AB}}{L_B}$, ignoring the $E^{(\text{d})}(E_{\alpha_A}^A, E_{\alpha_B}^B)$ and $\Sigma_{\alpha_A'\alpha_B'}(\omega)$ term in the denominator, and turn the summation over $\alpha_A',\alpha_B'$ into an integration, we find an imaginary self energy
\begin{equation}\label{seq:self-energy-ETH}
\begin{split}
\Sigma_{\alpha_A\alpha_B}(\omega)\approx & 2\pi i \Big[ \Big( \int d E^A \Omega_A(E^A)\Omega_B(\omega-E^A) \lambda_{AB}^2(\frac{E_{\alpha_A}^A+E^A}{2},E_{\alpha_A}^A-E^A, \frac{E_{\alpha_B}^B+\omega-E^A}{2}, E_{\alpha_B}^B-\omega+E^A) \Big) \\
&+ \Omega_B(\omega-E^A_{\alpha_A}) \lambda_{B}^2( E_{\alpha_B}^B, 0) + \Omega_A(\omega-E^B_{\alpha_B}) \lambda_{A}^2( E_{\alpha_A}^A, 0) \Big]\ ,
\end{split}
\end{equation}
where we have defined
\begin{equation}
\Omega_A(E)=\frac{1}{N_A}\sum_{\alpha_A}\delta(E-E_{\alpha_A}^A) \ ,\qquad \Omega_B(E)=\frac{1}{N_B}\sum_{\alpha_B}\delta(E-E_{\alpha_B}^B) 
\end{equation}
as the normalized density of states in subregions $A$ and $B$ ($\int \Omega_A(E) dE=\int \Omega_B(E) dE=1$). Note that since the range of energies in subregions $A$ ($B$) is proportional to $L_A$ ($L_B$), we have $\Omega_A(E)\propto 1/L_A$ and $\Omega_B(E)\propto 1/L_B$. Therefore, we find the value of the self energy $\Sigma_{\alpha_A\alpha_B}(\omega)$ is around the order of the boundary size $l_{AB}$. We therefore find the Green's function given by
\begin{equation}
G_{\alpha_A\alpha_B}(\omega)= \frac{1}{\omega-E_{\alpha_A}^A-E_{\alpha_B}^B-E^{(\text{d})}(E_{\alpha_A}^A, E_{\alpha_B}^B)-\Sigma_{\alpha_A\alpha_B}(\omega)}\ .
\end{equation}

On the other hand, it is known that the spectral weight is related to the eigenstates of Hamiltonian $H$ by
\begin{equation}\label{seq:Am-def}
A_{\alpha_A\alpha_B}(\omega)=2\text{Im} G_{\alpha_A\alpha_B}(\omega)=2\pi\sum_{\alpha}|\langle \alpha |\alpha_A,A;\alpha_B,B\rangle |^2\delta(\omega-E_{\alpha})=2\pi\sum_{\alpha}|u_{\alpha,\alpha_A,\alpha_B}|^2\delta(\omega-E_{\alpha})\ .
\end{equation}
Therefore, in the large $L_A,L_B$ limit, we approximately have (under the statistical average of $h^{\text{off}}$):
\begin{equation}\label{seq-uu-correlation1}
\begin{split}
 u_{\alpha,\alpha_A,\alpha_B}u_{\alpha,\alpha_A',\alpha_B'}^* &\approx \frac{\delta_{\alpha_A,\alpha_A'}\delta_{\alpha_B,\alpha_B'}}{\pi N \Omega(E_\alpha)} \text{Im} G_{\alpha_A\alpha_B}(E_\alpha) \\
 &= \frac{\delta_{\alpha_A,\alpha_A'}\delta_{\alpha_B,\alpha_B'}}{\pi N \Omega(E_\alpha)} \frac{|\Sigma_{\alpha_A\alpha_B}(E_\alpha)|}{[E_\alpha-E_{\alpha_A}^A-E_{\alpha_B}^B-E^{(\text{d})}(E_{\alpha_A}^A, E_{\alpha_B}^B)]^2+|\Sigma_{\alpha_A\alpha_B}(E_\alpha)|^2} \\ &\approx \frac{\delta_{\alpha_A,\alpha_A'}\delta_{\alpha_B,\alpha_B'}}{N\Omega(E_\alpha)} \delta (E_\alpha-E_{\alpha_A}^A-E_{\alpha_B}^B-E^{(\text{d})}(E_{\alpha_A}^A, E_{\alpha_B}^B)) \ ,
\end{split}
\end{equation}
where
\begin{equation}
\Omega(E)=\frac{1}{N}\sum_\alpha \delta(E-E_\alpha)
\end{equation}
is the density of states of the entire system. Note that the energy width of the delta function in Eq. (\ref{seq-uu-correlation1}) is $|\Sigma_{\alpha_A\alpha_B}(E_\alpha)|$, which is of order $l_{AB}$. It also has a dependence on $E_\alpha$, $E_{\alpha_A}^A$ and $E_{\alpha_B}^B$ (see Eq. (\ref{seq:self-energy-ETH})). In comparison, the ranges of $E_{\alpha_A}^A$ and $E_{\alpha_B}^B$ are proportional to $L_A$ and $L_B$. Therefore, the delta function approximation is legitimate when the subregion sizes $L_A$ and $L_B$ are large, in which case $l_{AB}\ll L_A$ and $l_{AB}\ll L_B$. The diagonal part $E^{(\text{d})}$ of the boundary term yields an order $l_{AB}$ contribution to the center position of the delta function.

If we take the approximation for $E^{(\text{d})}(E_{\alpha_A}^A, E_{\alpha_B}^B)$ in Eq. (\ref{seq-Ep-approx}), and assume $l_{AB}\ll L_A,L_B$, we find
\begin{equation}
\begin{split}
&\langle \alpha_A,A|\rho_A(\alpha)|\alpha_A',A\rangle=\sum_{\alpha_B} u_{\alpha,\alpha_A,\alpha_B}u_{\alpha,\alpha_A',\alpha_B}^* \\
&\approx \int dE^B \frac{\delta_{\alpha_A,\alpha_A'} N_B}{N\Omega(E_\alpha)}  \Omega_B(E_B) \delta \Big(E-E_{\alpha_A}^A-E^B- \frac{l_{AB}}{\epsilon_0^{AB}}(\frac{E_{\alpha_A}^A}{L_A}-\epsilon_0^A)(\frac{E^B}{L_B}-\epsilon_0^B)\Big) \\
&=\delta_{\alpha_A,\alpha_A'}\frac{\Omega_B\left(a_A(E_\alpha, E_{\alpha_A}^A)\right)}{N_A\Omega(E_\alpha)} \ ,
\end{split}
\end{equation}
where the function
\begin{equation}\label{seq-E-aA}
a_A(E_\alpha, E_{\alpha_A}^A)=\frac{E-E^A_{\alpha_A}+\frac{\epsilon_0^B }{\epsilon_0^{AB}}\frac{l_{AB}}{L_A}(E_{\alpha_A}^A-L_A\epsilon_0^A)}{1+\frac{l_{AB}}{\epsilon_0^{AB}L_B}(\frac{E_{\alpha_A}^A}{L_A}-\epsilon_0^A)}\ .
\end{equation}
When $l_{AB}\ll L_A\ll L_B$, to the linear order of $E^A_{\alpha_A}-L_A\epsilon^A_0$ (note that the average value of $ E^A_{\alpha_A}$ is $E^A_{av}=L_A\epsilon^A_0$), we approximately have
\begin{equation}
H_E^A(\alpha)=-\log \rho_A(\alpha)\approx \sum_{\alpha_A}\left( \beta^{(0)}_A(\alpha)+ \beta_A^{(1)}(\alpha)E^A_{\alpha_A}\right) |\alpha_A,A\rangle \langle \alpha_A,A|=\beta^{(0)}_A(\alpha) I_A +\beta_A^{(1)}(\alpha)(H_A-E^A_{av}) \ ,
\end{equation}
where
\begin{equation}
\beta^{(0)}_A(\alpha)=\log \left[\frac{N_A\Omega(E_\alpha)}{\Omega_B(E_\alpha-L_A\epsilon_0^A)}\right]\ ,\quad \beta^{(1)}_A(\alpha)=\Big(1+ \frac{E_\alpha-L_A\epsilon_0^A-L_B\epsilon_0^B}{\epsilon_0^{AB}}\frac{l_{AB}}{L_AL_B}\Big)\frac{\text{d}\log \Omega_B(E)}{\text{d}E}\Big|_{E=E_\alpha-L_A\epsilon_0^A}\ .
\end{equation}
Note that by definition in Eq. (\ref{seq-epAB}) and the fact that $\text{tr}(H_{AB})=0$, we have $L_A\epsilon_0^A+L_B\epsilon_0^B=\text{tr}(H)/N=E_{av}$ is the average value of the energy $E_\alpha$ of the entire system. Also, note that $L_A\epsilon_0^A=E^A_{av}$ is the mean value of subregion energy $E^A_{\alpha_A}$, so we can rewrite the coefficients as
\begin{equation}\label{seq-beta-ETH}
\beta^{(0)}_A(\alpha)=\log \left[\frac{N_A\Omega(E_\alpha)}{\Omega_B(E_\alpha-E^A_{av})}\right]\ ,\quad 
\beta^{(1)}_A(\alpha)=\Big(1+ \frac{E_\alpha-E_{av}}{\epsilon_0^{AB}}\frac{l_{AB}}{L_AL_B}\Big)\frac{\text{d}\log \Omega_B(E)}{\text{d}E}\Big|_{E=E_\alpha-E^A_{av}}\ .
\end{equation}
If we ignore all the terms to the linear order of $\frac{l_{AB}}{L_A}$, $\frac{l_{AB}}{L_B}$ and higher, we will have
\begin{equation}
\beta^{(0)}_A(\alpha)=\log \left[\frac{N_A\Omega(E_\alpha)}{\Omega_B(E_\alpha-E^A_{av})}\right]\ ,\qquad \beta^{(1)}_A(\alpha)\approx\frac{\text{d}\log \Omega_B(E)}{\text{d}E}\Big|_{E=E_\alpha-E^A_{av}}\ .
\end{equation}

\subsection{The case when the Hamiltonian is extremely nonlocal}
For completeness, we also discuss the case when the Hamiltonian $H$ is extremely nonlocal, in which case most terms are coupling subregions $A$ and $B$ and belong to $H_{AB}$, so we expect $||H_{AB}||\gg || H_0||$ in Eq. (\ref{seq-H-HAB}). We can then approximately regard $H_0=0$, and treat $H_{AB}$ as a fully random matrix (as a nonlocal chaotic system resembles a zero dimensional chaotic system). Accordingly, if we assume the random matrix $H_{AB}$ satisfies
\begin{equation}
\langle (H_{AB})_{\alpha_A'\alpha_B';\alpha_A\alpha_B} (H_{AB})_{\alpha_A\alpha_B;\alpha_A'\alpha_B'}\rangle =\frac{\lambda^2}{N}\ ,
\end{equation}
the SD equation gives the Green's function and spectral weight
\begin{equation}
G_{\alpha_A\alpha_B}(\omega)=\frac{\omega-\sqrt{\omega^2-4\lambda^2}}{2\lambda^2}\ ,\qquad A_{\alpha_A\alpha_B}(\omega)=\Theta(4\lambda^2-\omega^2) \frac{\sqrt{4\lambda^2-\omega^2}}{\lambda^2}\ ,
\end{equation}
where $\Theta(x)$ is the Heaviside step function. Note that $A_{\alpha_A\alpha_B}(\omega)$ has no dependence on the basis indices $\alpha_A,\alpha_B$. Therefore, the eigenstate wavefunction components have no obvious $\alpha_A,\alpha_B$ dependence, and we expect a uniform correlation
\begin{equation}\label{seq-uu-correlation2}
 u_{\alpha,\alpha_A,\alpha_B}u_{\alpha,\alpha_A'\alpha_B'}^*\approx\frac{\delta_{\alpha_A,\alpha_A'}\delta_{\alpha_B,\alpha_B'}}{N}\ .
\end{equation}

\section{Entanglement Hamiltonian of systems with multiple conserved quantities}\label{app:multipleQ}

In this section, we consider the case when there are multiple subregionally local or quaislocal conserved quantities. We assume the system in the entire region has linearly independent local and nonlocal conserved quantities $Q^{(n)}$ ($n\ge 1$) as shown in main text Eqs. (5) and (6), namely,
\begin{equation}\label{seq-Q-QAB}
Q^{(n)}|\alpha\rangle=q_\alpha^{(n)}|\alpha\rangle\ ,\qquad [Q^{(n)},Q^{(m)}]=0\ , \qquad Q^{(n)}=Q^{(n)}_A\otimes I_B+I_A\otimes Q^{(n)}_B+Q^{(n)}_{AB}\ ,
\end{equation}
where $Q^{(n)}_A$ ($Q^{(n)}_B$) has supports within subregion $A$ ($B$), and $Q^{(n)}_{AB}$ contains all the terms with supports across the two subregions, as we defined below the main text Eq. (6). Note that we have assumed the energy eigenstates $|\alpha\rangle$ are also simultaneous eigenstates of $Q^{(n)}$. We assume $Q^{(0)}=I$ is the trivial identity operator. The full Hamiltonian $H$ is given by a certain combination of $Q^{(n)}$ ($n\ge0$). Similarly, we denote the number of sites in subregion $A$ ($B$) as $L_A$ ($L_B$), and the number of sites adjacent to the boundary between subregions $A$ and $B$ as $L_{AB}$. The total system size is $L=L_A+L_B$.

Without loss of generality, we assume different $Q^{(n)}$ ($n\ge0$) are orthogonal, namely, their Frobenius inner product $(Q^{(n)},Q^{(m)})=\text{tr}(Q^{(n)}Q^{(m)})\propto \delta_{mn}$. In particular, this indicates that $\text{tr}(Q^{(n)})=\text{tr}(Q^{(n)}Q^{(0)})=0$ if $n\ge1$, so $Q^{(n)}$ ($n\ge1$) only contains traceless terms. Accordingly, the mean value of $q_\alpha^{(n)}$ for $n\ge1$ is zero. The total Hamiltonian $H_A$ is the linear combination of some $Q^{(n)}$.

If a conserved quantity $Q^{(n)}$ in Eq. (\ref{seq-Q-QAB}) satisfies the following order of magnitude bound as $L_A\rightarrow \infty$ and $L_B\rightarrow \infty$, respectively:
\begin{equation}\label{seq-quasilocal}
\frac{||Q^{(n)}_{AB}||}{||Q^{(n)}_{A}\otimes I_B||}=\mathcal{O}\left( \sqrt{\frac{l_{AB}}{L_A}}\right)\ ,\qquad \frac{||Q^{(n)}_{AB}||}{|| I_A\otimes Q^{(n)}_{B}||}=\mathcal{O}\left(\sqrt{\frac{l_{AB}}{L_B}}\right)\ ,
\end{equation}
we define $Q^{(n)}$ as a subregionally quasilocal conserved quantity in subregion $A$ and in subregion $B$, respectively. Here $||Q||=\sqrt{\text{tr}(QQ^\dag)}$ is the Frobenius norm of a matrix, and $\mathcal{O}(x)$ stands for up to order $x$. We note that $Q^{(n)}$ can be subregionally quasilocal in both subregions $A$ and $B$ (for instance, when $Q^{(n)}$ is extensive), or only subregionally quasilocal in one subregion $A$ or $B$ if only one condition in Eq. (\ref{seq-quasilocal}) is satisfied (for instance, if $Q^{(n)}$ is localized in one of the subregions). Furthermore, if $Q^{(n)}$ contains only terms with supports within a fixed finite size (independent of $L_A$, $L_B$), we say $Q^{(n)}$ is local. 

Eq. (\ref{seq-quasilocal}) can roughly be understood as follows: if $Q^{(n)}$ is extensive and consists of independent local product terms of similar order of magnitudes, and each term is localized around a site, one can see that $Q^{(n)}_{AB}$ contains order $l_{AB}$ number of independent product terms, while $Q^{(n)}_{A}\otimes I_B$ contains order $L_A$ number of independent product terms, so the ratio of their Frobenius norms is of order $\sqrt{\frac{l_{AB}}{L_A}}$, and similarly for subregion $B$. If $Q^{(n)}$ is instead localized (either in subregion $A$ or $B$), as long as it is not localized at the boundary, we expect $\frac{||Q^{(n)}_{AB}||}{||Q^{(n)}_{A}\otimes I_B||}\sim e^{-c L_A/l_{AB}}$ ($c>0$) if it localizes in $A$ (or similar for $B$). Eq. (\ref{seq-quasilocal}) is then an overestimation (for the corresponding localized subregion) and thus satisfied. In particular, we require the Hamiltonian $H$ to be subregionally (quasi)local.

By the definition of Eq. (\ref{seq-quasilocal}), if a set of conserved quantities $Q^{(n)}\in\text{Loc}$ are subregionally (quasi)local in subregion $A$ or $B$ and mutually commuting, one can show their subregion restrictions satisfy
\begin{equation}\label{seq-comm-error}
\frac{|| [H_A,Q^{(n)}_{A}] ||}{||H_A Q^{(n)}_{A}||}\sim \frac{|| [Q^{(n)}_{A},Q^{(m)}_{A}] ||}{||Q^{(n)}_{A} Q^{(m)}_{A}||}=\mathcal{O}\left( \frac{l_{AB}}{L_A}\right)\ ,\qquad \frac{|| [H_B,Q^{(n)}_{B}] ||}{||H_B Q^{(n)}_{B}||}\sim \frac{|| [Q^{(n)}_{B},Q^{(m)}_{B}] ||}{||Q^{(n)}_{B} Q^{(m)}_{B}||}=\mathcal{O}\left( \frac{l_{AB}}{L_B}\right)\ ,
\end{equation}
respectively. For extensive quantities, both equations in Eq. (\ref{seq-comm-error}) will be satisfied, which can be understood by noting that $[Q^{(n)}_{A},Q^{(m)}_{A}]$ contains up to $l_{AB}^2$ local product terms, while $Q^{(n)}_{A}Q^{(m)}_{A}$ contains roughly $L_A^2$ local product terms. For localized quantities in subregion $A$ ($B$), the left (right) equation in Eq. (\ref{seq-comm-error}) holds, and the error is overestimated, which would be $\mathcal{O}(e^{-cL_A/l_{AB}})$ ($\mathcal{O}(e^{-cL_B/l_{AB}})$) for some $c>0$. 

In some sense, Eq. (\ref{seq-comm-error}) can be viewed as an equivalent definition of subregionally quasilocal conserved quantities, which are mutually commuting in the entire system, but their subregion restrictions commute up to relative errors $\mathcal{O}\left( \frac{l_{AB}}{L_A}\right)$.

\subsection{Conserved quantities in the entanglement Hamiltonian}\label{sec:QA-diag}

We now discuss the effect of conserved quantities $Q^{(n)}$ in a system on the statistical average value of the wavefunction correlation $ u_{\alpha,\alpha_A,\alpha_B}u_{\alpha,\alpha_A',\alpha_B'}^*$ under the subregion eigenstate direct product basis $|\alpha_A,A\rangle\otimes|\alpha_B,B\rangle$, and further on the entanglement Hamiltonian. 

\subsubsection{Three cases of conserved quantities}
There are the following three cases of conserved quantities which we need to distinguish:

--- \emph{Case (i)}. If the conserved quantity $Q^{(n)}$ ($n\ge1$) is subregionally local or quasilocal, and is \emph{extensive} (i.e., containing local terms around all sites of the system), both equations in Eq. (\ref{seq-comm-error}) will be satisfied. In the large system size limit $l_{AB}\ll L_A,L_B$, 
one can approximately regard $[H_A,Q^{(n)}_{A}]=0$ and $[H_B,Q^{(n)}_{B}]=0$ as true, namely, $Q^{(n)}_A$ and $Q^{(n)}_B$ are approximate conserved quantities in subregions $A$ and $B$, respectively. Different subregionally (quasi)local $Q^{(n)}_A$ ($Q^{(n)}_{B}$) also approximately commute. We therefore assume that the subregion energy eigenstates approximately satisfy 
\begin{equation}
Q^{(n)}_{A}|\alpha_A,A\rangle=q_{A,\alpha_A}^{(n)}|\alpha_A,A\rangle\ ,\qquad Q^{(n)}_{B}|\alpha_B,B\rangle=q_{B,\alpha_B}^{(n)}|\alpha_B,B\rangle\ .
\end{equation}
Since $Q^{(n)}$ ($n\ge1$) is traceless, $Q^{(n)}_{A}$, $Q^{(n)}_{B}$ and $Q^{(n)}_{AB}$ should also be traceless, and thus the mean values of $q_{A,\alpha_A}^{(n)}$ and $q_{B,\alpha_B}^{(n)}$ should vanish. The non-commuting errors of $Q^{(n)}_{A}$ will be discussed in the next subsection \ref{sec:non-comm}.

We further assume that the boundary coupling term $Q^{(n)}_{AB}$ exhibit certain randomless in its off diagonal elements in the subregion eigenstate direct product basis $|\alpha_A,A\rangle\otimes|\alpha_B,B\rangle$, similar to $H_{AB}$ in Eq. (\ref{seq-HAB-ETH}). Then, in analogy to Eq. (\ref{seq-uu-correlation1}), under the random average of the off diagonal part of $Q^{(n)}_{AB}$, we expect the wavefunction correlation to satisfy $u_{\alpha,\alpha_A,\alpha_B}u_{\alpha,\alpha_A',\alpha_B'}^*\propto \delta_{\alpha_A,\alpha_A'}\delta_{\alpha_B,\alpha_B'}$, and to be large only if $|q_\alpha^{(n)}-q_{A,\alpha_A}^{(n)}-q_{B,\alpha_B}^{(n)}-q^{(n,\text{d})}_{\alpha_A,\alpha_B}|$ to be of order $l_{AB}$, where $q^{(n,\text{d})}_{\alpha_A,\alpha_B}$ is the diagonal element of $Q^{(n)}_{AB}$ (which is of order $l_{AB}$). Since $q_\alpha^{(n)}$, $q_{A,\alpha_A}^{(n)}$ and $q_{B,\alpha_B}^{(n)}$ are of order $L$, $L_A$ and $L_B$, respectively, in the $l_{AB}\ll L_A,L_B$ limit, we approximately have 
\begin{equation}
u_{\alpha,\alpha_A,\alpha_B}u_{\alpha,\alpha_A',\alpha_B'}^*\propto \delta_{\alpha_A,\alpha_A'}\delta_{\alpha_B,\alpha_B'}\delta(q_\alpha^{(n)}-q_{A,\alpha_A}^{(n)}-q_{B,\alpha_B}^{(n)}-q^{(n,\text{d})}_{\alpha_A,\alpha_B}). 
\end{equation}
Note that $Q^{(n)}_{AB}$ is traceless, the average value of $q^{(n,\text{d})}_{\alpha_A,\alpha_B}$ over all states is zero.

--- \emph{Case (ii)}. If the conserved quantity $Q^{(n)}$ ($n\ge1$) is subregionally (quasi)local, and is localized in subregion $A$ (e.g., in the many-body localization systems), one expects $|| I_A\otimes Q^{(n)}_B ||\sim || Q^{(n)}_{AB} ||\sim e^{-cL_A/l_{AB}} ||Q^{(n)}_A\otimes I_B ||$, for some number $c>0$ (the inverse of the localization length). Therefore, $Q^{(n)}$ is approximately equal to $Q^{(n)}_A\otimes I_B$, and one expects their eigenvalues to be almost equal. If we assume $Q^{(n)}$ is not the polynomial function of another different localized Hermitian conserved quantity (so that its eigenvalues are independent), this would indicate approximately a wavefunction of the product form $u_{\alpha,\alpha_A,\alpha_B}\sim \delta_{\alpha,\alpha_A}\zeta_{\alpha,\alpha_B}$, and thus
\begin{equation}
u_{\alpha,\alpha_A,\alpha_B}u_{\alpha,\alpha_A',\alpha_B'}^*\propto \delta_{\alpha_A,\alpha_A'}\zeta_{\alpha,\alpha_B}\zeta_{\alpha,\alpha_B'}^*\delta(q_\alpha^{(n)}-q_{A,\alpha_A}^{(n)})\ ,
\end{equation}
where the delta function has a width $\Delta_\alpha^{(n)}\propto e^{-c^{(n)} L_A/l_{AB}}$, and $\zeta_{\alpha,\alpha_B}$ is the wavefunction in subregion $B$ which is almost decoupled with subregion $A$ (normalized by $\sum_{\alpha_B} |\zeta_{\alpha,\alpha_B}|^2 =1$). 

A similar conclusion holds for conserved quantities localized in subregion $B$.

--- \emph{Case (iii)}. If the conserved quantity $Q^{(n)}$ is not subregionally quasilocal, in the large system size limit we will have $Q^{(n)}_{AB}$ comparable or even larger than $Q^{(n)}_A\otimes I_B$ and $I_A\otimes Q^{(n)}_B $. Therefore, $Q^{(n)}_A$ or $Q^{(n)}_B$ will not be approximate subregion conserved quantities. In this case, we expect the entire $Q^{(n)}$ matrix to be sufficiently random (due to the $Q^{(n)}_{AB}$ term) in the subregion eigenstate direct product basis $|\alpha_A,A\rangle\otimes|\alpha_B,B\rangle$, and won't contribute to the shape of the statistical average of the wavefunction correlation $ u_{\alpha,\alpha_A,\alpha_B}u_{\alpha,\alpha_A',\alpha_B'}^*$.

\subsubsection{Approximating the entanglement Hamiltonian}

With the arguments in the three above cases, we expect the eigenstate wavefunction correlation in the large system size limit to be approximately
\begin{equation}\label{seq-uu-correlation-Q}
 u_{\alpha,\alpha_A,\alpha_B}u_{\alpha,\alpha_A',\alpha_B'}^*=\frac{1}{N}\delta_{\alpha_A,\alpha_A'}\delta_{\alpha_B,\alpha_B'}\prod_{n\in\text{Loc}_1}\frac{1}{\Omega^{(n)}(q_\alpha^{(n)})}\delta(q_\alpha^{(n)}-q_{A,\alpha_A}^{(n)}-q_{B,\alpha_B}^{(n)}-q^{(n,\text{d})}_{\alpha_A,\alpha_B})\prod_{n\in\text{Loc}_2}\delta(q_\alpha^{(n)}-q_{A,\alpha_A}^{(n)})\ ,
\end{equation}
where 
\begin{equation}
\Omega^{(n)}(q)=\frac{1}{N}\sum_{\alpha}\delta(q-q_{\alpha}^{(n)}) 
\end{equation}
denotes the normalized density of states of the operator $Q^{(n)}$. Besides, $n\in\text{Loc}_1$ runs over all the subregionally (quasi)local conserved quantities $Q^{(n)}$ which are extensive, and $n\in\text{Loc}_2$ runs over all the subregionally (quasi)local conserved quantities $Q^{(n)}$ which are localized in subregion $A$ and are not the polynomial function of another localized conserved quantity. Generically, in the sets $\text{Loc}_1$ and $\text{Loc}_2$, we require $n\ge 1$ (excluding the identity operator), and require Eq. (\ref{seq-comm-error}) to be satisfied (see discussion in the next subsection \ref{sec:non-comm}). Note that when there is only one subregionally (quasi)local extensive conserved quantity, which has to be the Hamiltonian $H$, Eq. (\ref{seq-uu-correlation-Q}) reduces to Eq. (\ref{seq-uu-correlation1}).

This yields a reduced density matrix in subregion A (where we used the fact that the mean value of $q^{(n,\text{d})}_{\alpha_A,\alpha_B}$ is zero for $n\ge1$)
\begin{equation}\label{seq-approx-rhoA-Q}
\begin{split}
&\langle \alpha_A,A|\rho_A(\alpha)|\alpha_A',A\rangle=\sum_{\alpha_B} u_{\alpha,\alpha_A,\alpha_B}u_{\alpha,\alpha_A',\alpha_B}^* \\
&\approx \frac{\delta_{\alpha_A,\alpha_A'} N_B}{N} \prod_{n\in \text{Loc}_1}\int \frac{\Omega_B^{(n)}(q_B^{(n)})}{\Omega^{(n)}(q_\alpha^{(n)})} \delta(q_\alpha^{(n)}-q_{A,\alpha_A}^{(n)}-q_{B,\alpha_B}^{(n)}) dq_B^{(n)} \prod_{n\in\text{Loc}_2}\delta(q_\alpha^{(n)}-q_{A,\alpha_A}^{(n)})  \\
&= \frac{\delta_{\alpha_A,\alpha_A'}}{N_A}\prod_{n\in \text{Loc}_1}\frac{\Omega_B^{(n)}\left(a_A^{(n)}(q_\alpha^{(n)},q_{A,\alpha_A}^{(n)})\right)}{\Omega^{(n)}(q_\alpha^{(n)})} \prod_{n\in\text{Loc}_2}\delta(q_\alpha^{(n)}-q_{A,\alpha_A}^{(n)})\ ,
\end{split}
\end{equation}
where 
\begin{equation}
\Omega_B^{(n)}(q^{(n)}_B)=\frac{1}{N_B}\sum_{\alpha_B}\delta(q-q_{B,\alpha_B}^{(n)}) 
\end{equation}
is the normalized density of states of the subregion conserved quantity $Q^{(n)}_B$, and similar to Eq. (\ref{seq-E-aA}), one expects the function
\begin{equation}
a_A^{(n)}(q_\alpha^{(n)},q_{A,\alpha_A}^{(n)})\approx q_\alpha^{(n)}-q_{A,\alpha_A}^{(n)}
\end{equation}
in the limit $l_{AB}\ll L_A,L_B$.

As we have discussed in the previous subsection, for $n\in\text{Loc}_2$, the delta functions $\delta(q_\alpha^{(n)}-q_{A,\alpha_A}^{(n)})$ in Eq. (\ref{seq-approx-rhoA-Q}) has a width $\Delta_\alpha^{(n)} \propto e^{-c^{(n)} L_A/l_{AB}}$. We can therefore approximate it as Gaussian functions $\propto \frac{1}{\sqrt{2\pi}\Delta_\alpha^{(n)}}e^{-(q_\alpha^{(n)}-q_{A,\alpha_A}^{(n)})^2/2(\Delta_\alpha^{(n)})^2}$. If further we expand over $q_{A,\alpha_A}^{(n)}$ ($n\ge1$) to the first order (in the limit $l_{AB}\ll L_A\ll L_B$), we find an approximate entanglement Hamiltonian of subregion $A$ given by
\begin{equation}\label{seq-KA-conserv-0}
H_E^A(\alpha)\approx \beta^{(0)}_A(\alpha)I_A+\sum_{n\in\text{Loc}_1}\beta_A^{(n)}(\alpha)Q_A^{(n)} + \sum_{n\in\text{Loc}_2}\left[\beta_A^{(n)'}(\alpha)Q_A^{(n)} +\beta_A^{(n)''}(\alpha)\left(Q_A^{(n)}\right)^2 \right] \ .
\end{equation}
Similar to Eq. (\ref{seq-beta-ETH}), and note that the mean values of $q_\alpha^{(n)}$ and $q_{A,\alpha_A}^{(n)}$ are zero, we find the coefficients are approximately given by
\begin{equation}\label{seq:betacoefficients}
\begin{split}
&\beta_A^{(0)}(\alpha)=\log N_A+\sum_{n\in\text{Loc}_1}\log\frac{\Omega^{(n)}(q_\alpha^{(n)})}{\Omega_B^{(n)}(q_\alpha^{(n)})} +\sum_{n\in\text{Loc}_2}\left[ \frac{1}{2}\left( \frac{q_\alpha^{(n)}}{\Delta_\alpha^{(n)}} \right)^2 +\log \sqrt{2\pi}\Delta_\alpha^{(n)} \right]\ , \\
&\beta_A^{(n)}(\alpha)=b^{(n)}_\alpha\frac{\text{d}\log \Omega_B^{(n)}(q)}{\text{d}q}|_{q=q_\alpha^{(n)}}\ ,\ (n\in\text{Loc}_1)\ , \quad  \beta_A^{(n)'}(\alpha)=-\frac{q_\alpha^{(n)}}{\Delta_\alpha^{(n)}}\ ,\quad \beta_A^{(n)''}(\alpha)=\frac{1}{2(\Delta_\alpha^{(n)})^2}\ ,\ (n\in\text{Loc}_2)
\end{split}
\end{equation}
where $b^{(n)}_\alpha=1+\mathcal{O}\left(\frac{q_\alpha^{(n)}l_{AB}}{L_AL_B}\right)$ is a function which tends to $1$ in the limit $l_{AB}\ll L_A,L_B$. Note that $\left(Q_A^{(n)}\right)^2$ is also localized in subregion $A$ if $n\in\text{Loc}_2$. We can redefine a subregionally quasilocal set as
\begin{equation}\label{seq-Loc-set}
\{Q_A^{(n)}| n\in \text{Loc}\} =\{Q_A^{(n)}| n\in \text{Loc}_1\}\cup \{Q_A^{(n)}, (Q_A^{(n)})^2 | n\in \text{Loc}_2\}\ ,
\end{equation} 
and thus we can rewrite Eq. (\ref{seq-KA-conserv-0}) in the form

\begin{equation}\label{seq-KA-conserv}
H_E^A(\alpha)\approx \beta^{(0)}_A(\alpha)I_A+\sum_{n\in\text{Loc}}\beta_A^{(n)}(\alpha)Q_A^{(n)} \ .
\end{equation}
Note that the set $n\in\text{Loc}$ does not include $n=0$, which correspond to the trivial conserved quantity of the identity matrix. More generically, if the actual shape of the finite-width delta functions $\delta(q_\alpha^{(n)}-q_{A,\alpha_A}^{(n)})$ are not Gaussian but other functions, the polynomials (higher than square) of $Q_A^{(n)}$ with $n\in\text{Loc}_2$ may generically be included in the set $\text{Loc}$ in Eq. (\ref{seq-Loc-set}), which are by definition subregionally quasilocal and localized in subregion $A$.

\subsection{Non-commuting errors of the subregion conserved quantities}\label{sec:non-comm}

We now briefly discuss the non-commuting errors of the subregion conserved quantities, and the range of the set $\text{Loc}$ of subregionally (quasi)local conserved quantities. Eq. (\ref{seq-comm-error}) indicates that at finite system sizes, the subregionally quasi-local conserved quantities $Q^{(n)}_{A}$ always have non-commuting errors. In other words, under the subregion energy eigenbasis $|\alpha_A,A\rangle$, the quantities $Q^{(n)}_{A}$ in Eq. (\ref{seq-KA-conserv}) also have random off-diagonal elements in addition to the diagonal elements we discussed in subsection \ref{sec:QA-diag}. Eq. (\ref{seq-comm-error}) limits the magnitude of off-diagonal elements to
\begin{equation}\label{seq-QA-off}
(Q^{(n)}_{A})_{\alpha_A,\alpha_A'}=\mathcal{O}\left(\frac{l_{AB}\sqrt{L_A}}{\sqrt{N_A}} e^{-|q_{A,\alpha_A}^{(n)}-q_{A,\alpha_A'}^{(n)}|/q_0^{(n)}}\right)\ ,\qquad  (\alpha_A\neq \alpha_A')\ ,
\end{equation}
where $q_0^{(n)}$ is an order $1$ number, and the exponential decay is due to the (quasi)locality of $Q^{(n)}_{A}$. Note that $N_A^{-1/2}\sim e^{-S_A/2}$ where $S_A$ is the entropy of the state in subregion $A$, therefore, Eq. (\ref{seq-QA-off}) is in agreement with the estimations in the literature \cite{murthy2019}. Again, we note that this is an overestimation if $Q^{(n)}_{A}$ is localized around a site in subregion $A$.

Eq. (\ref{seq-QA-off}) ensures that $(Q^{(n)}_{A})^p$ for any power $p$ have off-diagonal elements up to order $\mathcal{O}\left(\frac{1}{\sqrt{N_A}}\right)$, and thus the factor $e^{-\beta_A^{(n)}(\alpha)Q_A^{(n)}}$ in the reduced density matrix $\rho_A(\alpha)$ is not far away from a diagonal matrix (or equivalently, the delta function in Eq. (\ref{seq-uu-correlation-Q}) holds up to order $1$).

Therefore, we conjecture that Eq. (\ref{seq-QA-off}), or equivalently, Eq. (\ref{seq-comm-error}), gives the criterion for $Q^{(n)}_{A}$ to contribute to the entanglement Hamiltonian (with nonzero weight $\beta_A^{(n)}$ in Eq. (\ref{seq-KA-conserv})), namely, the criterion for $n\in\text{Loc}$ (the set of mutually commuting subregionally (quasi)local conserved quantities). More explicitly, we conjecture that $Q^{(n)}_{A}$ will contribute to the entanglement Hamiltonians $H_E^A(\alpha)$ with a nonzero weight (i.e., $n\in \text{Loc}$) if and only if
\begin{equation}\label{seq-QA-criterion}
\frac{|| [H_A,Q^{(n)}_{A}] ||}{||H_A Q^{(n)}_{A}||}\sim \frac{|| [Q^{(n)}_{A},Q^{(m)}_{A}] ||}{||Q^{(n)}_{A} Q^{(m)}_{A}||}=\mathcal{O}\left( \frac{l_{AB}}{L_A}\right)\ ,\qquad (\forall \ n,m\in\text{Loc}),
\end{equation}
in the limit $l_{AB}\ll L_A$. This is nothing but a rewriting of Eq. (\ref{seq-comm-error}). As we will see (in Sec. \ref{sec:EHSM-free-2D}), this criterion of Eq. (\ref{seq-QA-criterion}) works intriguingly well for free fermions. For interacting models, Eq. (\ref{seq-QA-criterion}) does not seem to set a sharp boundary for the set $\text{Loc}$ (which may be limited to finite sizes of our numerical exact diagonalization).

\section{Diagonalization of the EHSM}\label{app:EHSMdiag}

When the entanglement Hamiltonians $H_E^{A}(\alpha)$ in subregion $A$ of an ensemble of the full region eigenstates $\alpha\in\Xi$ are known, all of which have the form of Eq. (\ref{seq-KA-conserv}), we can solve for the conserved quantities and their weights in the entanglement Hamiltonians. 

To do this, we can regard each entanglement Hamiltonian $H_E^{A}(\alpha)$ (written down in a certain Hilbert space basis $|j\rangle$) as a vector in the linear space of $N_A\times N_A$ matrices. More explicitly, we can define a matrix basis $|\zeta_{ij} )$ $(1\le i,j\le N_A)$ which represents an $N_A\times N_A$ matrix with matrix elements $\delta_{ii'}\delta_{jj'}$ in row $i'$ and column $j'$. Here half parenthesis instead of ket is used to denote the matrix basis, to avoid confusion with the quantum state basis. We can then rewrite the entanglement Hamiltonian $H_E^A(\alpha)$ as a vector $|H_E^A(\alpha) )=\sum_{i,j}H_{E,ij}^A(\alpha)|\zeta_{ij} )$, where $H_{E,ij}^A$ represent the matrix elements of $H_{E}^A$ in row $i$ and column $j$. Thus, the inner product between two matrices $P$ and $Q$ are exactly the Frobenius inner product, namely, $(Q|P)=\text{tr}(Q^\dag P)$.

As given in the main text Eq. (8), we then define the entanglement Hamiltonian superdensity matrix (EHSM) for the set of entanglement Hamiltonians:
\begin{equation}\label{seq-EHSM1}
R_A=\sum_{\alpha\in\Xi}\frac{w_\alpha}{N_A}|H_E^A(\alpha)) ( H_E^A(\alpha)|=\sum_{\alpha\in\Xi}\sum_{iji'j'}\frac{w_\alpha}{N_A}H_{E,ij}^A(\alpha)H_{E,i'j'}^{A*}(\alpha)|\zeta_{ij})(\zeta_{i'j'}|,
\end{equation}
where for generality we assume one is free to set a weight $w_\alpha > 0$ for each eigenstate $|\alpha\rangle$ in the ensemble $\Xi$, and $\sum_{\alpha\in\Xi}w_\alpha=1$. Note that the EHSM $R_A$ is a size $N_A^2\times N_A^2$ matrix. Assume the EHSM can be diagonalized into into
\begin{equation}\label{seq-EHSM2}
R_A=\sum_{n\ge0} p_{A,n}|\overline{Q}_A^{(n)})( \overline{Q}_A^{(n)}|\ ,
\end{equation}
where the eigenvectors $|\overline{Q}_A^{(n)})$ are orthonormal, namely, they satisfy $(\overline{Q}_A^{(m)}|\overline{Q}_A^{(n)})=\text{tr}(\overline{Q}_A^{(m)\dag}\overline{Q}_A^{(n)})=\delta_{mn}$. We note that the matrices $\overline{Q}_A^{(n)}$ as normalized eigenvectors here are not necessarily equal to $Q_A^{(n)}$ in Eq. (\ref{seq-KA-conserv}), although we expect that they span the same linear space of size $N_A\times N_A$ matrices. Generically, a physical local or quasilocal conserved quantity would have a Frobenius norm $||Q_A^{(n)}||\propto \sqrt{N_AL_A}$, therefore, we expect the physical conserved quantities $Q_A^{(n)}\sim \sqrt{N_AL_A}\  \overline{Q}_A^{(n)}$.

Generically, the EHSM is a big matrix of size $N_A^2\times N_A^2$. When the number of eigenstates $N_\Xi$ in the ensemble $\Xi$ is much smaller than $N_A^2$, an efficient way of diagonalizing $R_A$ is to diagonalize the following $N_\Xi\times N_\Xi$ entanglement correlation matrix $K_A$ with matrix elements
\begin{equation}\label{seq-SA}
K _{A,\alpha\beta}=\frac{\sqrt{w_\alpha w_\beta}}{N_A} (H_E^A(\alpha)|H_E^A(\beta))\ ,
\end{equation}
where $\alpha,\beta\in\Xi$. We now prove that the eigenvalues of the correlation matrix $K_A$ are the same as the nonzero eigenvalues of the EHSM $R_A$, and their eigenvectors have a simple relation. Assume the $n$-th eigenvector of $K_A$ is $v_n^A$ ($n\ge0$), which satisfies
\begin{equation}\label{seq-SA-eigenstate}
\sum_\beta K_{A,\alpha\beta}v_{n,\beta}^A=p_{A,n} v_{n,\alpha}^A\ ,\qquad  v_m^{A\dag}v_n=\sum_\alpha v_{m,\alpha}^{A*}v_{n,\alpha}^A=\delta_{mn}\ ,
\end{equation}
where we assume the eigenvalues $p_{A,n}$ are ranked in a descending order. We can then define a vector in the matrix space $|\overline{Q}_A^{(n)})=\frac{1}{\sqrt{N_A p_{A,n}}}\sum_\beta \sqrt{w_\beta}v_{n,\beta}^A|H_E^A(\beta))$, which has been normalized, namely, $(\overline{Q}_A^{(n)}|\overline{Q}_A^{(n)})=1$. One can then verify that $|\overline{Q}_A^{(n)})$ is an eigenvector of the EHSM $R_A$ with eigenvalue $p_{A,n}$:
\begin{equation}\label{seq-SA-RA-eigenstate}
\begin{split}
&R_A|\overline{Q}_A^{(n)})=\frac{1}{\sqrt{N_A p_{A,n}}}\sum_{\alpha\beta} w_\alpha \sqrt{w_\beta}v^A_{n,\beta} |H_E^A(\alpha))( H_E^A(\alpha)| H_E^A(\beta) ) =\frac{1}{\sqrt{N_A p_{A,n}}}\sum_{\alpha\beta} \sqrt{w_\alpha}|H_E^A(\alpha)\rangle K_{A,\alpha\beta} v^A_{n,\beta} \\
&=\frac{1}{\sqrt{N_A p_{A,n}}}\sum_\alpha p_{A,n} \sqrt{w_\alpha}v_{n,\alpha}^A|H_E^A(\alpha)\rangle =p_{A,n}|\overline{Q}_A^{(n)})\ ,
\end{split}
\end{equation}
where we have used Eq. (\ref{seq-SA-eigenstate}). Note that the rank of matrices $R_A$ and $K_A$ are both equal to $N_\Xi$, so they have equal number of nonzero eigenvalues. Therefore, we conclude that the eigenvalues $p_{A,n}$ of matrix $K_A$ are exactly equal to all the nonzero eigenvalues of $R_A$, and the relation between their corresponding eigenvectors are given by Eq. (\ref{seq-SA-RA-eigenstate}). Numerically, it is easier to diagonalize $K_A$ if the number of eigenstates $N_\Xi < N_A^2$, which is equivalent to the diagonalization of $R_A$.

\section{The EHSM of free fermions}\label{sec:EHSM-free}

In this section, we give the details on the diagonalization of the EHSM of the free fermion model given in main text Eq. (10), namely, 
\begin{equation}\label{seq-freefermion-H}
H=-t\sum_{\langle ij\rangle}(c_{\mathbf{r}_i}^\dag c_{\mathbf{r}_j}+h.c.)+\sum_{j} \mu_j c_{\mathbf{r}_j}^\dag c_{\mathbf{r}_j}\ ,
\end{equation}
which has one fermion degree of freedom per site $i$, where $t$ is the nearest neighbor hopping, and $\mu_i$ is a random potential distributed within the interval $[-W,W]$. 

\subsection{Method of calculation}

We assume the $l$-th single-particle eigenstate wavefunctions of the full region are $\phi_{l,i}$ ($1\le l\le L$), with the single-particle energy being $\epsilon_l$. Accordingly, the many-body eigenstates are generically given by
\begin{equation}
|\alpha\rangle=\prod_{l=1}^L (f^\dag_l)^{\eta_{\alpha,l}}|0\rangle\ ,\qquad f^\dag_l=\sum_i\phi_{l,i}c^\dag_{\mathbf{r}_i}\ ,
\end{equation}
where $f^\dag_l$ are the single-particle eigenstate fermion creation operators, $\eta_{\alpha,l}=0$ or $1$ is the occupation number of single-particle state $l$, and $|0\rangle$ is the particle number vacuum with zero fermions. According to Ref. \cite{peschel2003}, the entanglement Hamiltonian of the free fermion state $|\alpha\rangle$ in subregion $A$ is given by
\begin{equation}\label{seq-HE-free}
H_E^A(\alpha)=\gamma_A(\alpha)I_A+ \sum_{i,j\in A}\kappa_{A,ij}(\alpha)c^\dag_{\mathbf{r}_i} c_{\mathbf{r}_j}\ , 
\end{equation}
where the matrix $\kappa_A(\alpha)$ and the number $\gamma_A(\alpha)$ are defined by
\begin{equation}\label{seq-kappa-gamma}
\kappa_A(\alpha)=\log(\mathcal{C}_A(\alpha)^{-1}-I_A)\ ,\quad \gamma(\alpha)=-\text{tr}\log(I_A-\mathcal{C}_A(\alpha))\ ,
\end{equation}
in terms of the $L_A\times L_A$ 2-particle correlation matrix $\mathcal{C}_A(\alpha)$ with matrix elements
\begin{equation}\label{seq-2p-corr}
\mathcal{C}_{A,ij}(\alpha)=\langle \alpha|c^\dag_{\mathbf{r}_i}c_{\mathbf{r}_j}|\alpha\rangle=\sum_{l=1}^L \eta_{\alpha,l}\phi_{l,i}\phi_{l,j}^*\ ,\qquad (i,j\in A)
\end{equation}
Note that $H_E^A(\alpha)$ is a many-body Hamiltonian of size $N_A\times N_A$, while $\kappa_A(\alpha)$ and $\mathcal{C}_A(\alpha)$ are Hermitian matrices of size $L_A\times L_A$ (recall that $L_A$ is the number of sites, and $N_A=2^{L_A}$ in this model). In this case, the Frobenius inner products of entanglement Hamiltonians in Eq. (\ref{seq-SA}) are given by
\begin{equation}
\begin{split}
&\langle H_E^A(\alpha)|H_E^A(\beta)\rangle=\text{tr}(H_E^A(\alpha)H_E^A(\beta)) \\
&=\sum_{iji'j'\in A}(\kappa_{A}(\alpha))_{ij}(\kappa_{A}(\beta))_{i'j'}\text{tr}(c^\dag_i c_jc^\dag_{i'} c_{j'})+\gamma(\alpha) \sum_{ij\in A}(\kappa_{A}(\beta))_{ij}\text{tr}(c^\dag_i c_j)+\gamma(\beta) \sum_{ij\in A}(\kappa_{A}(\alpha))_{ij}\text{tr}(c^\dag_i c_j)+\gamma(\alpha)\gamma(\beta)\text{tr}I_A\\
&=\sum_{iji'j'\in A}(\kappa_{A}(\alpha))_{ij}(\kappa_{A}(\beta))_{i'j'} \frac{N_A}{4}(\delta_{ij}\delta_{i'j'}+\delta_{i'j}\delta_{ij'})+\gamma(\alpha)\sum_i \frac{N_A}{2} (\kappa_{A}(\beta))_{ii}+\gamma(\beta)\sum_i \frac{N_A}{2} (\kappa_{A}(\alpha))_{ii}+\gamma(\alpha)\gamma(\beta)N_A\\
&=N_A\Big\{\frac{1}{4}\text{tr}[\kappa_{A}(\alpha)\kappa_{A}(\beta)]+ \Big( \frac{1}{2}\text{tr}[\kappa_{A}(\alpha)]+\gamma(\alpha)\Big)\Big(\frac{1}{2}\text{tr}[\kappa_{A}(\beta)]+\gamma(\beta)\Big)\Big\}\ .
\end{split}
\end{equation}
This greatly simplifies the calculation of the matrix $K_A$ in Eq. (\ref{seq-SA}), and thus the diagonalization of the EHSM $R_A$ of free fermions.

\subsection{Numerical calculations}

We numerically diagonalize the EHSM of the free fermion Anderson model (\ref{seq-freefermion-H}) in both 1D and 2D (the lattices of which are illustrated in the main text Fig. 1):

(i) In the 2D case, the system is in a lattice with $L_x=60$ and $L_y=10$ sites in the $x$ and $y$ directions, with periodic boundary condition in both directions. We set the lattice constant in both directions to be $1$. The full 2D number of sites is thus $L=L_xL_y=600$. The 2D subregion $A$ is defined to be the region of sites within $x$ coordinate $1\le x_j\le L_{A,x}$ as shown in the main text Fig. 1a (here $(x_j,y_j)$ is the 2D coordinate of site $j$), which has number of sites $L_A=L_{A,x}L_y$. 

(ii) In the 1D case, we set the number of sites in the system to be $L=500$, with a periodic boundary condition. The 1D subregion $A$ is chosen to be the region of sites within the $x$ coordinate $1\le x_j\le L_A$.

In both cases, we calculate and diagonalize the EHSM for an ensemble $\Xi$ of randomly chosen $N_\Xi=1000$ eigenstates of the full system, and each of the eigenstates has an equal weight $w_\alpha=1/N_\Xi$ in the EHSM in Eq. (\ref{seq-EHSM1}). We note that if the 2-particle correlation matrix $\mathcal{C}(\alpha)$ in Eq. (\ref{seq-2p-corr}) has eigenvalues reaching $0$ or $1$, the entanglement Hamiltonian coefficients in Eq. (\ref{seq-kappa-gamma}) will encounter divergence. To avoid such numerical divergences, we relax the $0$ and $1$ eigenvalues of $\mathcal{C}(\alpha)$, if any, into $\delta$ and $1-\delta$, respectively, with $\delta$ being a sufficiently small positive number. In practice, we set $\delta=10^{-16}$. We note that $\mathcal{C}(\alpha)$ has almost $0$ or $1$ eigenvalues only when the single-particle eigenstates of the system are localized. We also verified that the numerical EHSM eigenvalues are insensitive to the cutoff $\delta$.

\begin{figure}[tbp]
\begin{center}
\includegraphics[width=6.8in]{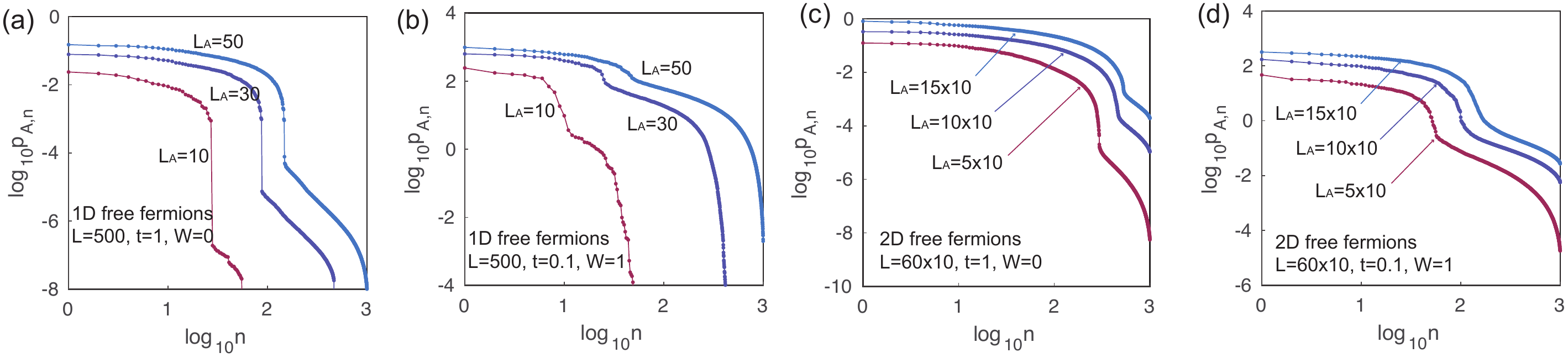}
\end{center}
\caption{The log-log plot ($\log_{10}p_{A,n}$ vs. $\log_{10}n$) of the EHSM eigenvalues $p_{A,n}$ for free fermion models in 1D ((a)-(b)) and in 2D ((c)-(d)), which are calculated for randomly chosen 1000 many-body eigenstates. The Hamiltonian is given by Eq. (\ref{seq-freefermion-H}), and the parameters are labeled in the panels. For 1D ((a)-(b)), the system size is $L=500$, and the $A$ subsystem sizes $L_A=5,10,15$ are considered. For 2D ((c)-(d)), the system size is given by $L=L_xL_y$, with $L_x=60$ and $L_y=10$ fixed, while the $A$ subsystem size is $L_A=L_{Ax}L_y$ with $L_{Ax}=5,10,15$ examined. In (a) and (c) when the single-particle states are delocalized, $p_{A,n}$ shows a sharp cutoff towards zero around $n=3 L_A$ and $3L_A\le n\le 7L_A$, respectively. In contrast, in (b) and (d) where the single-particle states are localized, $p_{A,n}$ shows a sharp cutoff towards zero around $n=L_A$.
}
\label{figS-free}
\end{figure}

The leading eigenvalues $p_{A,n}$ of the EHSM for several different parameters in 2D and 1D are shown in the main text Fig. 2 in a descending order (the values of $p_{A,n}/L_A$ are plotted). In 2D, we show the results for $A$ subregion sizes $L_{Ax}=5,10,15$ (corresponding to $L_A=50,100,150$), while in 1D, we show the results for $A$ subregion sizes $L_A=10,30,50$. The logarithm $\log_{10} p_{A,n}$ of the eigenvalues $p_{A,n}$ with respect to $\log_{10} n$ are shown in Fig. \ref{figS-free}. We now discuss the results for extended fermions and localized fermions, respectively.

\subsection{Conserved quantities of extended free fermions}

Fig. \ref{figS-free} (a) and (c) shows the results for $t=1$ and $W=0$, in 1D and 2D, respectively. In this case, the single-particle states of the entire system are delocalized plane waves. 
As shown in Fig. \ref{figS-free} (a) and (c), in this case with extended single-particle fermion eigenstates, we find the eigenvalues $p_{A,n}$ decays to $0$ around $n=zL_A$  (which corresponds to the sharp drop in the log-log plot in Fig. \ref{figS-free}), where $z=3$ for 1D, and $3\lesssim z\lesssim 7$ in 2D. Indeed, if we examine the subregion $A$ which has open boundary condition in the $x$ direction (and periodic boundary condition in the $y$ direction in 2D), we can approximately find $3L_A$ single-body conserved quantities in 1D and $3L_A$ to $7L_A$ single-body conserved quantities in 2D,  which agree well with the EHSM eigen-operators, as we will explain below.

\subsubsection{The 1D case}\label{sec:EHSM-free-1D}

We first examine the 1D model. Assume the $x$ coordinate of subregion $A$ of the 1D lattice ranges from $1$ to $L_{A}$. With an open boundary condition, the subregion $A$ eigenstates are standing waves with creation operators
\begin{equation}
\frac{c^\dag_{A,k_x}-c^\dag_{A,-k_x}}{\sqrt{2}}\ ,
\end{equation}
where $c^\dag_{A,k_x}=\frac{1}{\sqrt{L_A}}\sum_{j\in A} e^{-ik_x x_j} c^\dag_{x_i}$, and the momentum takes values $k_x=\frac{\pi m_x}{L_{A}+1}$, $m_x\in\mathbb{Z}_+$ and $1\le m_x\le L_A$. The Hamiltonian can thus be written as
\begin{equation}
\begin{split}
H_A&=-\sum_{k_x>0} 2t\cos k_x \left(\frac{c^\dag_{A,k_x}-c^\dag_{A,-k_x}}{\sqrt{2}} \right) \left(\frac{c_{A,k_x}-c_{A,-k_x}}{\sqrt{2}} \right)\\
&=- \sum_{k_x>0} t\cos k_x \left[ \left(c^\dag_{A,k_x}c_{A,k_x}+c^\dag_{A,-k_x}c_{A,-k_x} \right) - \left(c^\dag_{A,-k_x}c_{A,k_x}+c^\dag_{A,k_x}c_{A,-k_x} \right)\right]\ .
\end{split}
\end{equation}
One can then easily see the following quantities are conserved:
\begin{equation}
\widetilde{T}^A_{k_x}=c^\dag_{A,k_x}c_{A,k_x}+c^\dag_{A,-k_x}c_{A,-k_x}\ ,\qquad \widetilde{P}^A_{k_x}=c^\dag_{A,-k_x}c_{A,k_x}+c^\dag_{A,k_x}c_{A,-k_x}\ .
\end{equation}
By Fourier transformation, these conserved quantities can be linear recombined into the real space form:
\begin{equation}\label{seq-TP-free-1D}
\begin{split}
&T_{x}^A=\sum_{k_x}e^{ik_x x}\widetilde{T}^A_{k_x}\approx \sum_{x_i,x+x_i\in A} (c^\dag_{x_i+x}c_{x_i}+c^\dag_{x_i}c_{x_i+x} )\ , \qquad (0\le x< L_A) \\
&P_{x}^A=\sum_{k_x}e^{-ik_x x}\widetilde{P}^A_{k_x}\approx \sum_{x_i,x-x_i\in A} c^\dag_{x-x_i}c_{x_i}\ ,\qquad (2\le x\le 2L_A)
\end{split}
\end{equation}
which satisfy the approximately commuting criterion in Eq. (\ref{eq-QA-criterion}). 
In particular, we see that there are $L_A$ nonvanishing operators $T_{x}^A$ ($0\le x< L_A$), and approximately $2L_A$ nonvanishing operators $P_x^A$ ($2\le x\le 2L_A$).

\begin{figure}[tbp]
\begin{center}
\includegraphics[width=6.8in]{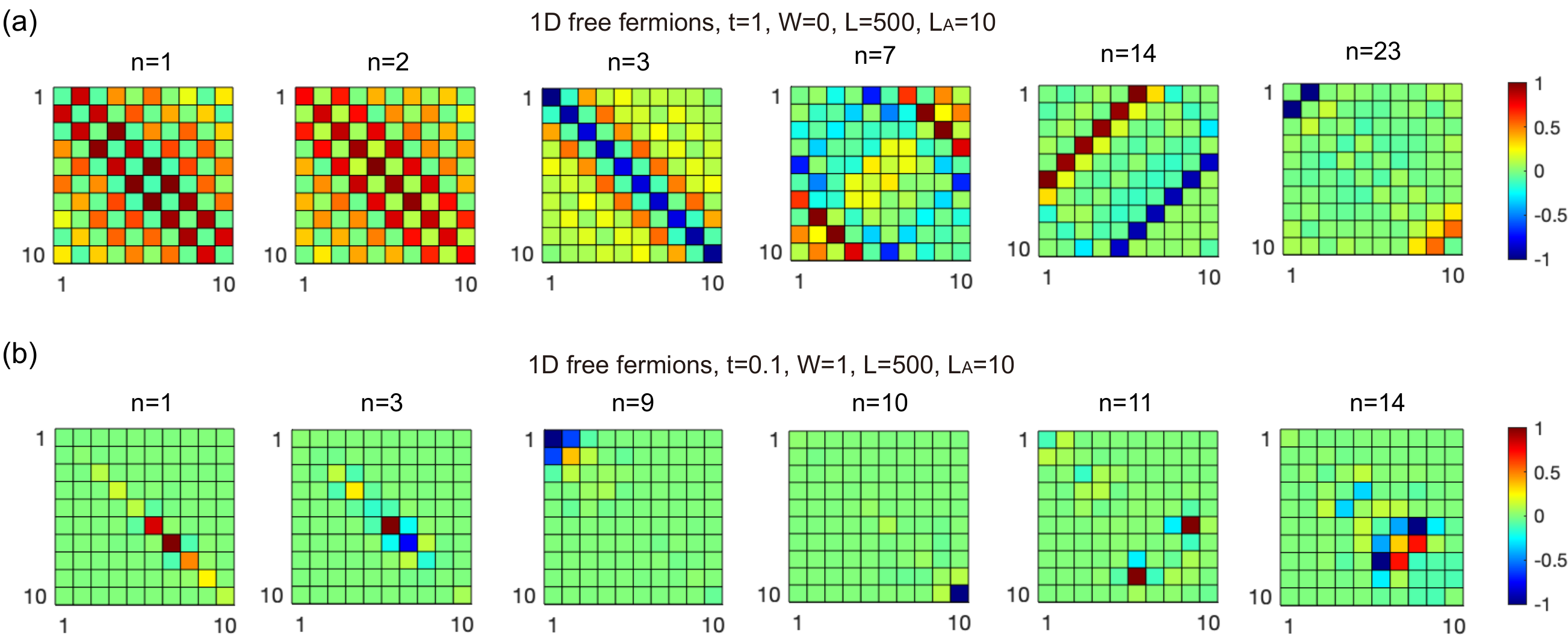}
\end{center}
\caption{The single-body matrices $\kappa_{A,ij}^{(n)}$ (as defined in Eq. (\ref{seq-Qn-free})) of eigen-operators $\overline{Q}^{(n)}$ ($n\ge 1$) for 1D free fermions, where the entire system size is $L=500$, and subregion size is $L_A=10$. The $x$ and $y$ axes of each panel give the row and column indices of the matrices $\kappa_{A,ij}^{(n)}$, with $i,j$ sorted along the lattice of subregion $A$ from the left to the right. The colorbar values for each $n$ are given in units of the maximal absolute value of matrix element $\kappa_{A,ij}^{(n)}$. (a) shows several examples for $t=1$, $W=0$ where the fermions are extended, while (b) shows a few examples for $t=0.1$ and $W=1$ where the fermions are localized.
}
\label{figS-free-Qn}
\end{figure}

The linear combinations of the $3L_A$ operators in Eq. (\ref{seq-TP-free-1D}) then give the $3L_A$ eigen-operators with nonzero EHSM eigenvalues $p_{A,n}$ ($n>0$) for 1D extended fermions. To see this, we investigate the numerically obtained EHSM eigen-operators $\overline{Q}^{(n)}$ of extended free fermions in 1D. From Eq. (\ref{seq-HE-free}), we know that $\overline{Q}^{(n)}$ is of the fermion bilinear form 
\begin{equation}\label{seq-Qn-free}
\overline{Q}_A^{(n)}=\gamma_A^{(n)}I_A+ \sum_{i,j\in A}\kappa_{A,ij}^{(n)}c^\dag_{\mathbf{r}_i} c_{\mathbf{r}_j}\ , 
\end{equation}
where $\gamma_A^{(n)}$ is some constant, and $\kappa_{A,ij}^{(n)}$ is a matrix of size $L_A\times L_A$. Numerically, we find that the $n=0$ quantity is dominantly $\overline{Q}_A^{(0)}\propto I_A$, while for $n\ge 1$ we approximately have $\text{tr}(\overline{Q}_A^{(n)})=0$. In Fig. \ref{figS-free-Qn}(a), we plot examples of the matrices $\kappa_{A,ij}^{(n)}$ ($n\ge 1$) for 1D free fermions with $t=1$, $W=0$, $L=500$ and $L_A=10$, where the horizontal and vertical axis are the row and column indices of the matrices $\kappa_{A,ij}^{(n)}$. We find $\overline{Q}_A^{(n)}$ with $1\le n\le L_A$ are approximately dominated by linear combinations of $T^A_{x}$ in Eq. (\ref{seq-TP-free-1D}), while $\overline{Q}^{(n)}$ with $L_A< n\le 3L_A$ are approximately dominated by linear combinations of $P^A_{x}$ in Eq. (\ref{seq-TP-free-1D}).

\subsubsection{The 2D case}\label{sec:EHSM-free-2D}

We now turn to the 2D case, which is more complicated. Assume the $x$ coordinate of subregion $A$ ranges from $1$ to $L_{A,x}$, and the $y$ coordinate is periodic with total length $L_y$. The total number of sites in subregion $A$ is $L_{A}=L_{A,x}L_y$. The subregion $A$ then has an open boundary condition in the $x$ direction and a periodic boundary condition in the $y$ direction. Therefore, the subregion $A$ eigenstates are standing waves in the $x$ direction, the creation operators of which are given by
\begin{equation}
\frac{c^\dag_{A,k_x,k_y}-c^\dag_{A,-k_x,k_y}}{\sqrt{2}}\ ,
\end{equation}
where $c^\dag_{A,k_x,k_y}=\frac{1}{\sqrt{L_A}}\sum_{j\in A} e^{-ik_x x_j-ik_y y_j} c^\dag_{x_i,y_i}$ is the momentum $\mathbf{k}$ eigenstate, $k_x=\frac{\pi m_x}{L_{A,x}+1}$, $k_y=\frac{2\pi m_y}{L_y}$, which take values $m_x\in\mathbb{Z}_+$, $1\le m_x\le L_{A,x}$ and $m_y\in \mathbb{Z}$, $0\le m_y\le L_y-1$.
Accordingly, the subregion $A$ Hamiltonian $H_A$ can be diagonalized into
\begin{equation}\label{seq-2D-free-HA}
\begin{split}
H_A&=\sum_{k_x>0}\sum_{k_y} \epsilon(\mathbf{k}) \left(\frac{c^\dag_{A,k_x,k_y}-c^\dag_{A,-k_x,k_y}}{\sqrt{2}} \right) \left(\frac{c_{A,k_x,k_y}-c_{A,-k_x,k_y}}{\sqrt{2}} \right)\\
&=\frac{1}{2} \sum_{k_x>0}\sum_{k_y} \epsilon(\mathbf{k}) \left[ \left(c^\dag_{A,k_x,k_y}c_{A,k_x,k_y}+c^\dag_{A,-k_x,k_y}c_{A,-k_x,k_y} \right) - \left(c^\dag_{A,-k_x,k_y}c_{A,k_x,k_y}+c^\dag_{A,k_x,k_y}c_{A,-k_x,k_y} \right)\right]\ ,
\end{split}
\end{equation}
where $\epsilon(\mathbf{k})=-2t(\cos k_x+\cos k_y)$.
Therefore, similar to the 1D case, one can prove the following quantities are conserved quantities and mutually commuting:
\begin{equation}\label{seq-TP-free-2D-0}
\begin{split}
\widetilde{T}^A_{k_x,k_y}=c^\dag_{A,k_x,k_y}c_{A,k_x,k_y}+c^\dag_{A,-k_x,k_y}c_{A,-k_x,k_y}\ ,\qquad \widetilde{P}^{A,1}_{k_x,k_y}=c^\dag_{A,-k_x,k_y}c_{A,k_x,k_y}+c^\dag_{A,k_x,k_y}c_{A,-k_x,k_y}\ .
\end{split}
\end{equation}
One can linear recombine these conserved quantities approximately into the following conserved quantities in the real space:
\begin{equation}\label{seq-TP-free-2D}
\begin{split}
&T_{x,y}^A=\sum_{k_x,k_y}e^{ik_x x+ik_y y}\widetilde{T}^A_{k_x,k_y}\approx \sum_{x_i,x+x_i\in A} (c^\dag_{x_i+x,y_i+y}c_{x_i,y_i}+c^\dag_{x_i,y_i+y}c_{x_i+x,y_i} )\ , \qquad (0\le x<L_{A,x},\ 0\le y<L_y) \\
&P_{x,y}^{A,1}=\sum_{k_x,k_y}e^{-ik_x x+ik_y y}\widetilde{P}^{A,1}_{k_x,k_y}\approx \sum_{x_i,x-x_i\in A} c^\dag_{x-x_i,y_i+y}c_{x_i,y_i}\ , \qquad (2\le x\le 2L_{A,x}, 0\le y <L_y)
\end{split}
\end{equation}
where all the fermion operators $c_\mathbf{r}^\dag$ and $c_\mathbf{r}$ are restricted within subregion $A$. Here $y_j$ identified with $y_j+L_y$, and $x,y\in \mathbb{Z}$. 
Therefore, for quantity $T^A_{x,y}$, we can take $0\le x\le L_{A,x}-1$, and $0\le y\le L_y-1$, which in total yields about $L_A=L_{A,x}L_y$ linearly independent quantities $T^A_{x,y}$. In contrast, for quantity $P^{A,1}_{x,y}$, the $x$ coordinate can take values $2\le 2L_{A,x}$, while $0\le y\le L_y-1$. Therefore, there are in total $2L_A-1$ linearly independent quantities $P^A_{x,y}$. These $3L_A$ quantities $T^A_{x,y}$ and $P^{A,1}_{x,y}$ satisfy the approximately commuting criterion in Eq. (\ref{eq-QA-criterion}), and are thus expected to contribute to the EHSM.


In addition, since the single-particle energy $\epsilon_{\mathbf{k}}$ is even in $k_x$ and $k_y$, there are two another sets of conserved quantities commuting with the 2D Hamiltonian $H_A$ in Eq. (\ref{seq-2D-free-HA}):
\begin{equation}
\begin{split}
&\widetilde{P}^{A,2}_{k_x,k_y}=c^\dag_{A,k_x,-k_y}c_{A,k_x,k_y}+c^\dag_{A,k_x,k_y}c_{A,k_x,-k_y}\ , \qquad 
\widetilde{P}^{A,3}_{k_x,k_y}=c^\dag_{A,-k_x,-k_y}c_{A,k_x,k_y}+c^\dag_{A,k_x,k_y}c_{A,-k_x,-k_y}\ .
\end{split}
\end{equation}
However, they do not commute with $\widetilde{T}^A_{k_x,k_y}$ and $\widetilde{P}^{A,1}_{k_x,k_y}$ in Eq. (\ref{seq-TP-free-2D-0}). Nevertheless, if we Fourier transform them, they can be rewritten as
\begin{equation}\label{seq-TP-free-2D-2}
\begin{split}
&P_{x,y}^{A,2}=\sum_{k_x,k_y}e^{ik_x x-ik_y y}\widetilde{P}^{A,2}_{k_x,k_y}\approx \sum_{x_i,x-x_i\in A} c^\dag_{x_i+x,y-y_i}c_{x_i,y_i}\ , \qquad (-L_{A,x}< x< L_{A,x},\ 0\le y<L_y) \\
&P_{x,y}^{A,3}=\sum_{k_x,k_y}e^{-ik_x x-ik_y y}\widetilde{P}^{A,3}_{k_x,k_y}\approx \sum_{x_i,x-x_i\in A} c^\dag_{x-x_i,y-y_i}c_{x_i,y_i}\ , \qquad (2\le x\le 2L_{A,x},\ 0\le y<L_y)
\end{split}
\end{equation}
Therefore, altogether we have $2L_A$ operators $P_{x,y}^{A,2}$, and $2L_A$ operators $P_{x,y}^{A,3}$.

\begin{figure}[tbp]
\begin{center}
\includegraphics[width=3in]{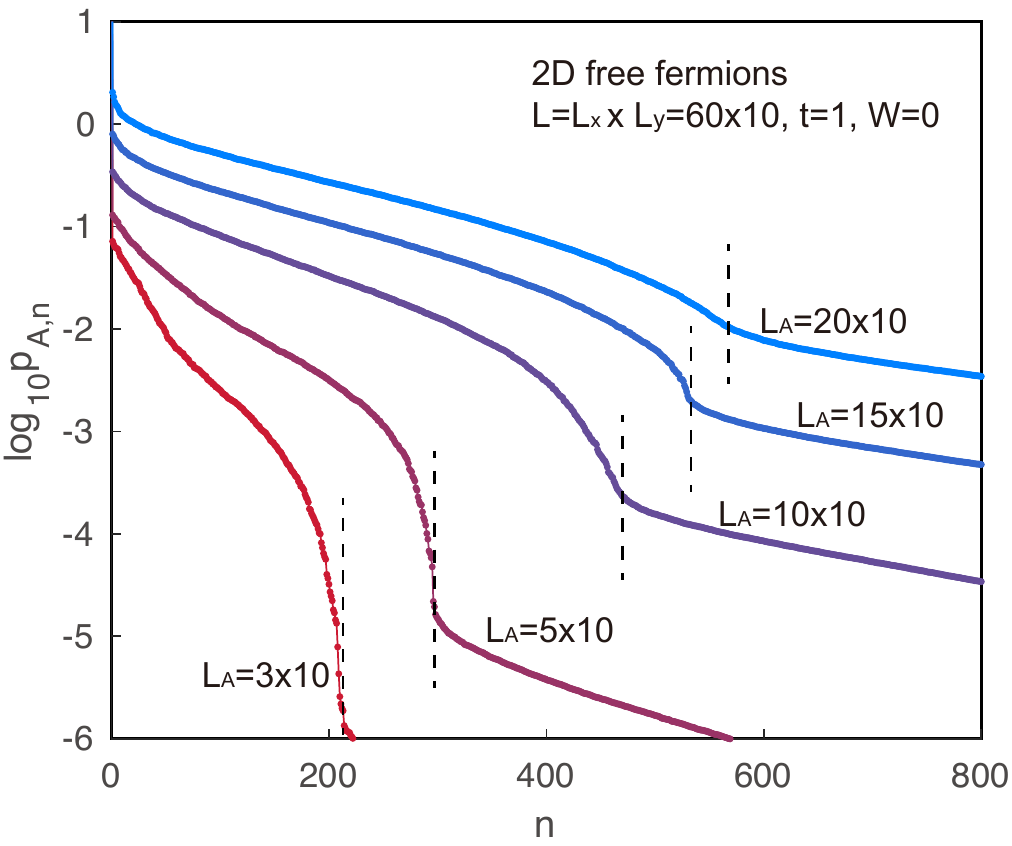}
\end{center}
\caption{The logarithm of the EHSM eigenvalues, $\log_{10} p_{A,n}$, plotted versus $n$ for 2D extended fermions. The full system size is $L=L_xL_y$, $L_x=60$, $L_y=10$, with periodic boundary condition in both directions. The subregion $A$ size is $L_A=L_{A,x}L_y$, with open boundary in the $x$ direction and periodic in the $y$ direction. The calculation is done for $L_{A,x}=3,5,10,15,20$, respectively. The dashed line shows the cutoff position of $p_{A,n}$ where it drops to almost zero (see also main text Fig. 2(a) for the plot of $p_{A,n}$). The cutoff approaches $7L_A$ when $L_{A,x}\ll L_y$, and approaches $3L_A$ when $L_{A,x}\gg L_y$.
}
\label{figS-2Dlogp}
\end{figure}

If one examines the commutation errors of the $4L_A$ additional operators in Eq. (\ref{seq-TP-free-2D-2}) with the $3L_A$ conserved operators in Eq. (\ref{seq-TP-free-2D}), one finds that
\begin{equation}\label{seq-2D-add-Q-1}
\frac{|| [T_{x,y}^{A},P_{x',y'}^{A,j}] ||}{||T_{x,y}^{A} P_{x',y'}^{A,j}||} \sim \frac{|| [P_{x,y}^{A,1},P_{x',y'}^{A,j}] ||}{||P_{x,y}^{A,1} P_{x',y'}^{A,j}||} \propto \frac{1}{\sqrt{L_A}}=\frac{1}{\sqrt{L_{A,x}L_y}}\ , \qquad (j=2,3)\ .
\end{equation}
This seems not satisfying the criterion Eq. (\ref{eq-QA-criterion}). However, by noting that the boundary between $A$ and $B$ in our setup has a size $l_{AB}=2L_y$, thus $\frac{l_{AB}}{L_A}=\frac{2}{L_{A,x}}$, we can rewrite the above equation as
\begin{equation}\label{seq-2D-add-Q-2}
\frac{|| [T_{x,y}^{A},P_{x',y'}^{A,j}] ||}{||T_{x,y}^{A} P_{x',y'}^{A,j}||} \sim \frac{|| [P_{x,y}^{A,1},P_{x',y'}^{A,j}] ||}{||P_{x,y}^{A,1} P_{x',y'}^{A,j}||} \propto \sqrt{\frac{L_{A,x}}{L_y}}\frac{l_{AB}}{L_A}\ , \qquad (j=2,3)\ .
\end{equation}
Therefore, for subregion $A$ with a fixed aspect ratio: 

(i) if $L_{A,x}\ll L_y$, the $4L_A$ additional operators $P_{x,y}^{A,j}$ ($j=2,3$) satisfy the criterion in Eq. (\ref{eq-QA-criterion}), and thus one would expect $7L_A$ approximately conserved quantities $Q_{A}^{(n)}$ with EHSM weights $p_{A,n}>0$.

(ii) if $L_{A,x}\gg L_y$, the $4L_A$ additional operators $P_{x,y}^{A,j}$ ($j=2,3$) would not satisfy Eq. (\ref{eq-QA-criterion}), in which case one expects only $3L_A$ approximately conserved quantities $Q_{A}^{(n)}$ (given by Eq. (\ref{seq-TP-free-2D}) with $p_{A,n}>0$. This is also the quasi-1D limit of the 2D system, and thus in agreement with the 1D case.

In Fig. \ref{figS-2Dlogp}, we plot the logarithm of EHSM eigenvalues $p_{A,n}$ for 2D extended free fermions with different subregion $A$ aspect ratios $L_{A,x}/L_y$. Indeed, as expected above, we find the cutoff of nonzero $p_{A,n}$ is at $zL_A$, with $z\rightarrow 3$ if $L_{A,x}\gg L_y$, and $z\rightarrow 7$ if $L_{A,x}\ll L_y$. Intriguingly, we see the criterion in Eq. (\ref{eq-QA-criterion}) for conserved quantities contributing to the EHSM works well.

\subsection{Conserved quantities of localized free fermions}

In Fig. \ref{figS-free} (b) and (d) (see also the main text Fig. 1 (b) and (d)), we set the parameters to $t=0.02$, $W=1$ (in 2D) and $t=0.1$, $W=1$ (in 1D), respectively, in which case the single-particle states are strongly localized. In this case, we find the EHSM eigenvalues $p_{A,n}$ has a sharp cutoff around $n\approx L_A$: the eigenvalues $p_{A,n}$ with $n>L_A$ become vanishingly small compared to those with $n<L_A$. This can be seen more clearly in the main text Fig. 1 (b) and (d), and can also be seen by noting the kink around $n=L_A$ in Fig. \ref{figS-free} (b) and (d). This is because in the strongly localized limit, the single-particle eigenstates are almost localized on each site, namely, the $l$-th eigenstate fermion operators $f^\dag_l\approx c_{\mathbf{r}_l}^\dag$. As a result, one expect no long range entanglement, and the entanglement Hamiltonian in Eq. (\ref{seq-HE-free}) almost only contains local fermion bilinear terms $c^\dag_{\mathbf{r}_l}c_{\mathbf{r}_l}$ ($l\in A$). This would only yield $L_A$ linearly independent eigen-operators $\overline{Q}^{(n)}_A$ with nonzero $p_{A,n}$ ($1\le n\le L_A$), which are approximately the linear combinations of $2c_{\mathbf{r}_l}^\dag c_{\mathbf{r}_l}-1$ (so written that it is traceless). Besides, numerically we find $\overline{Q}^{(0)}\propto I_A$.

In Fig. \ref{figS-free-Qn}(b), we have plotted the $\kappa_{A,ij}^{(n)}$ ($n\ge 1$) of the eigen-operators $\overline{Q}^{(n)}_A$ (defined in Eq. (\ref{seq-Qn-free})) for 1D free fermions with $t=1$, $W=0$, $L=500$ and $L_A=10$. In the panels, the $x$ and $y$ axis are the row and column indices of the matrices $\kappa_{A,ij}^{(n)}$. As one can see, for $1\le n\le L_A$, the eigen-operators $\overline{Q}^{(n)}_A$ are linear combinations of the occupation numbers of single-particle localized wavefunctions. Two examples with $n>L_A$ ($n=11,14$) are also shown in Fig. \ref{figS-free-Qn}(b), which become less localized. Accordingly, their EHSM eigenvalues $p_{A,n}$ are vanishingly small compared to those of $n\le L_A$.

\section{The EHSM of the 1D XYZ model in a magnetic field}\label{app:XYZ}

\subsection{EHSM eigenvalues and entanglement entropies}

We numerically study the EHSM of the 1D XYZ model, to which we can add either uniform or disordered magnetic fields. The model Hamiltonian is given by the main text Eq. (11), which we rewrite here:
\begin{equation}\label{seq-XYZ}
H=\sum_{j=1}^L\Big[J_x\sigma_{j,x}\sigma_{j+1,x}+ J_y\sigma_{j,y}\sigma_{j+1,y} +J_z\sigma_{j,z}\sigma_{j+1,z} +(\mathbf{B}+\delta\mathbf{B}_{j})\cdot\bm{\sigma}_j\Big]\ .
\end{equation}
Here $\sigma_{j,\nu}$ ($\nu=x,y,z$) are the Pauli matrices on site $j$. We have added both a uniform magnetic field $B$, and a random magnetic field $\delta\mathbf{B}_j$ with each component $\delta B_{j,\nu}$ independently randomly distributed in the interval $[-B_{R,\nu},B_{R,\nu}]$ ($\nu=x,y,z$), in which $\mathbf{B}_R$ is given. We perform the exact diagonalization of model (\ref{seq-XYZ}) for a 1D lattice with $L=14$ sites with periodic boundary condition. We then calculate the EHSM in subregion $A$ of sizes $L_A$ up $7$ (half of the system size). 

In calculating the entanglement Hamiltonians, an entanglement Hamiltonian may have a diverging part if its corresponding reduced density matrix has an exactly zero eigenvalue. This may happen when the Hilbert space is fragmented into non-communicating subspaces. Such exactly zero eigenvalues usually does not occur in $\rho_A$ if the system is delocalized, in which case such fragmented Hilbert subspace would involve both subregions $A$ and $B$, thus is not a closed subspace within the Hilbert space of subregion $A$. However, if the system is localized, one may end up with almost or exact zero eigenvalues in $\rho_A$. In this case, we substitute the (almost) zero eigenvalues by a small number $\epsilon>0$ to avoid divergence. We find the behaviors of the EHSM eigenvalues are rather insensitive to the small number cutoff $\epsilon$. Here we take $\epsilon$ to $10^{-16}$.

In the main text Fig. 3, we diagonalize the EHSM for the ensemble $\Xi$ containing all the $N$ eigenstates of the full region, with equal weights $w_\alpha=1/N$ for all the eigenstates, and plot the EHSM eigenvalues $p_{A,n}$ with respect to $n$. The six panels of main text Fig. 3 correspond to six different representative sets of parameters (labeled at the top of the panels, see also below), and the sizes of  subregion $A$ we examined are $L_A=4,5,6,7$. Fig. \ref{figS-logp-logn} shows the log-log plot of the main text Fig. 3, namely, $\text{log}_{10}p_{A,n}$ as a function of $\text{log}_{10} n$. 


Fig. \ref{figS-SA} shows the subregion $A$ (size $L_A=7$) entanglement entropies 
\begin{equation}
S_A(\alpha)=-\text{tr}(\rho_A(\alpha)\log \rho_A(\alpha))
\end{equation}
of all the eigenstates $|\alpha\rangle$ of the full region plotted versus the eigenstate energies $E_\alpha$. The parameters of the 6 panels are the same as those in the main text Fig. 3 (and Fig. \ref{figS-logp-logn}).

\begin{figure}[tbp]
\begin{center}
\includegraphics[width=6in]{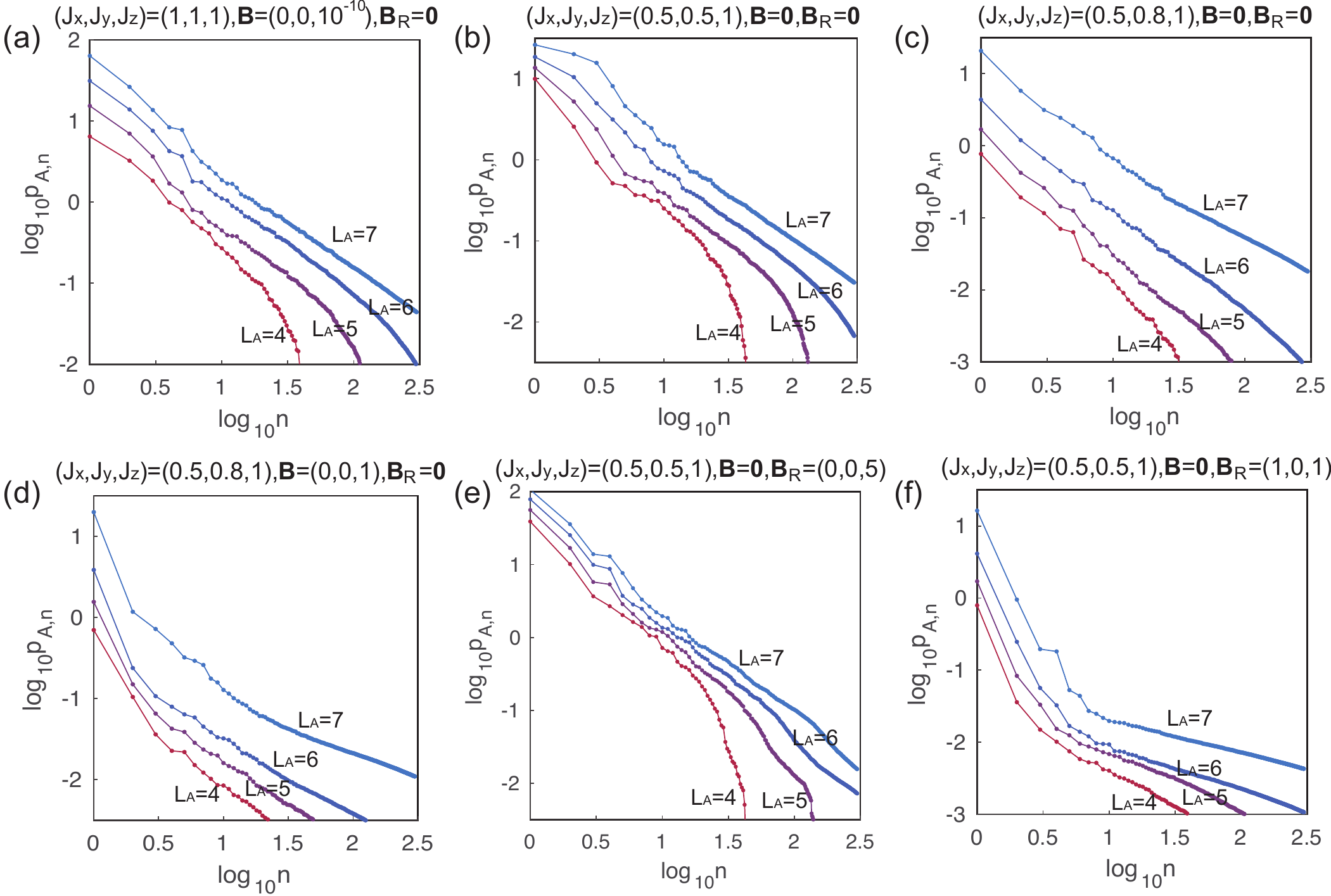}
\end{center}
\caption{The log-log plot of $\text{log}_{10}p_{A,n}$ vs. $\text{log}_{10} n$ for the EHSM of the 1D XYZ model, calculated for an ensemble $\Xi$ of all the eigenstates $|\alpha\rangle$ of the full system with equal weights $w_\alpha$. The full system size is $L=14$, and subregion $A$ has size $L_A=4,5,6,7$. The parameters in each panel are the same as those in each panel of the main text Fig. 3, namely, this figure is the log-log plot of the main text Fig. 3. In panels (a)-(c) where the magnetic fields are zero, the XYZ model is translationally invariant and known to be integrable. Within the system size studied here, we find $p_{A,n}$ decays approximately as $p_{A,n}\propto n^{-s}$, where the exponent $s\approx 1$. In panel (e) where there is a random magnetic field in the $z$ direction, the state is known to be in the MBL phase (see Fig. \ref{figS-SA} for entanglement entropy evidence), which has localized quasilocal conserved quantities. Accordingly, we find $p_{A,n}$ also decays approximately in power law as $p_{A,n}\propto n^{-s}$, but the exponent $s\approx 1.5\sim 2$ is larger than that in (a)-(c).
}
\label{figS-logp-logn}
\end{figure}

\begin{figure}[tbp]
\begin{center}
\includegraphics[width=6.5in]{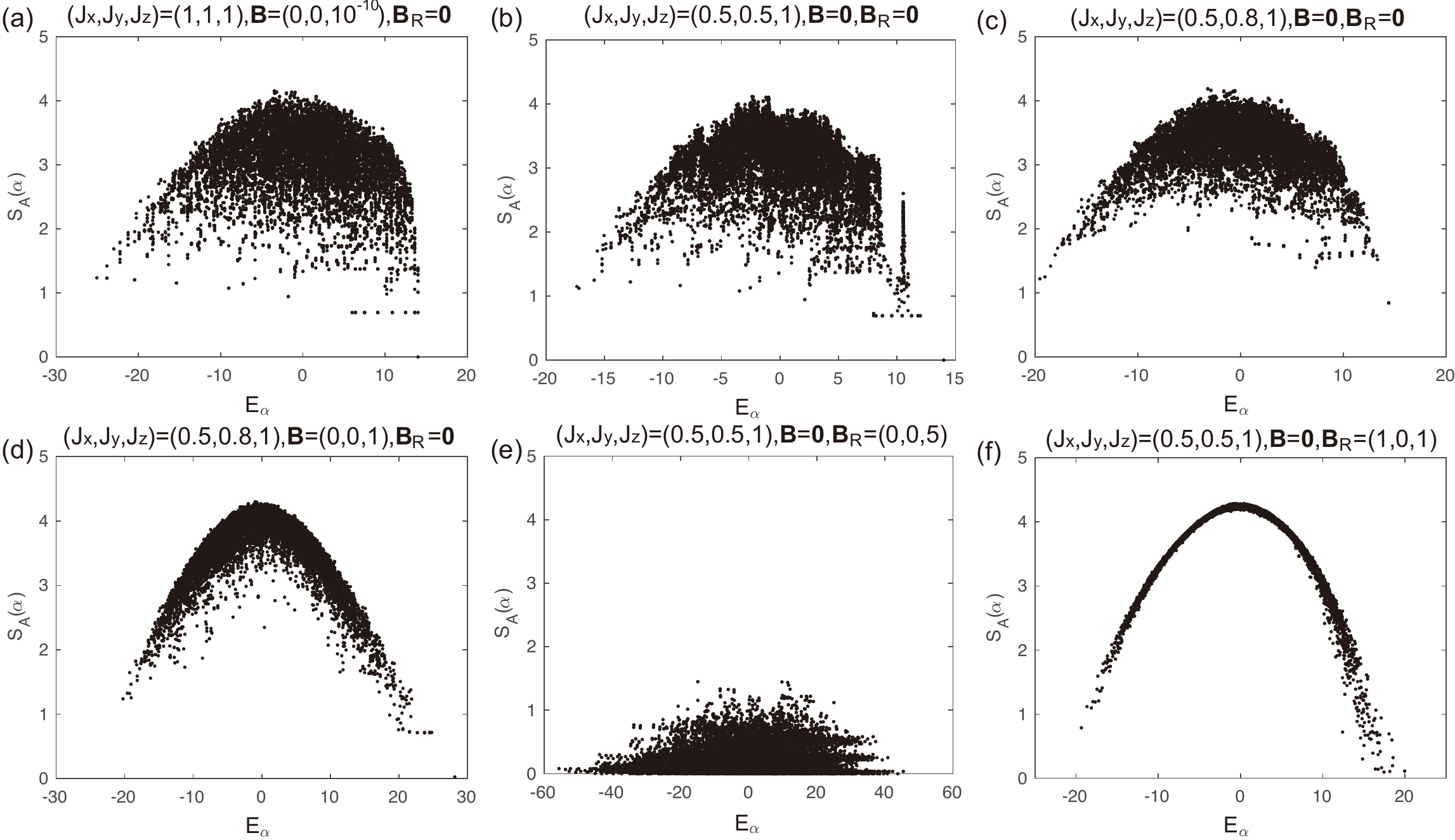}
\end{center}
\caption{The entanglement entropy of all the eigenstates, where the full system size $L=14$, and subregion $A$ has a size $L_A=7$. The parameters in each panel is chosen the same as those in each panel of the main text Fig. 3.
}
\label{figS-SA}
\end{figure}

We first briefly describe the properties of the six sets of parameters in the six panels of the main text Fig. 3 (a)-(f) (as well as Fig. \ref{figS-logp-logn} (a)-(f) and Fig. \ref{figS-SA} (a)-(f)):

(a) $(J_x,J_y,J_z)=(1,1,1), \mathbf{B}=(0,0,10^{-10}),\mathbf{B}_R=(0,0,0)$. This is the isotropic case with equal spin couplings in all directions, known as the XXX model. In our calculations, we have added a very small magnetic field in the $z$ direction to pin the energy eigenstates also into eigenstates of $\sum_{j}\sigma_{j,z}$, an obvious conserved quantity. The XXX model is known to be integrable (exactly solvable) by the Bethe ansatz, and a class of local and quasilocal conserved quantities have been derived in \cite{grabowski1995,ilievski2015}.

(b) $(J_x,J_y,J_z)=(0.5,0.5,1), \mathbf{B}=(0,0,0),\mathbf{B}_R=(0,0,0)$. This set of parameters with $J_x=J_y$ gives the XXZ model, which is also integrable, and has a class of local conserved quantities \cite{grabowski1995}.

(c) $(J_x,J_y,J_z)=(0.5,0.8,1), \mathbf{B}=(0,0,0),\mathbf{B}_R=(0,0,0)$. This is the generic XYZ model with three spin couplings unequal. The model is still integrable, and a class of local conserved quantities can be found \cite{grabowski1995}.

(d) $(J_x,J_y,J_z)=(0.5,0.8,1), \mathbf{B}=(0,0,1),\mathbf{B}_R=(0,0,0)$. This is the generic XYZ model in a uniform magnetic field $\mathbf{B}$. It is proved that local conserved quantities do not exist for such a model \cite{shiraishi2019}. However, this does not rule out the existence of quasilocal conserved quantities.

(e) $(J_x,J_y,J_z)=(0.5,0.5,1), \mathbf{B}=(0,0,0),\mathbf{B}_R=(0,0,5)$. This set of parameters give an XXZ model with a random magnetic field in the $z$ direction. This model is expected to be in the MBL phase when the random magnetic field $B_{R,z}$ is above a threshold. The MBL phase is argued to have numerous localized (quasi)local conserved quantities, making the system (approximately) integrable.

(f) $(J_x,J_y,J_z)=(0.5,0.5,1), \mathbf{B}=(0,0,0),\mathbf{B}_R=(1,0,1)$. This is the XXZ model with independent random magnetic fields in the $x$ and the $z$ direction. In this case, we find the model is fully chaotic: the level spacing statistics shows the Wigner-Dyson statistics of the gaussian orthogonal ensemble (GOE), and the entanglement entropy of all the eigenstates show a perfect volume law (Fig. \ref{figS-SA}(f)). Accordingly, we find only $p_{A,0}$ and $p_{A,1}$ are obviously nonzero (main text Fig. 3(f)), which correspond well to the only two local subregion conserved quantities of the trivial identity matrix $I_A$ and the subregion Hamiltonian $H_A$.

In the cases (a)-(d), as shown in Fig. \ref{figS-SA}(a)-(d), the majority eigenstates show a volume law entanglement entropy, and this is due to the existence of extended quasiparticle states in the system. In case (e) (Fig. \ref{figS-SA}(e)) where the system shows many-body localization, most eigenstates have small entanglement entropy due to the area law nature of the states. While in the fully chaotic case (f) (Fig. \ref{figS-SA}(f)), the eigenstates show perfect volume law entanglement entropies.

From the log-log plot of the EHSM eigenvalues $p_{A,n}$ vs. $n$ in Fig. \ref{figS-logp-logn}, we find that within the limited system size we studied, $p_{A,n}$ of integrable systems approximately decay in a power law as $p_{A,n}\propto n^{-s}$. For parameters in Fig. \ref{figS-logp-logn}(a)-(c) where the XYZ model is known to be analytically integrable, we find approximately $p_{A,n}\propto n^{-s}$, with the exponent $s\approx 1$. For Fig. \ref{figS-logp-logn}(e) which is in the MBL phase, we also see $p_{A,n}\propto n^{-s}$ for a considerable range of $n$, with $s\approx 1.5\sim 2$. More examples of MBL phase is shown in Fig. \ref{figS-MBL}, where we see that the decaying exponent $s$ has no obvious dependence on the parameters (generically around $s\approx 1.5\sim 2.5$), as long as the system is in the MBL phase. When the random magnetic field increases (Fig. \ref{figS-MBL}(b)), $p_{A,n}$ deviates more from the power-law decaying behavior, possibly because the system is closer to a non-interacting system (dominated by random fields). 

Overall, for interacting integrable models, within the small system sizes we studied, we find power-law decay is a good fit for the EHSM eigenvalues $p_{A,n}$. Enlarging the system size for interacting models is numerically difficult, and we leave the study of larger system sizes in the future.

In contrast, in Fig.  \ref{figS-logp-logn} (f) which is fully chaotic, the decaying behavior of $p_{A,n}$ clearly deviates from a simple power-law decay. In the main text Fig. 3, one can see that only $p_{A,0}$ and $p_{A,1}$ are large, and we find their eigen-operators approximately give the identity $I_A$ and subregion Hamiltonian $H_A$.

\begin{figure}[tbp]
\begin{center}
\includegraphics[width=6.8in]{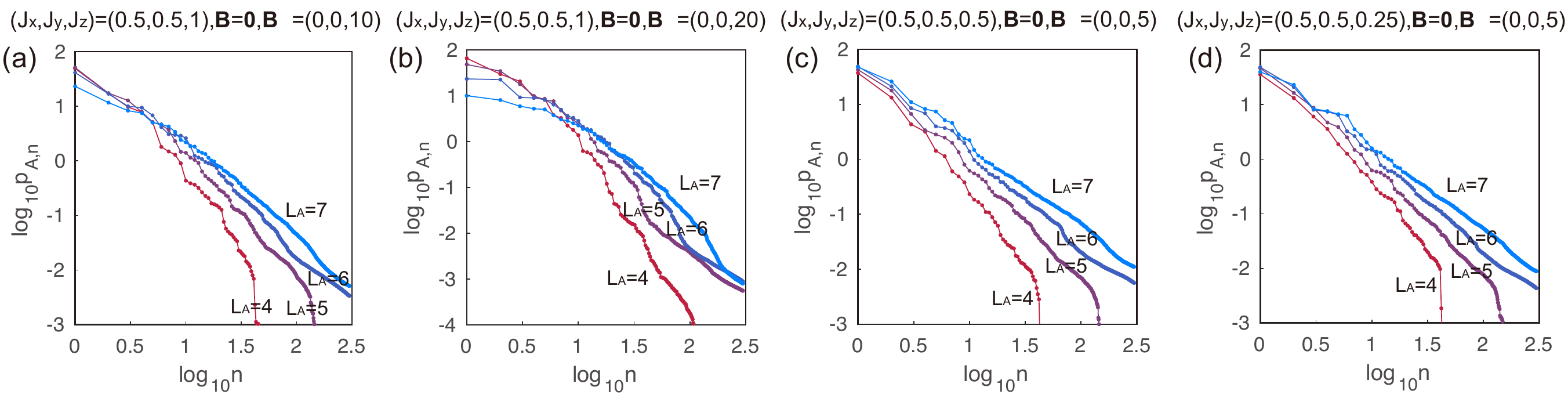}
\end{center}
\caption{More examples of log-log plot of the EHSM eigenvalues for systems in the MBL phase, where the full system size $L=14$, and subregion $A$ has a size $L_A=7$. The parameters are labeled in each panel, which are all in the MBL phase.
}
\label{figS-MBL}
\end{figure}

\subsection{How well the EHSM eigen-operators are conserved quantities}\label{app:XYZ-commu}

In this subsection, we test how well the EHSM eigenvectors (eigen-operators) $\overline{Q}_A^{(n)}$ (which are matrices in the Hilbert space) are conserved quantities in subregion A. To examine this, for each normalized eigen-operator $\overline{Q}_A^{(n)}$, we define a commutator-anticommutator ratio
\begin{equation}
r_A^{(n)}=\frac{\text{tr}\left( -\left[ \overline{Q}_A^{(n)}, H_A\right]^2 \right)}{\text{tr}\left( \left\{ \overline{Q}_A^{(n)}, H_A\right\}^2 \right)}\ ,
\end{equation}
where $H_A$ is the Hamiltonian in subregion $A$ as defined in main text Eq. (2), while $[A,B]=AB-BA$ and $\{A,B\}=AB+BA$ stand for commutator and anticommutator, respectively. For Hermitian operators $\overline{Q}_A^{(n)}$, one has $r_A^{(n)}\ge0$. If $r_A^{(n)}$ is close to zero, $\overline{Q}_A^{(n)}$ will be a good conserved quantity of subregion $A$.


In Tab. \ref{Tab-rn} below, we list the commutator-anticommutator ratio of the first 7 EHSM eigen-operators $\overline{Q}_A^{(n)}$ ($0\le n\le 6$) for the XYZ model with six groups of parameters given in the main text Fig. 3 (see also Fig. \ref{figS-SA}), where the total system size $L=14$ and subsystem size $L_A=7$. As we can see, all the ratios $r_A^{(n)}$ are close to zero, indicating they are indeed approximate subregion $A$ conserved quantities.

\begin{table}[htbp]
  \centering
  \begin{tabular}{c|c|c|c|c|c|c|c|c}
  \hline
Fig. 3 label & XYZ model parameters & $r_A^{(0)}$ &$r_A^{(1)}$  & $r_A^{(2)}$ & $r_A^{(3)}$ & $r_A^{(4)}$ & $r_A^{(5)}$ & $r_A^{(6)}$ \\
  \hline
(a) & $(J_x,J_y,J_z)=(1,1,1), \mathbf{B}=(0,0,10^{-10}),\mathbf{B}_R=(0,0,0)$  &0.0001 & 0.0014 & 0.0010 & 0.0067  &0.0356 & 0.0443 & 0.1615 \\
  \hline
(b) & $(J_x,J_y,J_z)=(0.5,0.5,1), \mathbf{B}=(0,0,0),\mathbf{B}_R=(0,0,0)$  &0.0001 & 0.0011 & 0.0011 & 0.0018  &0.0476 & 0.0804 & 0.0240 \\
  \hline
(c) & $(J_x,J_y,J_z)=(0.5,0.8,1), \mathbf{B}=(0,0,0),\mathbf{B}_R=(0,0,0)$  &0.0004 & 0.0012 & 0.0423 & 0.0356  &0.0799 & 0.0188 & 0.0314 \\
  \hline
(d) & $(J_x,J_y,J_z)=(0.5,0.8,1), \mathbf{B}=(0,0,1),\mathbf{B}_R=(0,0,0)$  &0.0001 & 0.0005 & 0.0010 & 0.0035  &0.0025 & 0.0145 & 0.0143 \\
  \hline
(e) & $(J_x,J_y,J_z)=(0.5,0.5,1), \mathbf{B}=(0,0,0),\mathbf{B}_R=(0,0,5)$  &0.0000 & 0.0001 & 0.0001 & 0.0002  &0.0005 & 0.0004 & 0.0002 \\
  \hline
(f) & $(J_x,J_y,J_z)=(0.5,0.5,1), \mathbf{B}=(0,0,0),\mathbf{B}_R=(1,0,1)$  &0.0002 & 0.0008 & 0.0048 & 0.0092  &0.0294 & 0.0166 & 0.0190 \\
  \hline
  \end{tabular}
  \caption{The commutator-anticommutator ratio $r_A^{(n)}$ for the leading $7$ EHSM eigen-operators $\overline{Q}_A^{(n)}$ of the XYZ model, where the parameters are as labeled in the panels (a)-(f) of main text Fig. 3 (also Fig. \ref{figS-SA}), and the full system and subsystem sizes are $L=14$ and $L_A=7$. }\label{Tab-rn}
\end{table}

\subsection{Extracted subregion conserved quantities for the XXZ model}\label{app:XYZ-Qn}

In this subsection, we discuss how the EHSM eigen-operators $\overline{Q}_A^{(n)}$ look like for the XXX model (main text Fig. 3(a)) and the XXZ model (main text Fig. 3(b)). Recall that the eigen-operators $\overline{Q}_A^{(n)}$ are sorted in the order of descending EHSM eigenvalues $p_{A,n}$ ($n\ge0$).

\begin{table}[htbp]
  \centering
  \begin{tabular}{c|c|c|c|c|c|c|c|c|c|c|}
  \hline
 Model parameter & \multicolumn{5}{c|}{XXX model (main text Fig. 3(a))} &  \multicolumn{5}{c|}{XXZ model (main text Fig. 3(b))} \\
  \hline
Overlap $\xi$ with & $\overline{Q}_A^{(0)}$ &$\overline{Q}_A^{(1)}$  & $\overline{Q}_A^{(2)}$ & $\overline{Q}_A^{(3)}$ & $\overline{Q}_A^{(4)}$ & $\overline{Q}_A^{(0)}$ &$\overline{Q}_A^{(1)}$  & $\overline{Q}_A^{(2)}$ & $\overline{Q}_A^{(3)}$ & $\overline{Q}_A^{(4)}$ \\
  \hline
  $I_A$  &-0.995 & 0.010 &-0.041 &0.040  &0.027 &0.989&0.047&-0.035&-0.088&-0.011  \\
  \hline
  $H_A$  & 0.043 &-0.026 & -0.963 & 0.061 & 0.091 &-0.077&0.895&-0.283&-0.210&0.082  \\
  \hline
 $\sum_{i}\sigma_{z,i}$  & 0.012 & 0.979 & -0.023 & 0.024 & -0.0156 &0.059&0.355&0.768&0.463&0.098 \\
  \hline
  $\sum_{i}\sigma_{x,i}$  & 0 & 0 &0 & -0.0002 & -0.0003 &0&0&0&0&0  \\
  \hline
 $\sum_{i}\sigma_{z,i}\sigma_{z,i+1}$  & 0.020 & -0.027 & -0.551 & 0.151 & -0.217 &-0.079&0.725&-0.113&-0.350&0.244 \\
  \hline
 $\sum_{i}\sigma_{x,i}\sigma_{x,i+1}$  &0.027 &-0.009 &-0.558  & -0.023 &0.187 &-0.015&0.372&-0.234&0.092&-0.144  \\
  \hline
 $\sum_{i}\sigma_{y,i}\sigma_{y,i+1}$ &0.027 &-0.009 &-0.558  & -0.022 &0.187  &-0.015&0.372&-0.234&0.092&-0.144 \\
  \hline
 $\sum_{i}\sigma_{z,i}\sigma_{z,i+2}$  &0.011 &-0.018 & 0.033 &0.516  &-0.152 &-0.039&-0.006&0.324&-0.515&-0.007 \\
  \hline
 $\sum_{i}\sigma_{x,i}\sigma_{x,i+2}$  &0.019 &-0.001 & 0.005 & 0.327 & 0.210 &0.006&-0.012&0.022&-0.051&0.192 \\
  \hline
  $\sum_{i}\sigma_{y,i}\sigma_{y,i+2}$ &0.019 &-0.001 & 0.005 & 0.327 & 0.210 &0.006&-0.012&0.023&-0.050&0.192  \\
  \hline
  $\sum_{i}\sigma_{z,i}\sigma_{z,i+3}$  &0.021 &-0.015 &0.022  &0.321 &-0.263 &-0.040&-0.024&0.251&-0.361&-0.265  \\
  \hline
  $\sum_{i}\sigma_{x,i}\sigma_{x,i+3}$  &0.027 &-0.0003 &-0.0007 &0.170  & 0.058 &-0.006&0.001&0.003&0.001&0.020 \\
  \hline
  $\sum_{i}\sigma_{y,i}\sigma_{y,i+3}$  &0.027 &-0.0003 &-0.0008 &0.170  & 0.058 &-0.006&0.001&0.003&0.001&0.020  \\
  \hline
  $\sum_{i}\sigma_{z,i}\sigma_{z,i+4}$  &0.008 &-0.014 &0.002  &0.261 &-0.236 &-0.017&-0.018&0.174&-0.264&-0.203  \\
  \hline
  $\sum_{i}\sigma_{x,i}\sigma_{x,i+4}$  &0.013 &-0.001 &-0.015  &0.143 &0.005 &0.003&0.003&0&-0.012&0.005  \\
  \hline
  $\sum_{i}\sigma_{y,i}\sigma_{y,i+4}$  &0.013 &-0.001 &-0.015  &0.143 &0.005 &0.003&0.003&0&-0.012&0.005  \\
  \hline
  $\sum_{i}\sigma_{z,i}\sigma_{z,i+1}\sigma_{z,i+2}\sigma_{z,i+3}$  & -0.022 &-0.001 & -0.010 & 0.067 &-0.099 &0.039&0.035&-0.006&-0.112&0.009  \\
  \hline
  $\sum_{i}\sigma_{z,i}\sigma_{x,i+1}\sigma_{x,i+2}\sigma_{z,i+3}$  &-0.004 &-0.0006 & -0.033 & -0.077 & -0.230 &-0.001&0.043&-0.038&0.032&-0.438 \\
  \hline
  $\sum_{i}\sigma_{z,i}\sigma_{y,i+1}\sigma_{y,i+2}\sigma_{z,i+3}$  &-0.004 &-0.0006 & -0.033 & -0.077 & -0.230 &-0.001&0.043&-0.038&0.032&-0.438 \\
  \hline
  $\sum_{i}\sigma_{x,i}\sigma_{z,i+1}\sigma_{z,i+2}\sigma_{x,i+3}$  & -0.004 &-0.0001 & -0.044 & -0.093 &-0.215 &0.002&0.017&-0.015&0.010&-0.080 \\
  \hline
  $\sum_{i}\sigma_{y,i}\sigma_{z,i+1}\sigma_{z,i+2}\sigma_{y,i+3}$  & -0.004 &-0.0001 & -0.044 & -0.093 &-0.215 &0.002&0.017&-0.015&0.010&-0.080 \\
  \hline
  $\sum_{i}\sigma_{x,i}\sigma_{y,i+1}\sigma_{y,i+2}\sigma_{x,i+3}$  & -0.004 &-0.0004 & -0.047 & -0.093 &-0.215 &0.001&0.031&-0.025&0.023&-0.172 \\
  \hline
  $\sum_{i}\sigma_{y,i}\sigma_{x,i+1}\sigma_{x,i+2}\sigma_{y,i+3}$  & -0.004 &-0.0004 & -0.047 & -0.093 &-0.215  &0.001&0.031&-0.025&0.023&-0.172 \\
  \hline
  $P_3$  & 0 & 0 & 0 & 0 & 0 &0&0&0&0&0  \\
  \hline
  $P^{\text{orth}}_4$  & 0.018 & -0.007 & -0.107 & 0.095 & -0.506 &-0.001&0.082&-0.075&0.066&-0.681  \\
  \hline
  \end{tabular}
  \caption{The overlap of EHSM eigen-operators $\overline{Q}_A^{(n)}$ with various operators in subregion $A$, where the model parameters are given by the main text Fig. 3(a) (the zero field XXX model with $(J_x,J_y,J_z)=(1,1,1)$) and the main text Fig. 3(b) (the zero field XXZ model with $(J_x,J_y,J_z)=(0.5,0.5,1)$), respectively. The full system size is $L=14$, and the subregion $A$ size is $L_A=7$.}\label{Tab-XXZ}
\end{table}

In Tab. \ref{Tab-XXZ}, we calculate the overlap between the numerical EHSM eigen-operators $\overline{Q}_A^{(n)}$ and various operators $M_A$ in subregion $A$, which is defined as
\begin{equation}
\xi(\overline{Q}_A^{(n)},M_A)=\frac{\text{tr}(\overline{Q}_A^{(n)}M_A)}{|| M_A ||}\ .
\end{equation}
Note that we have normalized $|| \overline{Q}_A^{(n)} ||=1$. The parameters are as defined in the main text Fig. 3(a) (the XXX model) and in the main text Fig. 3(b) (the XXZ model), in both cases the magnetic field is zero.

In particular, we examine the overlaps of $\overline{Q}_A^{(n)}$ with the known analytical local conserved quantities $P_n$ ($n=3,4$) \cite{tetelman1981,grabowski1995} generated by a boost operator $K$, as defined below. We first define the $3\times3$ matrix $J=\text{diag}(J_x,J_y,J_z)$. We can the rewrite the XYZ model without magnetic field Hamiltonian $H$ and define the Boost operator $K$ as
\begin{equation}
H=\sum_j  \bm{\sigma}_j\cdot (J\bm{\sigma}_{j+1})\ ,\qquad  K=\sum_j  j \bm{\sigma}_j\cdot (J\bm{\sigma}_{j+1})\ .
\end{equation}
Accordingly, a series of local conserved quantities are given by $P_3=c_3[K,H]$, and $P_n=c_n[K,P_{n-1}]$, where $c_n$ are only number factors which we choose for convenience. Note that $P_n$ is generically $n$-supported, namely, all the terms in $P_n$ are supported by no more than $n$ neighboring sites. Here we only study the first two conserved quantities derived in this way, which are explicitly
\begin{equation}\label{seq:P3}
P_3=\frac{1}{2}[K,H]=\sum_j  (J \bm{\sigma}_j)\cdot [\bm{\sigma}_{j+1}\times (J\bm{\sigma}_{j+2})]\ ,
\end{equation}
and
\begin{equation}\label{seq:P4}
\begin{split}
P_4=\frac{1}{4}[K,P_3] =&\sum_{j,\nu}\Big[\sum_\mu|\epsilon_{\mu\nu\lambda}|J_\mu J_\lambda \sigma_{j,\mu}\sigma_{\nu,j+1}(J_\mu\sigma_{\nu,j+2}\sigma_{\mu,j+3}-J_\nu \sigma_{\mu,j+2}\sigma_{\nu,j+3}) \\ 
&+\sum_{\mu\neq \nu} J_\mu^2J_\nu \sigma_{\nu,j}\sigma_{\nu,j+1} +J_xJ_yJ_z \sigma_{\nu,j-1}\sigma_{\nu,j+1} \Big]\ ,
\end{split}
\end{equation}
where $\epsilon_{\mu\nu\lambda}$ is the Levi-Civita symbol. 
We note that $P_3$ is orthogonal to $H_A$ ($\text{tr}(P_3H_A)=0$), but $P_4$ is not orthogonal to the physical Hamiltonian $H_A$, namely, $\text{tr}(H_A P_4)\neq0$. Therefore, we define a conserved quantity $P^{\text{orth}}_4$ orthogonal to $H_A$ as
\begin{equation}\label{seq-P4o}
P^{\text{orth}}_4=P_4-\frac{\text{tr}(P_4H_A)}{|| H_A ||^2}H_A\ .
\end{equation}
Besides, we have $\text{tr}(P_3P^{\text{orth}}_4)=0$.

To a good approximation, we find generically $\overline{Q}_A^{(0)}\propto I_A$ in all the cases. For the XXZ models shown in Tab. \ref{Tab-XXZ}, the first two nontrivial conserved quantities $\overline{Q}_A^{(1)}$ and $\overline{Q}_A^{(2)}$ are almost the linear combinations of the subregion Hamiltonian $H_A$ and the total $z$-direction spin $\sum_i\sigma_{z,i}$. We find the 3rd conserved quantity to be approximately
\begin{equation}
\overline{Q}_A^{(3)}\approx \sum_j\sum_{\ell\ge1}[\zeta_z(l)\sigma_{z,j}\sigma_{z,j+l}+\zeta_\perp(l)(\sigma_{x,j}\sigma_{x,j+l}+\sigma_{y,j}\sigma_{y,j+l})]+\zeta' \sum_j\sigma_{z,j}\ ,
\end{equation}
where $\zeta_{z}(l)$ and $\zeta_{\perp}(l)$ decay as $l$ grows, and $\zeta'$ is some constant. For the example of the XXX model, $\zeta'\approx0$. The 4th conserved quantity $\overline{Q}_A^{(4)}$ is dominated by 4-support operators. Accordingly, it has a major overlap with the 4-supported local conserved quantity $P_4^{\text{orth}}$ in Eq. (\ref{seq-P4o}).

In particular, we note that none of the conserved quantities $\overline{Q}_A^{(n)}$ have a nonzero overlap with the local conserved quantity $P_3$.

\section{Fitting the behaviors of EHSM spectra}
We have shown that the free fermion EHSM spectra have a cutoff in $n$ where $p_{A,n}$ vanishes, while for the (small size) integrable interacting XYZ model, the EHSM spectra decays exponentially without a clear cutoff. This might be because unlike free models where the entanglement Hamiltonians are single-body terms, interacting models allow more many-body terms in their entanglement Hamiltonians, which we leave for the future studies. In this section, we show that these EHSM spectra decaying behaviors fit certain probability distributions of the weights $\overline{\beta}^{(n)}_A(\alpha)$ of conserved quantities in the entanglement Hamiltonians.

We first rewrite the main text Eq. (7) in terms of a set of Frobenius orthonormal eigen-operators $\overline{Q}_A^{(n)}$ into
\begin{equation}
H_{E}^A(\alpha)=\sum_{n} \overline{\beta}^{(n)}_A(\alpha) \overline{Q}_A^{(n)}\ .
\end{equation}
For states $|\alpha\rangle$ in an ensemble $\Xi$ (with uniform weights $w_\alpha=\frac{1}{N_\Xi}$), we assume $\overline{\beta}^{(n)}_A(\alpha)$ satisfies a Gaussian random distribution with mean value $\beta_{0}^{(n)}$ and standard deviation $\sigma^{(n)}$, namely,
\begin{equation}
\langle \overline{\beta}^{(n)}_A(\alpha)\rangle_\Xi=\beta_{0}^{(n)}\ ,\qquad \sqrt{\langle \left(\overline{\beta}^{(n)}_A(\alpha)-\beta_{0}^{(n)}\right)^2\rangle_\Xi }=\sigma^{(n)}\ .
\end{equation}
We can then calculate the EHSM eigenvalues $p_{A,n}$ of such an ensemble, which can be easily obtained by diagonalizing the correlation matrix defined in Eq. (\ref{seq-SA}), which has matrix elements here $K _{A,\alpha\alpha'}=\frac{1}{N_\Xi N_A} \sum_n \overline{\beta}^{(n)}_A(\alpha)  \overline{\beta}^{(n)}_A(\alpha')$. 

Numerically, we find the mean value $\beta_0^{(n)}$ does not qualitatively affect the behavior of the EHSM eigenvalues $p_{A,n}$ for $n>0$, but only mainly affect the value of $p_{A,0}$. Since we are interested in the decaying behavior of $p_{A,n}$ at $n>0$, hereafter we shall set $\beta_0^{(n)}=0$.

\subsection{Free fermions}\label{sec:fitting-r-free}

First, we find the free fermion EHSM spectra near their cutoffs in the main text Fig. 2 can be roughly fitted by a random distribution of $\overline{\beta}^{(n)}_A(\alpha)$ with standard deviations
\begin{equation}\label{seq-free-deviation}
\sigma^{(n)}= \sigma^{(0)}\left(1-\frac{n}{zL_A}\right)^r\ ,\qquad (r\ge 0)
\end{equation}
where $zL_A$ is the number of nonzero $p_{A,n}$ in the EHSM spectra ($z$ is an order $1$ number). Fig. \ref{figS-freefit} (a)-(c) show three examples of EHSM spectra for $zL_A=100$, $N_\Xi=1000$ and standard deviations in Eq. (\ref{seq-free-deviation}), where the exponent $r=0,0.5$ and $1$, respectively. The most prominent feature is that the eigenvalues $p_{A,n}\propto (1-\frac{n}{zL_A})^{2r}$ near $n=zL_A$ (see for instance, the inset of Fig. \ref{figS-freefit}) (c). By comparing with the free fermion EHSM spectra in main text Fig. 2, we find the situations fit roughly with the following parameters:

\begin{figure}[tbp]
\begin{center}
\includegraphics[width=6.8in]{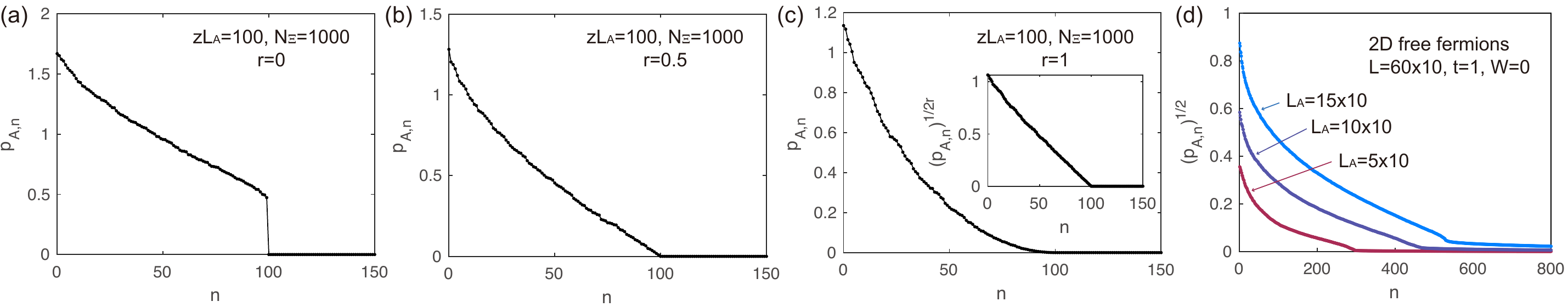}
\end{center}
\caption{(a)-(c) The EHSM spectrum for $\overline{\beta}^{(n)}_A(\alpha)$ with a standard deviation given by Eq. (\ref{seq-free-deviation}), where the cutoff of nonzero $p_{A,n}$ is set as $zL_A=100$, and the exponent $r=0,0.5$ and $1$ in (a)-(c), respectively. The inset in (c) shows that $(p_{A,n})^{1/2r}$ is linear in $(1-\frac{n}{zL_A})$ near $n=zL_A$. (d) $(p_{A,n})^{1/2}$ plotted versus $n$ for the 2D extended fermions with parameters given in the main text Fig. 2(a). The linear dispersion near the cutoff $p_{A,n}$ indicates an effective exponent $r=1$. 
}
\label{figS-freefit}
\end{figure}

(i) For delocalized fermions in $D=1,2$ dimensions, the exponent $r\approx 0.5D$. In the 1D case as shown in the main text Fig. 2(c), we find $r\approx 0.5$, as $p_{A,n}$ is roughly linear in $(1-\frac{n}{zL_A})$ as $n$ approaches the cutoff $zL_A$ (similar to Fig. \ref{figS-freefit}(b)), with $z=3$ in 1D. In the 2D case shown in main text Fig. 2(a), $r\approx 1$, which can be seen more clearly from Fig. \ref{figS-freefit}(d), where $(p_{A,n})^{1/2}$ is roughly linear in $(1-\frac{n}{zL_A})$ near the cutoff at $n=zL_A$. The cutoff in 2D is roughly at $z\approx 5$ if the subregion has $L_{A,x}\lesssim L_{A,y}$ ($L_{A,x}$ is the size perpendicular to the subregion boundary), while is reduced towards $z\rightarrow 3$ when $L_{A,x}> L_{A,y}$ which is more 1D-like.

(ii) For localized fermions in $D=1,2$ dimensions, we find the exponent is roughly $r\approx 0.5 (D-1)$, and the cutoff is at $z=1$ independent of the spatial dimension. In 1D (main text Fig. 2(d)), the sharp edge of $p_{A,n}$ dropping towards zero resembles the spectra in Fig. \ref{figS-freefit}(a), suggesting $r\approx 0$ in the strong disorder limit. In 2D shown in the main text Fig. 2(b), $p_{A,n}$ tends to zero linearly near $n=L_A$, which indicates an exponent $r=0.5$.

\subsection{Interacting integrable XYZ models}\label{sec:fitting-s-XYZ}

We now turn to the fitting of the EHSM spectrum of the integrable interacting XYZ models, which has a power-law decaying behavior $p_{A,n}\propto n^{-s}$. This behavior is well-fitted by assuming the standard deviation of $\overline{\beta}^{(n)}_A(\alpha)$ is also power-law decaying:
\begin{equation}\label{seq-interaction-deviation}
\sigma^{(n)}=\sigma^{(0)}n^{-s/2}\ ,\qquad (n>0, s>0)\ .
\end{equation} 
In Fig. \ref{figS-pfit} (a) and (b), we have plotted the EHSM spectrum for the $\overline{\beta}^{(n)}_A(\alpha)$ standard deviations given by Eq. (\ref{seq-interaction-deviation}) with the exponent $s=1$ and $s=2$, respectively. In the inset log-log plots, one can clearly see that $p_{A,n}$ decays in power law as $n^{-s}$. Such a behavior is the same as the actual EHSM spectra calculated for the integrable XYZ model parameters (see Fig. \ref{figS-logp-logn}).

\begin{figure}[tbp]
\begin{center}
\includegraphics[width=3.5in]{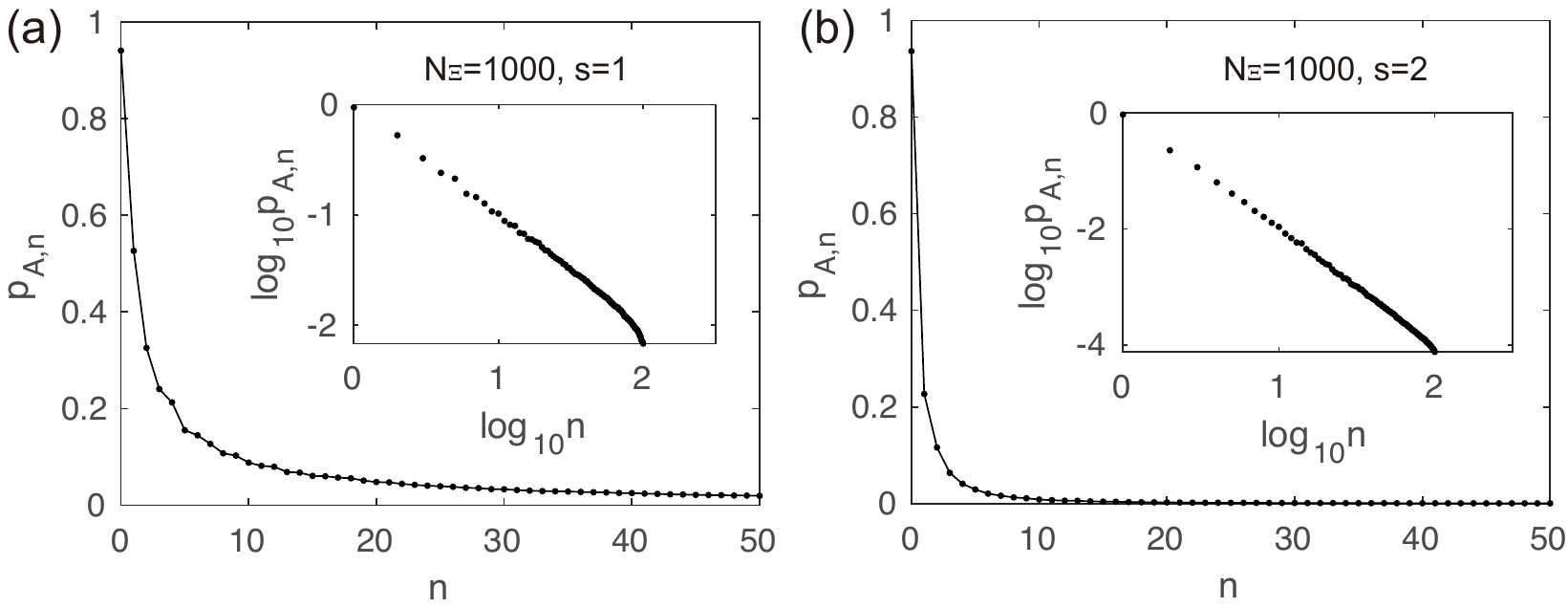}
\end{center}
\caption{The EHSM spectrum by assuming $\overline{\beta}^{(n)}_A(\alpha)$ has a standard deviation given by Eq. (\ref{seq-interaction-deviation}), where the exponent $s=1$ in (a) and $s=2$ in (b). From the inset log-log plot, it is clear that the spectrum $p_{A,n}$ is power-law decaying as $n^{-s}$.
}
\label{figS-pfit}
\end{figure}

\section{EHSM from time-evolution of non-eigenstates}\label{sec:Tstate}

In some cases, one only knows the time evolution $|\psi(\tau)\rangle$ of a non-eigenstate $|\psi(0)\rangle$ from time $0$ to $T$. If the time $T$ is long enough, one may choose a set of random energies $\{\widetilde{E}_\alpha\}$ ($\alpha\in\Xi$), and define a set of  approximate ``eigenstates"
\begin{equation}\label{seq-T-state}
|\widetilde{\alpha}\rangle_T=\frac{1}{\mathcal{N}_T(\alpha)}\int_0^{T}d\tau e^{i\widetilde{E}_{\alpha}\tau}|\psi(\tau)\rangle\ ,
\end{equation}
where $\mathcal{N}_T(\alpha)$ is the normalization factor. An exact energy eigenstate $|\alpha'\rangle$ of energy $E_{\alpha'}$ are expected to have an overlap $|\langle \alpha'|\widetilde{\alpha}\rangle|^2 \propto \frac{\sin^2[(\widetilde{E}_{\alpha}-E_{\alpha'})T/2]}{(\widetilde{E}_{\alpha}-E_{\alpha'})^2}$. Therefore, as $T\rightarrow\infty$, the state $|\widetilde{\alpha}\rangle_T$ is expected to resemble the energy eigenstate with energy closest to $\widetilde{E}_\alpha$. We can then calculate the EHSM spectrum for the set of states $|\widetilde{\alpha}\rangle_T$ generated from $|\psi(\tau)\rangle$.

Here we do the calculations for the 1D XYZ model with the same parameters $L=14,L_A\le7$ as studied in the main text Fig. 3, and we choose the initial state $|\psi(0)\rangle$ to be a tensor product state of randomly chosen on-site spin states. We then generate $N_{\Xi}=500$ random energies $\widetilde{E}_{\alpha}$ within the energy range of the model's spectrum, and diagonalize the EHSM of states $|\widetilde{\alpha}\rangle_T$ in Eq. (\ref{seq-T-state}). 

Fig. \ref{figS-T-evolution} shows the EHSM for different time period $T$ (from $10$ to $10000$) starting from the same the initial state $|\psi(0)\rangle$, where the parameters are in the extended integrable phase ((a)-(e)), in the MBL phase ((f)-(j)), and in the fully chaotic phase ((k)-(o)), respectively. We find that the EHSM spectrum stabilizes as $T\sim 100$. As expected, in the integrable cases ((a)-(j)), a power-law tail of the EHSM eigenvalues exist, while in the fully chaotic case ((k)-(o)), only $p_{A,0}$ and $p_{A,1}$ are significantly nonzero.

However, we note that the EHSM eigen-operators obtained in this way are less mutually commuting than those obtained from exact eigenstates. This is because $|\widetilde{\alpha}\rangle_T$ are only approximate eigenstates, for which the entanglement Hamiltonians would be less commuting with the physical Hamiltonian. Unless $T$ approaches the order of the Hilbert space dimension $N=d^L$ (here $N=2^{14}\approx 16000$), the states $|\widetilde{\alpha}\rangle_T$ would not be able to reproduce accurate enough subregionally (quasi)local conserved quantities. The only exception is the physical Hamiltonian, which emerge as the first nontrivial subregionally (quasi)local conserved quantity pretty accurately at small $T$ ($\gtrsim 10$).

In reality, the time evolution of non-eigenstates may be numerically calculated less costly by trotterization (i.e., by dividing time $T$ into small steps). This may provide a more efficient way to observe the behavior of the EHSM spectrum, which we showed in Fig. \ref{figS-T-evolution} requires less time $T$. However, for the recovery of subregionally conserved quantities, $T\sim d^L$ might be required, for which the error of trotterization will become large.

\begin{figure}[tbp]
\begin{center}
\includegraphics[width=6.8in]{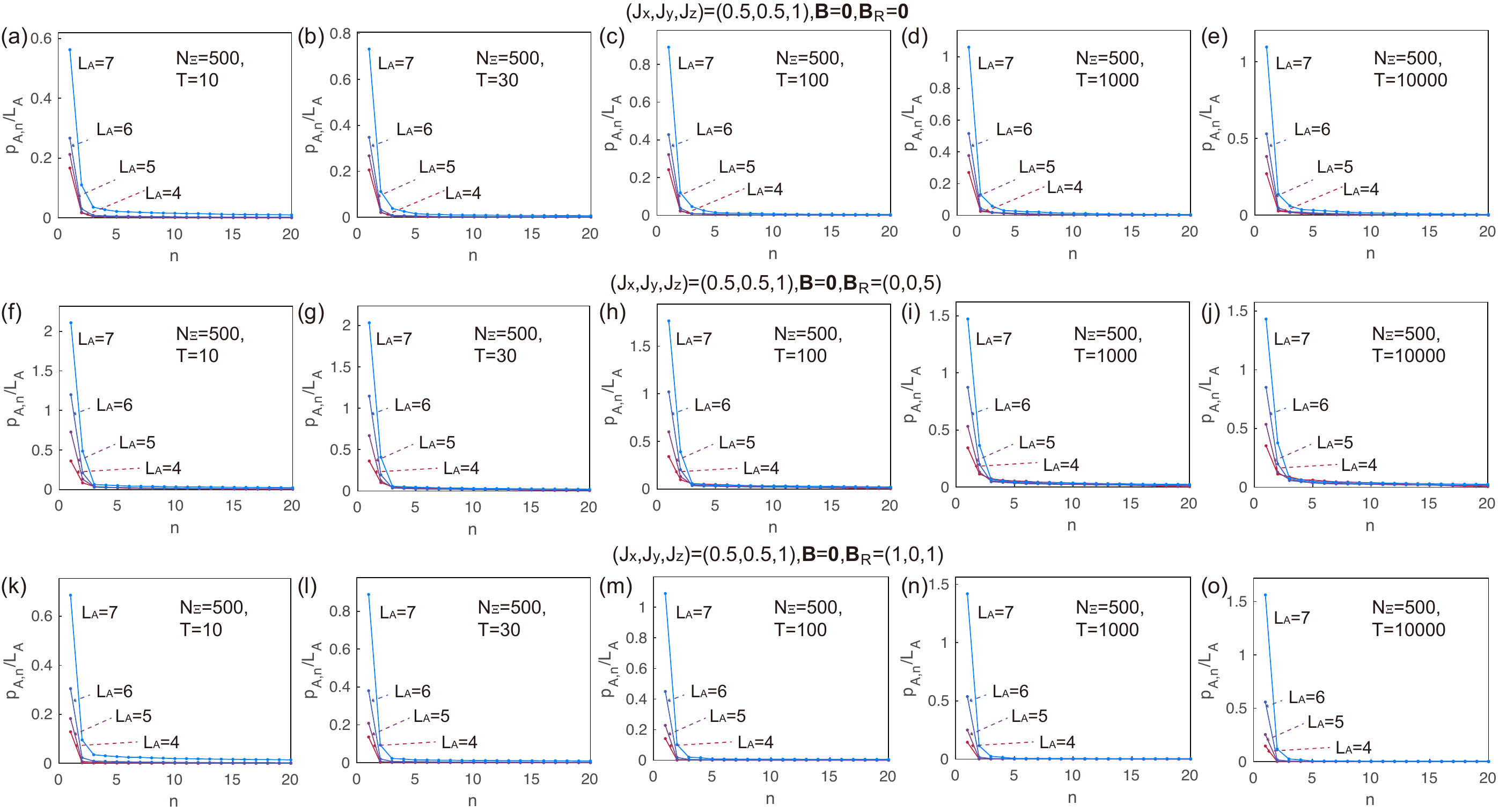}
\end{center}
\caption{The EHSM spectrum of the 1D XYZ model calculated from the approximate ``eigenstates" in Eq. (\ref{seq-T-state}) by the time-evolution of some non-eigenstate. The time $T$ varies from $10$ to $10000$. The initial state is chosen to be a tensor product of random local spins. The model parameters are given on top of each row and are the same within each row. Row 1 ((a)-(e)) is in the extended integrable phase, row 2 ((f)-(j)) is in the MBL phase, and row 3 ((k)-(o)) is in the fully chaotic phase. The number of random energies $\widetilde{E}_\alpha$ is $N_\Xi=500$, and $\widetilde{E}_\alpha$ are randomly chosen within an energy window smaller than the energy range of the model.
}
\label{figS-T-evolution}
\end{figure}

\end{widetext}

\bibliography{ent_ref}

\end{document}